\documentclass[onecolumn,a4paper]{IEEEtran}
 
\usepackage{amsmath,amssymb,color,graphicx,caption,subcaption,comment,tikz,url,latexsym} %amsthm
\usepackage{pgfplots}
\usepackage{verbatim}
\usetikzlibrary{decorations.markings}
\usepackage{cite}
\usepackage{enumerate}
\usepackage{hyperref}

\usepgfplotslibrary{units}
\usetikzlibrary{spy,backgrounds}
\usepackage{pgfplotstable}
\newtheorem{theorem}{Theorem}
\newtheorem{corollary}{Corollary}[theorem]
\newtheorem{lemma}[theorem]{Lemma}
\newtheorem{proposition}[theorem]{Proposition}
\newtheorem{definition}{Definition}

\newtheorem{remark}{Remark}

\newcommand{\mbf}[1]{\mathbf{#1}}
\newcommand{\set}[1]{\mathcal{#1}}

\renewcommand{\Pr}{\mathbb{P}}

\newcommand{\M}{{M}}
\renewcommand{\d}{\mbf{d}}
\newcommand{\Pe}{{\mathsf{P}_\text{e}}}
\newcommand{\R}{R}
\newcommand{\F}{{F}}

\newcommand{\Ss}{\mathcal{S}  \backslash \{s\}}

\newcommand{\w}{\textnormal{w}}
\newcommand{\s}{\textnormal{s}}
\newcommand{\Kw}{\mathcal{K}_\w}
\newcommand{\Ks}{\mathcal{K}_\s}

\newcommand{\rt}[1]{{\color{black}{#1}}}
\newcommand{\mw}[1]{{\color{black}{#1}}}
\newcommand{\ssb}[1]{{\color{black}{#1}}}

\allowdisplaybreaks[4] % Allow align environment to page break.
\begin{document}
\title{Noisy Broadcast Networks with Receiver Caching}
%\title{Cache-Aided Broadcast Channels}

\author{Shirin Saeedi Bidokhti, Mich\`{e}le Wigger and Roy Timo
\thanks{
S.~Saeedi~Bidokhti is with the Department of Electrical Engineering at Stanford University, saeedi@stanford.edu.  S.~Saeedi~Bidokhti is supported by the Swiss National Science Foundation fellowship no. 158487.} 
\thanks{M.~Wigger is with  CNRS, LTCI,  Telecom ParisTech, Universit\'e Paris-Saclay, 75013 Paris, michele.wigger@telecom-paristech.fr}
\thanks{R.~Timo is with the Institute for Communications Engineering at the Technical University of Munich, roy.timo@tum.de. R.~Timo is supported by the Alexander von Humboldt Foundation.}
\thanks{\mw{Parts of the material in this paper have been presented at the  \emph{IEEE International Symposium on Wireless Communications Systems (ISWCS)}, Bruxelles, Belgium, August 2015 \cite{timowigger-2015-1}, and will be presented at the \emph{2016 IEEE International Symposium on Information Theory}, Barcelona, Spain \cite{saeeditimowiggerisit} and at the \emph{International Symposium on Turbo Codes \& Iterative Information Processing}, Brest, France, September~2016} \cite{saeediwiggertimoturbo}.}
}

\maketitle

\begin{abstract}
\ssb{
We study noisy broadcast networks with local cache memories at the receivers, where the transmitter can pre-store information even before learning the receivers' requests. We mostly focus on packet-erasure broadcast networks with two \ssb{disjoint} sets of receivers: a set of weak receivers with all-equal erasure probabilities and  equal cache sizes and a set of strong receivers with all-equal erasure probabilities and no cache memories. We present lower and upper bounds on the \emph{capacity-memory tradeoff} of this network. The lower bound is achieved by a new   \emph{joint cache-channel coding} idea and significantly improves \ssb{on schemes that are} based on separate cache-channel coding.    We discuss how this coding idea could be extended to  more general discrete memoryless broadcast channels and to unequal cache sizes. \mw{Our upper bound holds  for all stochastically degraded broadcast channels.}}

\mw{For the described packet-erasure broadcast network, our lower and upper bounds are tight when there is a single weak receiver (and any number of strong receivers) and the cache memory size does not exceed a given threshold.  When there are a single weak receiver, a single strong receiver, and two files, then we can strengthen our upper and lower bounds so as they coincide  over a wide regime of cache sizes.} 

Finally, we completely characterise the rate-memory tradeoff for general discrete-memoryless broadcast channels with arbitrary cache memory sizes and arbitrary (asymmetric) rates when all receivers always demand exactly the same file. 
%(Here again, it is assumed that the demand is unknown during the cache placement.) The achievability follows again by a joint cache-channel coding scheme.

\end{abstract}

%%%%
%%%%
%%%%
%%%%

\section{Introduction} 

We address a one-to-many broadcast \rt{communications problem where many users demand files} from a single server during \emph{peak-traffic} times --- \rt{periods of high network congestion.} To improve network performance, the server can pre-place information in local cache memories \rt{near users at the network edge.} \rt{This pre-placement of information is called the \emph{caching communications phase}, and it occurs during off-peak times when the communications rate is not a limiting network resource.} The server typically does not know in advance  which files the users will demand, so it can try to cache information that is likely to be useful for many users during the \emph{delivery communications phase} (the peak-traffic time when the users demand files from the server). \rt{For example, researchers at Huawei Laboratories~\cite{Bastug-Feb-2016-A} recently used machine learning techniques to predict user behavior and proactively cache data to improve user request satisfaction ratios and reduce backhaul loads during the delivery phase.} 

The above caching problem is \rt{particularly} relevant to video-streaming services in mobile networks. Here network operators pre-place information in clients' caches (or, on servers near the clients) to improve latency and throughput during peak-traffic times. The network operator does not know in advance which movies the clients will request, and thus the cached information cannot depend on the clients' specific demands. It is now widely expected that there will be a nine-fold increase in mobile data traffic by 2020, and around 60 percent of this traffic will be mobile video~\cite{Ericsson-Mobility-Report}. Smart data caching strategies, new bandwidth allocations, reduced cell sizes and new radio-access technologies will all be needed to meet these growing demands~\cite{Andrews-June-2014-A}. 

\rt{The information-theoretic aspects of cache-aided} communications have received significant attention in recent years~\cite{maddahali_niesen_2014-1, chenfanletaief-2014, mohammadi_gunduz-2016, wan_tuninetti_piantanida-2016, wan_tuninetti_piantanida-2015,  wang_lim_gastpar_2015,ji_tulino_llorca_caire_2015, sengupta_tandon_clancy-2015, tian-2015,  ghasemi_ramamoorthy, wanglimgastpar-2016, huang_wang_ding_yang_zhang_2015, wang_xian_liu_2015, ugurawansezgin-2015, hassanzadeherkipllorcatulino-2015, ghorbelkobayashiyang-2015, zhangelia-2015,maddahaliniesen-2015-1, parksimeoneshamai-2016, azarisimeonespagnolinitulino-2016,naderializadehmaddahaliavestimehr-2016,shariatpanahimotaharikhalaj-2015, jicairemolish-2014, golrezaeietal-2013,sahraeigastpar-2016, hachemkaramchandanidiggavi-2015, hachemkaramchandanidiggavi-2014,liulau-2015,baszczyszyngiovanidis-2014,Timo-Mar-2016-C}. Maddah-Ali and Niesen \cite{maddahali_niesen_2014-1} considered a one-to-many broadcast problem where the receivers have independent caches of equal sizes and the delivery phase takes place \rt{over a noiseless broadcast communications link}. \ssb{They showed} that a smart design of the cache contents \rt{enables the server to send} coded (XOR-ed) data during the delivery phase that can simultaneously \rt{meet the demands of} multiple receivers with a single transmission. This \emph{coded caching scheme}, by simultaneously satisfying user demands, allows the server to reduce the delivery rate beyond the obvious \emph{local caching gain} (the data rate that each receiver can immediately retrieve from its cache without using coded caching). Intuitively, this \rt{performance improvement} occurs because the receivers can profit from other receivers' caches, and was thus termed \cite{maddahali_niesen_2014-1} \emph{global caching gain}. 

\ssb{Improved caching and delivery strategies for the Maddah-Ali and Niesen model were presented in~\cite{chenfanletaief-2014, mohammadi_gunduz-2016, wan_tuninetti_piantanida-2016}.  Fundamental converse (lower) bounds on the total required delivery rate \rt{were} presented in  \cite{maddahali_niesen_2014-1, tian-2015, ghasemi_ramamoorthy, sengupta_tandon_clancy-2015,  wang_lim_gastpar_2015, ji_tulino_llorca_caire_2015, wanglimgastpar-2016}. 
It was shown in \cite{wan_tuninetti_piantanida-2015} that the coded caching scheme of Maddah-Ali and Niesen is optimal among schemes that have uncoded caching placement when there are more files than users.

The Maddah-Ali and Niesen model  considers a \emph{worst-case scenario}, meaning that the goal is to satisfy all possible user demands. The caching problem has also been studied in \emph{average-case scenarios}  \cite{ji_tulino_llorca_caire_2015, wang_lim_gastpar_2015, wanglimgastpar-2016} where the receivers' demands follow a given probability distribution and the delivery rate is averaged over this demand distribution.}

%The lower bounds in \cite{maddahali_niesen_2014-1,chenfanletaief-2014, ghasemi_ramamoorthy,sengupta_tandon_clancy-2015,tian-2015} apply to \emph{worst-case scenarios} where the delivery rate needs to suffice for all possible receiver demands. The lower bounds in  \cite{ji_tulino_llorca_caire_2015, wang_lim_gastpar_2015, wanglimgastpar-2016} are derived for \emph{average-case scenarios} (and also apply to worst-case scenarios).% where the receivers' demands follow a given probability distribution and the delivery rate is averaged over this demand distribution. 

 In contrast to~\cite{maddahali_niesen_2014-1}, we \rt{will} assume \rt{in this paper} that the delivery phase  takes place over a noisy broadcast channel (BC), \rt{and we will see that} further global caching gains can be achieved by \emph{joint cache-channel coding}. Intuitively, when the BC is noisy the cache content not only determines \emph{what} to transmit but also \emph{how} to transmit it.  %\rt{Other works~\cite{Timo-Mar-2016-C,wang_lim_gastpar_2015} on data caching have taken a source coding perspective, where the files have a probabilistic dependence and are reconstructed with respect to a fidelity criterion.} 
 
\rt{We will focus on packet erasure broadcast channels that provide a first order model\footnote{Here we can assume that bit-level errors within a packet are handled on a link-by-link basis using physical layer error-correction techniques, and packets arrive at the users promptly or are lost due to, for example, buffer overflows.} of packet losses in congested networks. The importance of including a noisy} channel model for the delivery phase was also observed in \cite{timowigger-2015-1, huang_wang_ding_yang_zhang_2015, wang_xian_liu_2015, ugurawansezgin-2015, hassanzadeherkipllorcatulino-2015, ghorbelkobayashiyang-2015,zhangelia-2015,maddahaliniesen-2015-1, parksimeoneshamai-2016, azarisimeonespagnolinitulino-2016,naderializadehmaddahaliavestimehr-2016,liulau-2015}. For example, \cite{ghorbelkobayashiyang-2015} and \cite{zhangelia-2015} illustrate interesting \rt{interplay} between feedback or channel state information with caching, and \cite{maddahaliniesen-2015-1} and \cite{naderializadehmaddahaliavestimehr-2016} show that   caches at the transmitter-side and receiver-side allow for load-balancing and interference mitigation  in noisy interference networks. The works in \cite{ghorbelkobayashiyang-2015, zhangelia-2015, maddahaliniesen-2015-1,naderializadehmaddahaliavestimehr-2016} focus on the high signal-to-noise ratio regime. 

Our main interest in this paper is \rt{to characterize some of the fundamental} \emph{rate-memory tradeoffs} \rt{for cache-aided broadcast networks}; \rt{that is,} \rt{we wish to determine the set of} rates at which messages can be reliably communicated \rt{for given} cache sizes. We focus on a  worst-case (worst-demand) setup and consider two different communication scenarios.

\begin{figure}[h!]
\begin{center}
\includegraphics[width=0.49\textwidth]{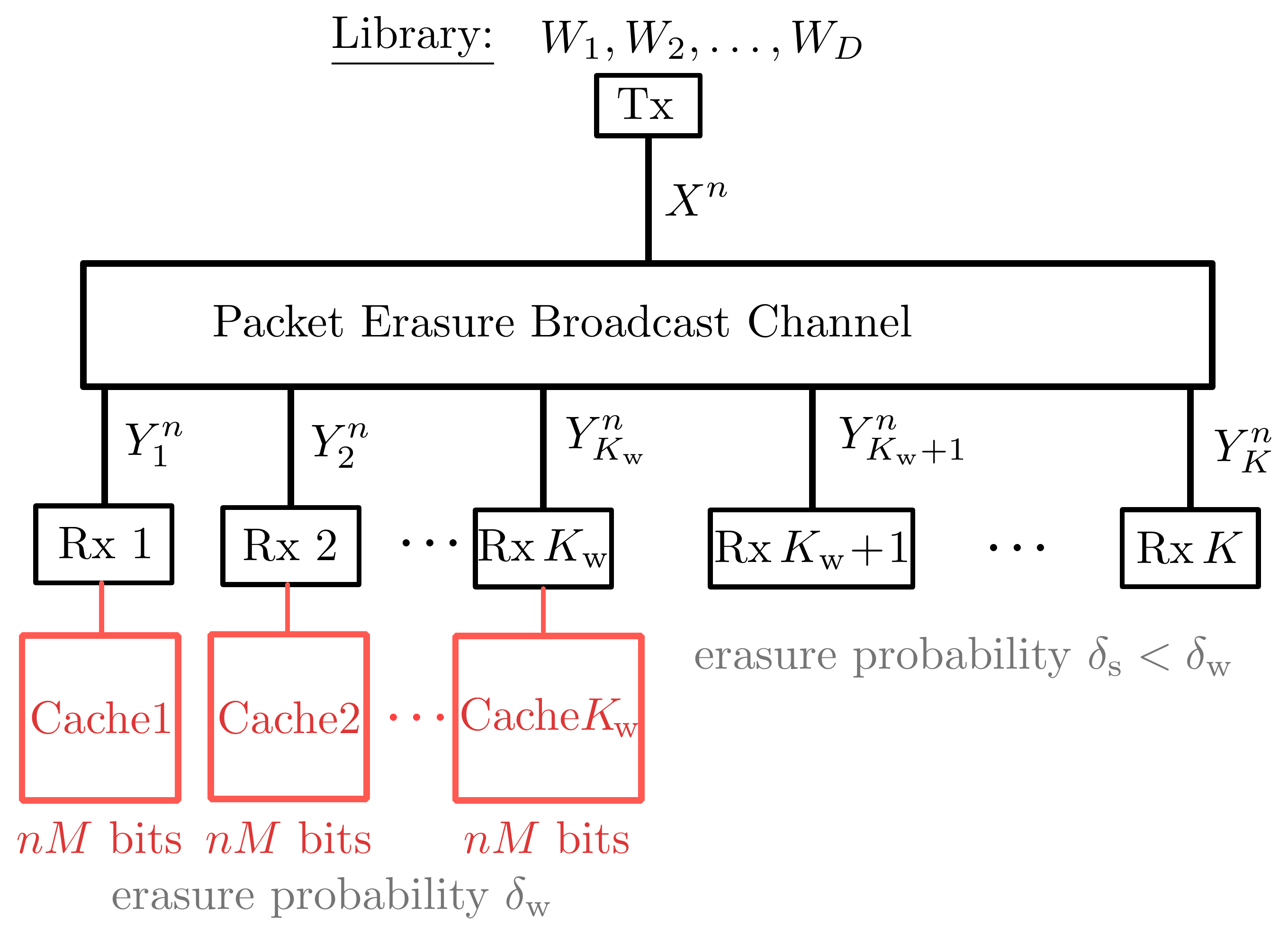}
\caption{$K$ user packet-erasure BC with $K_{\w}$ weak and $K_{\s}$ strong receivers and where the weak receivers have \ssb{cache memories}.}
\label{fig:general_model}
\end{center}
\end{figure}

\emph{Scenario 1:} \rt{We assume} that the receivers' demands \rt{are} arbitrary\footnote{\rt{Each user can choose any file from the server.}} \rt{and the} messages are \rt{all of equal rate}. We focus on the packet-erasure BC \rt{illustrated} in Figure~\ref{fig:general_model}, \rt{and divide the $K$ receivers into two sets: 
\begin{itemize}
\item A set of $K_\w$ \emph{weak receivers} with equal ``large'' BC erasure probabilities $\delta_\w > 0$. These receivers are each equipped with an individual cache of equal memory $M$. 
\item A set of $K_\s = K - K_\w$ \emph{strong receivers} with equal ``small''  BC erasure probabilities $\delta_\s \geq 0$ with $\delta_\s\leq \delta_\w$. These receivers are not provided with caches. 
\end{itemize} 
}
This scenario is motivated by previous studies~\cite{timowigger-2015-1,huang_wang_ding_yang_zhang_2015} \rt{that} showed \rt{the benefit of prioritizing cache placements near weaker receivers}. In practical systems, this means that \rt{telecommunications} operators with a limited number of caches might first place caches at houses that are further away from an optical fiber access point. \rt{Or, they} might place caches at \rt{pico or femto} base stations \rt{in heterogenous networks} that are located in areas with notoriously bad coverage. \rt{Scenario 1 also} arises as part of a more complex \rt{system model} in which \rt{every} receiver is equipped with a cache. \rt{Suppose, for example, that} the stronger receivers want to \rt{decode} additional data that will never be demanded by the weak receivers  (see also \rt{Section}~\ref{sec:extension}). This additional data might represent, for example, a higher resolution of  a video. A practical solution in this case is to separate transmission of files from the two libraries  \cite{sahraeigastpar-2016,hachemkaramchandanidiggavi-2015,hachemkaramchandanidiggavi-2014}: A first transmission sends the files that are of interest to all receivers, and a second transmission sends only  files from the additional library to the strong receivers. The question is now how to divide the cache memory between the two transmissions. Based on the results we obtain in this paper, we propose to assign all the cache memory at the strong receivers to the second transmission, because through a careful design of the first transmission scheme, the strong receivers can already benefit  from the weak receivers' caches without accessing their own cache memories.

%We also discuss in Section~\ref{sec:extension} how our coding ideas extend  to unequal message rates and to general discrete memoryless BCs (DMBCs).

\rt{The fundamental rate versus cache memory tradeoff of interest in Scenario 1 is the largest rate~$R$ at which all messages can be reliably transmitted (in the usual Shannon sense) for \emph{a given} cache size $\M$ at the weak receivers. This largest rate is called the \emph{capacity-memory tradeoff} and will be denoted by $C(M)$.}
 
We present general \rt{achievable (lower)} and \rt{converse (upper)} bounds on the capacity-memory tradeoff $C(M)$. Our \rt{achievable} bound \rt{on $C(M)$} is based on a joint cache-channel coding scheme that builds on the \rt{`piggyback' coding idea in \cite{wuwigger-2016} (see \rt{S}ubsection~\ref{sec:piggyback}). The basic idea of piggyback coding is to carry messages to strong receivers on the back of messages to the weak receivers. These messages can be carried for ``free'' if the server pre-places appropriate ``message side information'' in the weak receivers' caches.} \rt{We will see that} joint cache-channel coding provides substantial gains over separate cache-channel coding \rt{with} Maddah-Ali and Niesen's coded caching scheme \cite{maddahali_niesen_2014-1} \rt{and} a capacity achieving scheme for the packet-erasure BC. \rt{For example, if the library has $D$ messages of rate $R$ and $M$ is smaller than \rt{approximately} $\frac{D R (\delta_\w-\delta_\s)}{K_\w+K_\s}$ (i.e., a `small-memory' regime)}, \rt{then}
\begin{equation}\label{eq:ll}
C(\M) \geq R_0 + \frac{\M}{D} \cdot \gamma_{\textnormal{local}}\cdot \gamma_{\textnormal{global,sep}}\cdot \gamma_{\textnormal{global,joint}}.
\end{equation}
\rt{Here} $R_0$ represents the capacity of the network without caches\rt{, and $\gamma_{\textnormal{local}} \leq 1,$ $\gamma_{\textnormal{global,sep}} \geq 1$ and $\gamma_{\textnormal{global,joint}} \geq 1$ are constants (depending on the number of receivers and erasure probabilities).} The exact expressions of the three \rt{constants} are presented later  in \eqref{eq:gains}, but they have the following interpretations. If we  employ  a standard capacity-achieving coding scheme to transmit the parts of the demanded messages that are not already stored in the intended receivers' caches, then we achieve a rate equal to $R_0  + \frac{\M}{D} \cdot \gamma_{\textnormal{local}}$. This strategy achieves only a local caching gain, and hence the subscript ``local" for this factor. When some receivers in the network have no caches ($K_\s \geq 1$), then $\gamma_{\textnormal{local}}< 1$. 
 A scheme that combines  Maddah-Ali and Niesen's coded caching algorithm with a capacity-achieving scheme for the packet-erasure BC attains the lower bound $R_0 + \frac{\M}{D} \cdot \gamma_{\textnormal{local}}\cdot \gamma_{\textnormal{global,sep}}$.
 The factor $\gamma_{\textnormal{global,sep}}$ thus describes the global caching gain obtained by a separate cache-channel coding scheme, and hence the subscript ``global,sep". Whenever $K_\w>1$, we have $\gamma_{\textnormal{global,sep}}>1$. Finally, our joint cache-channel coding scheme achieves the lower bound \eqref{eq:ll}, and  the parameter $\gamma_{\textnormal{global,joint}}$ describes this scheme's gain over the previous separate cache-channel coding. In other words, the factor
%We conclude that  in the regime of small cache sizes the 
%parameter, \eqref{eq:joint_gain},
\begin{equation}\label{eq:jg}
\gamma_{\textnormal{global,joint}}=1 +  \frac{2 K_\w}{1+K_\w} \cdot \frac{K_\s(1-\delta_\w)}{K_\w (1- \delta_\s)}
\end{equation} describes the further global caching gain that is possible using our joint cache-channel coding scheme (that was not achievable with the aforementioned separate cache-channel coding scheme).
%describes  the benefit of our joint cache-channel coding over this last separate cache channel coding scheme. The subscript ``global,joint" indicates that it corresponds to a global caching gain and is achieved with a joint cache-channel coding scheme. 
By \eqref{eq:jg}, the improvement of our joint cache-channel coding scheme over the separate cache-channel coding scheme is not bounded for small cache sizes. In particular, it is strictly increasing in the number of strong receivers~$K_\s$. 

Our general lower and upper bounds match for 
\begin{subequations} \label{eq:match}
\begin{equation} 
K_\w=1\qquad \textnormal{and}\qquad \M\leq \F D \frac{ (1-\delta_\s)(\delta_\w-\delta_\s)}{K_\s(1-\delta_\w)+(1-\delta_\s)}\cdot 
\end{equation}

For the special case $K_\w=K_\w=1$ and $D=2$, we present a refined lower bound on $C(\M)$ as well as a refined upper bound. The idea is to cache also  the XOR of a part of the two messages in the library, similarly to \cite[Appendix]{maddahali_niesen_2014-1}. Our refined bounds coincide when 
\begin{equation}
K_\w=K_\s=1;  \qquad  D=2; \qquad \delta_\w=\delta_\s
\end{equation} 
and 
\begin{equation}
K_\w=K_\s=1;  \qquad  D=2; \qquad \M\geq \F((1-\delta_\s)+(\delta_\w-\delta_\s)).
\end{equation} 
\end{subequations}

%We will also describe how to extend our results to more general cache assignments or more general channels.

\textit{Scenario~2:} In our second scenario  (section~\ref{sec:common}) we allow for general discrete memoryless broadcast channels (DMBCs),  arbitrary cache sizes $\{M_k\}_{k=1}^K$, and non-equal rates of the various messages $R_1, \ldots, R_D$. However, we impose that each receiver demands exactly the same message. For this scenario we completely characterize the entire rate-memorry tradeoff $(R_, \ldots, R_D, \M_1, \ldots, \M_k)$. 
The remainder of this paper is organised as follows. In Sections~\ref{sec:definition} and \ref{sec:preliminaries}, we state the problem setup and some auxiliary results that are helpful in the design of our joint cache-channel coding scheme. Section \ref{sec:results} summarizes our main results for the first scenario.  We describe and analyze  our joint source-cache channel coding scheme in  Section~\ref{sec:joint_cache_channel} and sketch how our scheme  can be extended to more general scenarios with arbitrary cache sizes and arbitrary DMBCs  in Section~\ref{sec:extension}. In Section~\ref{sec:general_upperbound}, we prove  an upper bound on the capacity-memory tradeoff of general (stochastically) degraded BCs with caches at the receivers. Our second scenario is discussed in section~\ref{sec:common}. %Finally, We conclude in Section~\ref{sec:conclude}.}% that describes setup and results for our second scenario.
%\texttt{To be added} }

\section{Problem Definition}\label{sec:definition}

%%%
%%%
\subsection{Notation} 
Random variables are identified by uppercase letters, e.g. $A$, their alphabets by matching calligraphic font, e.g. $\set{A}$, and elements of an alphabet by lowercase letters, e.g. $a \in \set{A}$. We also use uppercase letters for deterministic quantities like rate $R$, capacity $C$, number of users $K$, memory size $M$, and number of files in the library $D$. 

The Cartesian product of $\set{A}$ and $\set{A}'$ is $\set{A} \times \set{A}'$, and the $n$-fold Cartesian product of $\set{A}$ is $\set{A}^n$. Vectors are identified by bold font symbols, e.g., $\mathbf{a}$, and matrices by the font $\mathsf{A}$. 
We use the shorthand notation $A^n$ for the tuple $(A_1,\ldots, A_n)$. LHS and RHS stand for left-hand side and right-hand side. 

Finally, we use the notation $W_1\bigoplus W_2$ to denote the bitwise XOR over the binary strings corresponding to the messages $W_1$ and $W_2$, which are assumed to be of equal length.

\subsection{Message and channel models}

Consider a broadcast channel (BC) with a single transmitter and $K$ receivers as depicted in Figure~\ref{fig:general_model}. We have two sets of receivers: $K_{\textnormal{w}}$ weak receivers that statistically have a bad channel and $K_{\textnormal{s}}=K-K_{\textnormal{w}}$ strong receivers that statistically have a good channel. (The meaning of good and bad channels will be explained shortly.) For convenience of notation, we assume that the first $K_{\w}$~receivers are weak and the subsequent $K_\s$ receivers are strong, and we define the sets 
\begin{equation*} 
\mathcal{K}_\textnormal{w} :=\{1,\ldots, K_\textnormal{w}\}
\end{equation*}
and
\begin{equation*} 
\mathcal{K}_\textnormal{s} :=\{K_\textnormal{w}+1,\ldots, K\}.
\end{equation*}

We model the channel from the transmitter to the receivers by a memoryles
\emph{packet-erasure BC} with 
 input alphabet 
 \begin{equation*}
 \set{X} := \{0,1\}^{F}
 \end{equation*} and equal output alphabet at all receivers 
 \begin{equation*}
 \set{Y} := \set{X} \cup \{\Delta\}.
 \end{equation*} Here $F \geq 0$ is a fixed positive integer, and each input symbol $x \in \set{X}$ is an $F$-bit packet. The output erasure symbol~$\Delta$ models loss of a packet at a given receiver. Each receiver~$k\in{\mw{\set{K}:=\{1,\ldots, K\}}}$ observes the erasure symbol $\Delta$ with a given probability $\delta_k\geq 0$, and it observes an output $y_k$ equal to the input, $y_k=x$, with probability~$1-\delta_k$. 
 The marginal transition laws\footnote{As will become clear in the following, for our problem setup only this marginal transition law is relevant, but not the joint transition law.} of the memoryless BC are thus described by
 \begin{equation}\label{eq:channel}
\Pr[Y_k = y_k |X = x] = 
\left\{
\begin{array}{cl}
1 - \delta_k & \text{ if } y_k = x\\
\delta_k & \text{ if } y_k = \Delta\\
0 & \text{ otherwise}
\end{array}
\right.
\quad \forall\ k.
\end{equation}
{We will assume throughout that
\begin{equation}\label{eq:same_weak}
\delta_i =
\left\{
\begin{array}{ll}
\delta_\w & \text{ if } i \in \set{K}_{\w}\\
\delta_\s & \text{ if } i \in \set{K}_{\s}
\end{array}
\right.
\end{equation} %{eq:same_strong}
for fixed erasure probabilities\footnote{Though, in principle, we allow $\delta_{s}=\delta_{\w}$, our main interest will be $\delta_\s< \delta_\w$.} $0 < \delta_\s \leq \delta_w \leq 1$. \ssb{Since $\delta_\s \leq \delta_w$, the weak receivers have statistically worse channels than the strong receivers, hence the distinction between good and bad channels}. In the sequel, we will assume that each weak receiver is provided with a cache memory of size $n\M$ bits. The strong receivers are not  provided with cache memories. We explain shortly how the cache memory at the weak receivers can be exploited.}

%%%
%%%
%%%

\subsection{Message library and receiver demands}

The transmitter has access to a library with $D\geq K$ messages
\begin{equation}\label{Eqn:Messages}
W_1,\ldots, W_D.
\end{equation}
 These messages are all independent of each other and each of them is uniformly distributed over the message set $\{1,\ldots, \lfloor 2^{n\R}  \rfloor\}$, where $\R \geq 0$ is the rate of each message and $n$ the blocklength of transmission. %We represent a particular combination of receivers' demands by a tuple $\d = (d_1,\ldots, d_K) \in \{1,\ldots, D\}^K.$ 

Each receiver will demand  (i.e., request and download) exactly one of these messages. %, irrespective of the messages demanded at the other receivers. 
Let 
\[\mathcal{D}:=\{1,\ldots, D\}.
\] We denote the demand of receiver~1 by $d_1\in\set{D}$, the demand of receiver~2 by $d_2\in\set{D}$,  etc., to indicate that receiver~1 desires message~$W_{d_1}$, receiver~2 desires message~$W_{d_2}$, and so on. For most of the time in this manuscript we assume that the 
demand vector 
\begin{equation}
\d:= (d_1,\ldots, d_K)
\end{equation}
can take on any value in $\set{D}^K.$

Communication takes place in two phases: a first \emph{caching phase} where information is stored in the weak receivers' cache memories and a subsequent \emph{delivery phase} where the demanded messages are delivered to all the receivers.  The next two subsections detail these two communication phases.

%%%
%%%
%%%

\subsection{Caching phase} 
During the first communication phase the transmitter sends caching information $V_i$ to each weak receiver~$i\in\set{K}_\w$, who then stores this information in its cache memory. The strong receivers do not take part in the caching phase.

{The demand vector $\d$ is unknown to the transmitter and receivers during the caching phase, and, therefore, the cached information~$V_i$  cannot depend on the users' specific demands~$\d$. Instead, $V_i$ is a function of the entire library:}
\begin{equation*}
V_i := g_i(W_1, \ldots, W_D)
\qquad 
i\in\set{K}_{\w}
\end{equation*}
for some function 
\begin{equation}\label{eq:caching}
g_i \colon \big\{1,\ldots, \lfloor 2^{n\R} \rfloor \big\}^D\to \set{V},\qquad i\in\set{K}_{\w},
\end{equation}
where 
\begin{equation*}
 \set{V}:=\{1,\ldots, \lfloor 2^{n\M}\rfloor \}.
\end{equation*}
The caching phase occurs during a low-congestion period. We therefore assume that this phase incurs  no erasures or other types of errors, and each weak receiver~$i\in\set{K}_{\w}$ can store  $V_i$ in its cache memory.

%%%
%%%
%%%

\subsection{Delivery phase}
{The transmitter is provided with the demand vector $\d$, and it communicates the corresponding messages $W_{d_1}, \ldots, W_{d_K}$ over the packet-erasure BC.  The entire demand vector $\mathbf{d}$ is assumed to be known to the transmitter and all receivers\footnote{It takes only $\lceil \log(D)\rceil$ bits to describe the demand vector $\mathbf{d}$. The demand vector can thus be revealed to all terminals using zero transmission rate.}.
}

Depending on the demand vector $\d$, the transmitter chooses an encoding function 
\begin{equation}\label{eq:encoding}
f_\d\colon 
\{1,\ldots, \lfloor 2^{n\R}\rfloor \}^D
\to 
\set{X}^n
\end{equation}
and it sends
\begin{equation}
X^n = f_\d(W_1,\ldots,W_D),
\end{equation}
over the packet-erasure BC. 

Each receiver $k\in\{1,\ldots, K\}$ observes $Y_k^n$ according to the memoryless transition law in \eqref{eq:channel}. {Each weak receiver attempts to reconstruct its desired message from it channel output, cache contents and demand vector $\d$. Similarly, each strong receiver attempts to reconstructs its desired message from its channel output and the demand vector $\d$. More formally,  
\begin{subequations}\label{eq:decoding}
\begin{equation}
\hat{W}_i := 
\left\{
\begin{array}{ll}
\varphi_{i,\d}(Y_i^n, V_i) & \text{ if } i\in\set{K}_\w\\[8pt] 
\varphi_{i,\d}(Y_j^n)        & \text{ if } i\in\set{K}_\s
\end{array}
\right.
\end{equation}
where
\begin{equation}
\varphi_{i,\d} \colon \set{Y}^n \times \set{V} \to \{1,\ldots, \lfloor 2^{nR}\rfloor \}
\qquad i\in\set{K}_{\w}
\end{equation}
and
\begin{equation}
\varphi_{j,\d}\colon \set{Y}^n \to \{1,\ldots, \lfloor 2^{nR}\rfloor \}
\qquad j\in\set{K}_{\s}.
\end{equation}
\end{subequations} 
}
%%%
%%%
%%%

\subsection{\mw{Capacity-memory tradeoff}}%Achievable rate-memory tuples}
{An error is said to occur whenever}
\begin{equation}
\hat{W}_k \neq W_{d_k} 
\quad 
\textnormal{for some}
\quad 
k\in\{1,\ldots, K\}.
\end{equation} 
For a given demand vector $\mathbf{d}$ the probability of error is thus
\begin{equation}
\Pe(\mathbf{d}) := 
\Pr\bigg[
\bigcup_{k = 1}^K
 \hat{W}_k \neq W_{d_k}
\bigg].
\end{equation}
We consider a worst-case probability of error over all feasible demand vectors: 
\begin{equation}
\Pe^{\!\!\textnormal{worst}}:= \max_{\d \in \set{D}^K} \Pe(\mathbf{d}).\label{def:pe}
\end{equation}
\ssb{In definitions \eqref{eq:caching}-\eqref{def:pe}, we sometimes add a superscript $(n)$ to emphasise  the dependency on the blocklength $n$.}

We say that a rate-memory pair $(R, \M)$ is  \emph{achievable} if for every $\epsilon >  0$ there exists a sufficiently large blocklength $n$ and {caching, encoding and decoding functions as in \eqref{eq:caching},} \eqref{eq:encoding} and \eqref{eq:decoding} such that $\Pe^{\!\!\textnormal{worst}}<\epsilon$. The main problem of interest in this paper is to determine the following {capacity versus cache memory tradeoff.}

\begin{definition}
Given cache memory size $\M$, we define the \emph{capacity-memory tradeoff} ${C}(\M)$ as the supremum of all rates $R$ such that the rate-memory pair $(R,\M)$ is achievable.
\end{definition}

%%%
%%%
%%%

\subsection{Trivial and non-trivial memory sizes}

The capacity-memory tradeoff $C(\M)$ is trivially upper bounded by the capacity of the packet-erasure BC to the strong receivers only (see Proposition~\ref{Prop:CapacityRegion} in Section~\ref{sec:capacity} ahead):
\begin{equation}
C(\M) \leq \F\left(\frac{1-\delta_\s}{K_\s}\right)
\quad\ 
\forall\ \M \geq 0.
\end{equation}
{For $\M = D \F (1-\delta_\s) / K_\s$, this upper bound is also achievable because the weak receivers can store the entire library in their caches and the transmitter \ssb{thus needs to only} serve the strong receivers during the delivery phase.} Therefore,
\begin{equation}\label{eq:ClargeM}
C(\M)= \F \left(\frac{1-\delta_\s}{K_\s}\right), 
\qquad  
\forall\  \M \geq D \F \left(\frac{1-\delta_\s}{K_\s}\right).
\end{equation}
{We will henceforth restrict attention to nontrivial cache memories}
\begin{equation*}
\M\in \bigg[0, \;\;D   \F \frac{(1-\delta_\s)}{K_\s}\bigg].
\end{equation*}

%%%%
%%%%
%%%%
%%%%

\section{Previous Related Results} \label{sec:preliminaries}%Preliminaries for {the} Joint Cache-Channel Coding Scheme}\label{sec:preliminaries}

\ssb{This section reviews capacity results and coding schemes for three related scenarios that form the basis for our new bounds on the capacity-memory tradeoff $C(M)$ and the joint cache-channel coding scheme that we present in Section~\ref{sec:joint_cache_channel} ahead.}
%{This section lays the foundation for the joint cache-channel coding scheme described in the sequel. The scheme will be built on three key ideas:} 1) the capacity of the packet-erasure BC; 2) a coding idea that we recently introduced for the BC with feedback \cite{wuwigger-2016} and that we term here  \emph{piggyback} coding; and 3) the \emph{coded caching} idea that Maddah-Ali and Niesen used for cache-aided communication over a common noise-free pipe \cite{maddahali_niesen_2014-1}. {The purpose of this section is to review and explain these ideas.}

%%%
%%%
%%%

\subsection{Capacity of packet-erasure BCs}\label{sec:capacity}

\begin{figure}[h!]
\begin{center}
\includegraphics[width=0.42\textwidth]{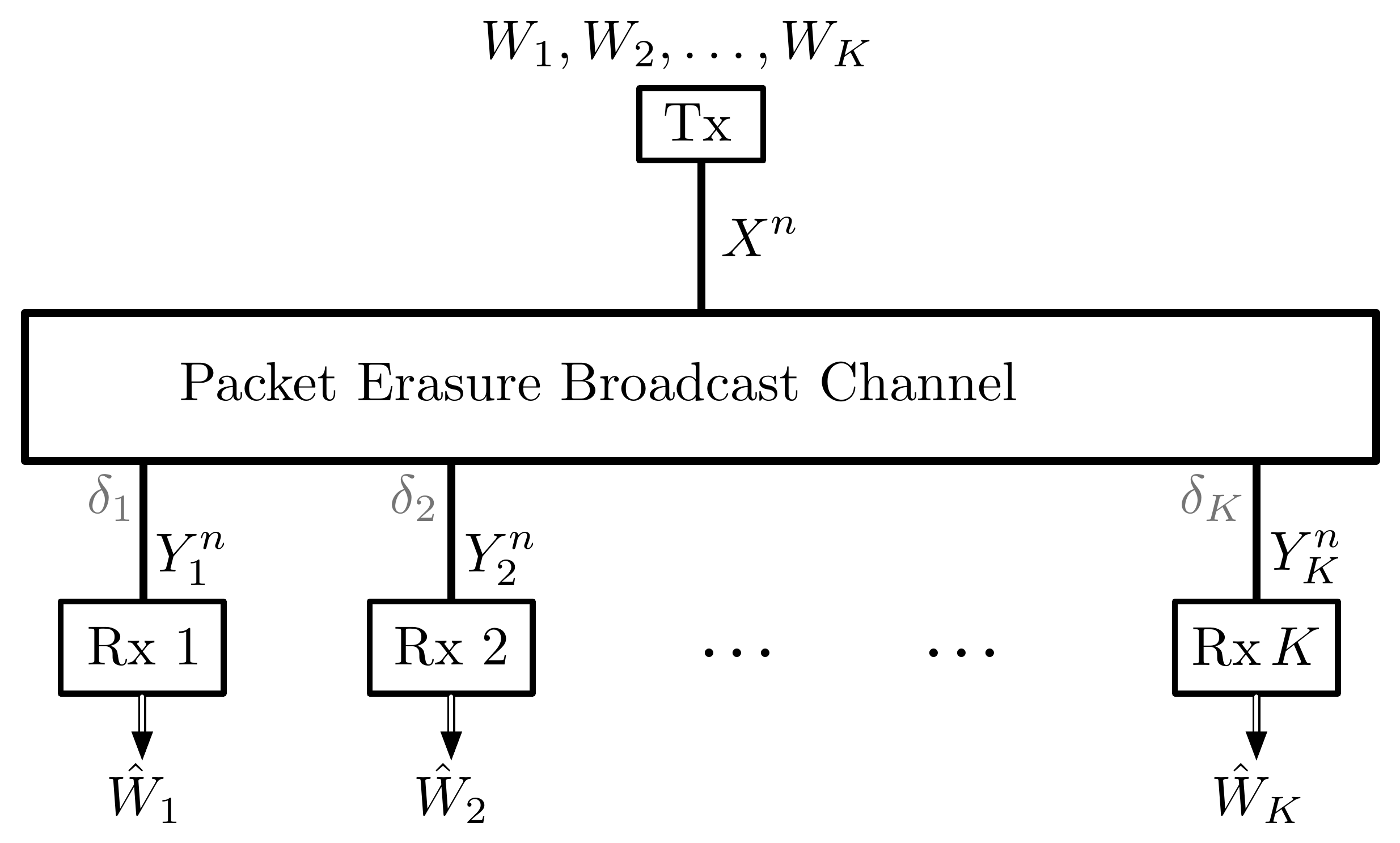}
\caption{Standard $K$-user packet-erasure BC \ssb{with {arbitrary erasure probabilities} and no caches.}}
\label{fig:BC}
\end{center}
\end{figure}

Temporarily consider the $K$ receiver packet-erasure BC illustrated in Figure~\ref{fig:BC}. \ssb{The BC is characterised by \eqref{eq:channel} with arbitrary erasure probabilities $0 < \delta_1,\delta_2,\ldots,\delta_K \leq 1$.} Suppose that the are no caches (i.e., $M = 0$), and that each receiver~$k$ wishes to learn an independent message $W_{k}$ that is uniformly distributed over the set $\{1, \ldots, \lfloor 2^{nR_k}\rfloor\}$. \ssb{Notice that, in contrast to previous sections, messages have different rates, and it is a priori known which message is intended for which receiver.}  Let $\hat{W}_k$ denote receiver~$k$'s {reconstruction} of message $W_k$.

The capacity region of this standard packet erasure-BC is achieved by time-sharing capacity-achieving point-to-point codes~\cite{Urbanke-1999-A}. The point-to-point capacity {of the channel from the transmitter to receiver~$k\in\{1,\ldots, K\}$ is}
\begin{equation}
C_k= (1-\delta_k) F.
\end{equation}

\ssb{\begin{proposition}\label{Prop:CapacityRegion} 
The capacity region of the packet-erasure BC to $K$ receivers with erasure probabilities $\delta_1, \delta_2, \ldots, \delta_K$  is the closure of the set of nonnegative rate-tuples~$(R_1, \ldots, R_K)$ that satisfy~\cite{Urbanke-1999-A}
\begin{equation*}%\label{eq:capaBC}
 \sum_{k=1}^K \frac{R_k}{(1-\delta_k)F}\leq 1 .
\end{equation*}
\end{proposition}}

%%%%
%%%%
%%%%
%%%%

%%%
%%%
%%%

\subsection{Coded caching over a BC with noise-free common bit-pipe}
\label{sub:MAN}
\begin{figure}[t!]
\begin{center}
\includegraphics[width=0.4\textwidth]{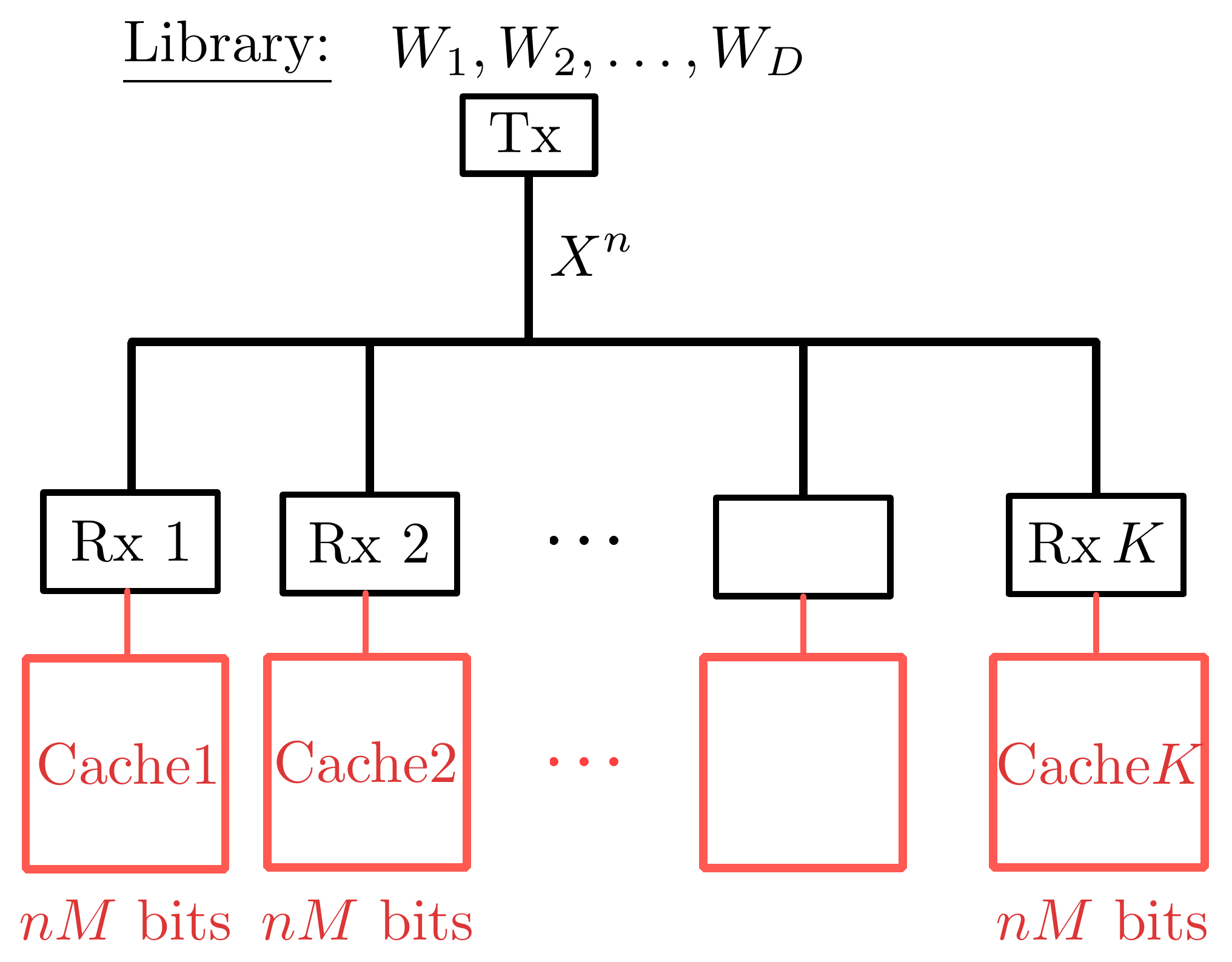}
\caption{BC with a common noise-free pipe of rate $F$ to all $\tilde{K}$ receivers, which all have a cache memory of $n\M$ bits.}
\label{fig:pipe}
\end{center}
\end{figure}

We \ssb{briefly explain} the Maddah-Ali and Niesen's \emph{\mw{coded} caching scheme} in~\cite{maddahali_niesen_2014-1}. The scheme has two parameters: 
\begin{itemize}
\item a positive integer $\tilde{K}$ {(representing the number of users with caches)} and 
\item an index $\tilde{t}\in\{1,\ldots, \tilde{K}\}$.
\end{itemize} 
Coded caching applies to the {noiseless} communication scenario {illustrated} in Figure~\ref{fig:pipe}. This scenario coincides with our original scenario in Section~\ref{sec:definition} in the special case where
\begin{equation*}
K_\w=K=\tilde{K} \qquad \textnormal{and} \qquad \delta_\w=0.
\end{equation*}
As in~\eqref{Eqn:Messages}, the messages $(W_1,W_2,\ldots,W_D)$ depicted in Figure~\ref{fig:pipe} have a common rate $R$. 

{It will be convenient to describe coded caching using the following methods:} 
\begin{itemize}
\item \emph{Method}~\texttt{Ca} describing the caching phase for the setup in Figure~\ref{fig:pipe}. 
\item \emph{Method}~\texttt{En} describing the delivery-phase encoding.
\item \emph{Methods}~$\{\texttt{De}_k;\ k = 1,2,\ldots,\tilde{K}\}$ describing the delivery-phase decoding at each user.
\end{itemize}

\subsubsection{{Preliminaries}}  
Let 
\begin{equation*}
\set{G}_{1}, \ldots, \set{G}_{\binom{\tilde{K}}{\tilde{t}} }
\end{equation*}
denote the $\binom{\tilde{K}}{\tilde{t}}$ subsets of~$\{1,\ldots,\tilde{K}\}$ of size $\tilde t$. Split each message $W_d$ into $\tilde{K}$ choose $\tilde{t}$ independent submessages, 
\begin{equation*}
W_d = \bigg\{ W_{d, \set{G}_\ell}  \colon \ \ell=1,  \ldots, \binom{\tilde{K}}{\tilde t} \bigg\}.
\end{equation*}
Each of these submessages  is of equal rate 
\begin{equation}\label{eq:rate_1}
{R}_{\textnormal{sub}}:= R\ \binom{\tilde{K}}{\tilde{t}}^{-1}.
\end{equation}

\subsubsection{{Method~\texttt{Ca}}}

{This method takes the entire library $W_1, \ldots, W_D$ as an input, and it outputs the cache contents $V_1,V_2,\ldots,V_{\tilde{K}}$ where} 
%\begin{IEEEeqnarray}{rCl}\label{eq:cache_content_MN}
%\lefteqn{V_k = \big\{ W_{d, \set{G}_\ell} \colon \ d\in\{1,\ldots, D\}\; \textnormal{and}\; k \in \set{G}_{\ell}\big\},}\nonumber \\
%& & \hspace{4cm} \quad k\in\{1,\ldots, \tilde{K}\}
%\end{IEEEeqnarray}
\begin{align}\label{eq:cache_content_MN}
{V_k = \big\{ W_{d, \set{G}_\ell} \colon \ d\in\{1,\ldots, D\}\; \textnormal{and}\; k \in \set{G}_{\ell}\big\},}
\quad k\in\{1,\ldots, \tilde{K}\}.
\end{align}
In other words, during the caching phase, the tuple
\begin{equation*}
\Big(W_{1,\set{G}_\ell},\ W_{2, \set{G}_\ell},\ \ldots, \ W_{D,\set{G}_\ell}\Big)
\end{equation*} 
is stored in the cache memory of every receiver in $\set{G}_\ell$.

\subsubsection{{Method~\texttt{En}}} 
This method takes the entire library $W_1,$ $\ldots, W_D$ and the demand vector $\d$ as inputs, and it outputs 
\begin{equation} 
\big\{W_{\textnormal{XOR},\mathcal{S}} \colon\quad  \set{S} \subseteq \{1,\ldots, \tilde{K}\} \; \textnormal{and} \; |\set{S}|=\tilde{t}+1\big\},
\end{equation} 
where
\begin{equation}\label{eq:xor}
W_{\textnormal{XOR},\mathcal{S}}:=\bigoplus_{s \in \set{S}} W_{d_s, \Ss}.
\end{equation}

\subsubsection{{Methods~\texttt{De$_k$} (for $k=1,\ldots,\tilde{K})$}}

This method takes as inputs the demand vector $\d$; the XOR-messages $\{W_{\set{S}} \colon k\in \set{S} \}$ produced by method~\texttt{En}; and the cache content $V_k$ produced by method~\texttt{Ca}. It outputs the $\tilde{K} \choose \tilde{t}$-tuple reconstruction 
\begin{equation}
\hat{W}_{d_k}:= \Big(\hat{W}_{d_k, \set{G}_1}, \ldots, \hat{W}_{d_k, \set{G}_{\binom{\tilde{K}}{\tilde{t}}}} \Big),
\end{equation} 
where 
\begin{equation}\label{eq:MN_decoding}
\hat{W}_{d_k, \set{G}_\ell}= \begin{cases}  W_{d_k, \set{G}_\ell}& \textnormal{if }  k\in\set{G}_\ell\\
\left( \bigoplus_{{s} \in \set{G}_\ell} W_{d_{{s}},\set{G}_\ell \cup \{k\} \backslash\{s\}} \right) \bigoplus  W_{\textnormal{XOR},\mathcal{G}_\ell \cup \{k\}} \qquad &\textnormal{if } k\notin \set{G}_\ell.
\end{cases}
\end{equation}
Notice that all the (XOR) messages on the right are inputs of this method, because they are either part of the cache content~$V_k$ produced by method \texttt{Ca} or part of the XOR messages $\{W_{\textnormal{XOR},\set{S}} \colon k\in \set{S} \}$ produced by method~\texttt{En}.

\subsubsection{{Analysis}}

We {now} analyse the three methods above {for} Figure~\ref{fig:pipe}. 

\begin{lemma}\label{lem:MN}
Consider the scenario in Figure~\ref{fig:pipe}. The XOR messages $\{W_{\textnormal{XOR},\set{S}}\}$ produced by method \texttt{En} can be sent over the \mw{common noise-free pipe} if and only if the \mw{rate of the pipe} satisfies
\begin{equation}\label{eq:Maddah_Niesen_rate}
F\geq {{\tilde K}\choose{\tilde t+1}} R_{\textnormal{{sub}}}= R\ \frac{\tilde K-\tilde t}{\tilde t+1}. 
\end{equation}
Moreover, each receiver~$k\in\{1,\ldots,\tilde{K}\}$ can store the cache content $V_k$ if, and only if, the cache memory size satisfies
\begin{equation}\label{eq:rate_memory}
\M \geq D{\tilde K-1 \choose \tilde t-1} {R}_{\textnormal{{sub}}} =D \frac{\tilde t}{\tilde K} R. 
\end{equation}
\end{lemma}
\begin{IEEEproof}
 Inequality~\eqref{eq:Maddah_Niesen_rate} follows because there are  $\binom{\tilde K}{\tilde t+1}$ XOR messages  $W_{\set{S}}$ of rate {$R_{\textnormal{{sub}}}$} and by~\eqref{eq:rate_1}. Inequality~\eqref{eq:rate_memory} follows 
because each $V_ k$ produced by method \texttt{Ca}, consists of {$D\binom{\tilde{K}-1}{\tilde{t} - 1}$} messages of rate $R_{\textnormal{{sub}}}$, see~\eqref{eq:cache_content_MN}. 
\end{IEEEproof}

%%%
%%%
%%%

\subsection{{Separate cache and channel coding for packet-erasure BCs (and the proof of Proposition~\ref{prop:separate})}}\label{sec:separate} 

Starting from Maddah-Ali and Niesen's \mw{coded-caching scheme} one readily obtains a separation-based coding scheme for the packet-erasure BC with caches described in Section~\ref{sec:definition}. Details are as follows.

Choose $\tilde{K}=K_\w$ and an arbitrary $\tilde{t}\in\{1,\ldots, K_\w\}$. For the caching phase: Apply  {Method \texttt{Ca}} to the library $W_1, \ldots, W_D$;  {take the resulting $V_1,\ldots,V_\w$; and store each $V_k$ the cache memories of receiver~$k$}.  The delivery-phase transmitter  proceeds in two steps: 
 \begin{itemize}
 \item[\emph{T1:}]  {The transmitter} applies method \texttt{En} to the library $W_1,W_2,$ $\ldots,W_D$ and demand vector $\d_\w:=(d_1,d_2,\ldots, d_{K_\w})$.   
 \item[\emph{T2:}]  {The transmitter} sends the XOR-messages produced in step T1 together with the messages that are demanded by receivers in~$\Ks$ using a capacity-achieving scheme for the packet-erasure BC. 
 \end{itemize}
 The strong receivers decode their intended messages using an optimal decoding method for the packet-erasure BC. The weak receivers decode in two steps:
 \begin{itemize}
 \item [\emph{R1:}] Each weak receiver uses an optimal decoder for the packet-erasure BC to recover all XOR-messages generated by method \texttt{En}.  
 \item[\emph{R2:}] It applies method \texttt{De} to the XOR messages decoded in  step~R1. 
 \end{itemize}
 
This {separate cache and channel coding scheme} can be easily analysed using Lemma~\ref{lem:MN} and Proposition~\ref{Prop:CapacityRegion}, and {from this analysis one obtains Proposition~\ref{prop:separate} in Section~\ref{sec:results}.}

%%%%
%%%%
%%%%
%%%%

\subsection{``Piggyback'' coding for BCs with message side information}\label{sec:piggyback}

%%%
%%%
%%%

\begin{figure}[h!]
\begin{center}
\includegraphics[width=0.26\textwidth]{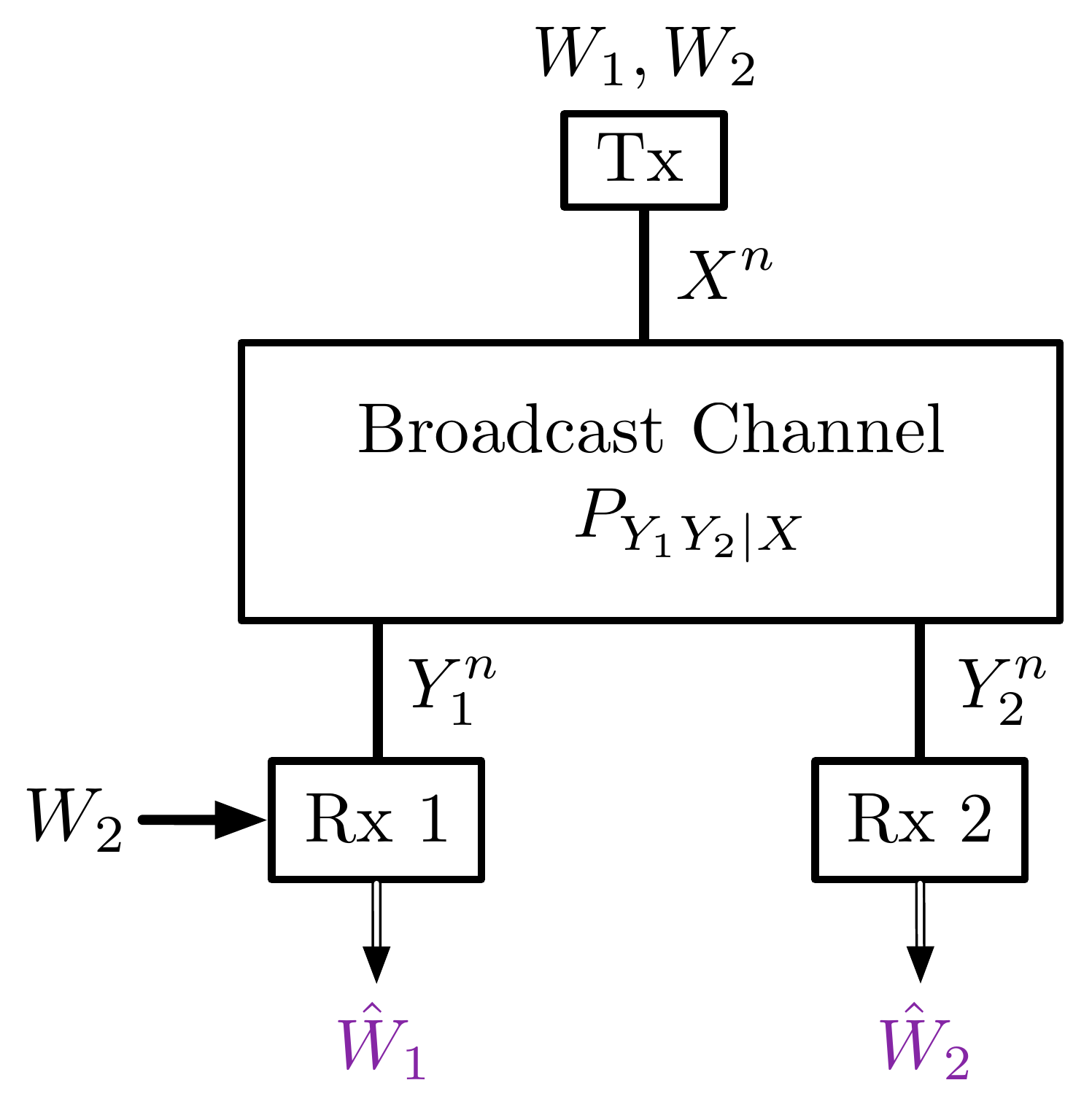}
\caption{Standard two-user BC with message side-information at receiver~2.}
\label{fig:piggyback_model}
\end{center}
\end{figure}

Consider the two-user BC with message side-information  \cite{bracherwigger-IT2015,kramershamai07} {illustrated} in Figure~\ref{fig:piggyback_model}. 

The transmitter observes two independent messages $W_1$ and $W_2$ of rates $R_1$ and $R_2$. Message $W_1$ is intended  for receiver~1 and message~$W_2$ for receiver~2. {Suppose that receiver~1 is given $W_2$ prior to communications and the BC has arbitrary transition probabilities $P_{Y_1Y_2|X}$.} The capacity region of this BC with message side-information at receiver~1 {can be derived}\footnote{{Kramer and Shamai assume in~\cite[Thm.~3]{kramershamai07} that} receiver~1 not only needs to decode message $W_1$ but also $W_2$. However, since receiver~1 has message $W_2$  as side information, this additional requirement is not a limitation {and the two} setups have identical capacity regions.} from~\cite[Thm.~3]{kramershamai07}. 

{We now present a specific random} coding scheme for the BC in Figure~\ref{fig:piggyback_model}, which we call \emph{piggyback coding}.

\subsubsection{{Code Construction}}

\begin{figure}[h!]
\begin{center}
\includegraphics[width=0.42\textwidth]{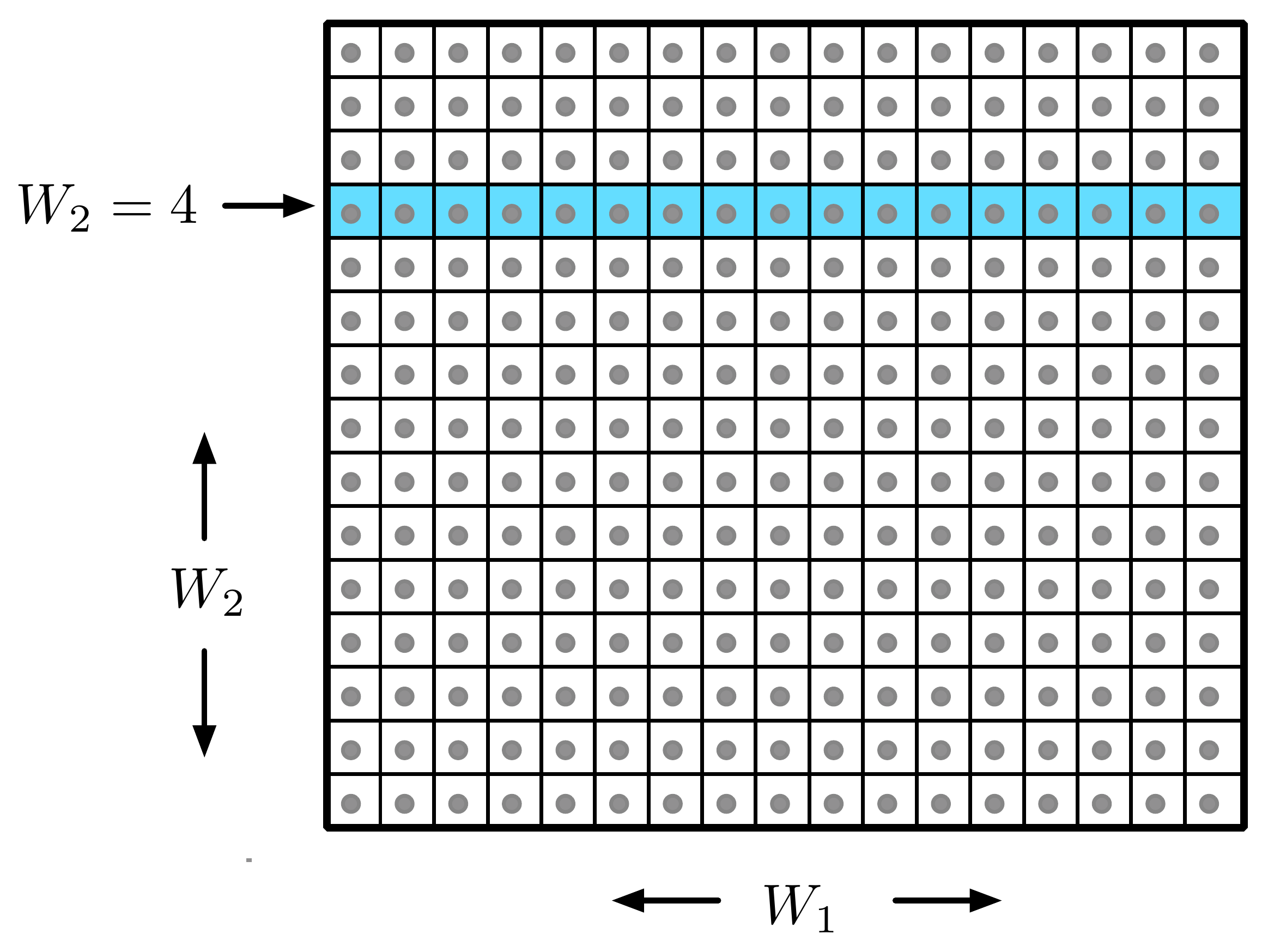}
\end{center} 
\caption{Codebook~$\mathcal{C}$ where the dots represent the codewords $x^n(w_1, w_2)$ arranged in a matrix array. Columns encode message $W_1$ and rows encode message $W_2$. Receiver~1 knows the value of message $W_2$ and thus can restrict its decoding to a single row of the codebook.}
\label{fig:product_code}
\end{figure}

Fix a distribution $P_X$ on the input alphabet of the BC, a small $\epsilon>0$ and a large blocklength $n$. {Randomly generate} a codebook \ssb{$\mathcal{C}$} with $\lfloor 2^{n R_1}\rfloor\times \lfloor 2^{nR_2}\rfloor$ codewords of length $n$ by independently picking each entry of each codeword {using} $P_X$. We place the codewords into a matrix with $\lfloor2^{nR_1}\rfloor$ columns and $\lfloor 2^{nR_2}\rfloor$ rows, and denote the codeword in column~$w_1$ and row~$w_2$ by $x^n(w_1,w_2)$. Figure~\ref{fig:product_code} sketches the codebook: Each dot represents a codeword; message $W_1$ determines the column of the codeword to pick and $W_2$ \ssb{determines} the row. {The codebook is given to the transmitter and both receivers.}

\subsubsection{{Encoding}} 

Given that the transmitter wishes to send messages $W_1=w_1$ and $W_2=w_2$, it transmits the codeword $x^n(w_1,w_2)$ over the channel.

\subsubsection{{Decoding at receiver~1 (the receiver with side information)}} 

{Since receiver~1 knows $W_2=w_2$, it can restrict attention to the $w_2$-th row of the codeword.} For example if $W_2=4$, then it needs only to consider the codewords (dots) that lie in the highlighted row of Figure~\ref{fig:product_code}.

Given that receiver~1 observes the channel outputs $Y_1^n = y_1^n$, it looks for a unique index $\hat{w}_1\in\{1,\ldots,\lfloor  2^{nR_1}\rfloor\}$ satisfying
\begin{equation*}
(x^n(\hat{w}_1,w_2), y_1^n) \in\mathcal{T}_{\epsilon}^n(P_{XY_1}),
\end{equation*}
where $\mathcal{T}_{\epsilon}^n(P_{XY_1})$ denotes the typical set as defined in \cite{elgamalkim10}. If there is no such index $\hat{w}_1$, then receiver~1 declares an error.

\subsubsection{{Decoding at receiver~2 (the receiver without side information)}}
 
Receiver~2 will {attempt to} decode both messages $W_1$ and $W_2$. Given that it observes channel outputs $Y_2^n=y_2^n$, 
it looks for the unique pair of indices $(\hat{w}_1, \hat{w}_2)\in\{0,\ldots, 2^{nR_1}-1\}\times \{0,\ldots, 2^{nR_2}-1\}$ satisfying
\begin{equation*}
(x^n(\hat{w}_1,{\hat{w}_2}), y_2^n) \in\mathcal{T}_{\epsilon}^n(P_{XY_2}).
\end{equation*}
If there is no such pair {of indices}, then receiver~2 declares an error.

\subsubsection{{Analysis \& discussion}} 

By the \emph{covering lemma} \cite{elgamalkim10} and because receiver~1 restricts attention to a single row in the codebook, the probability of decoding error at receiver 1, $\Pr[{\hat{W}_1\neq W_1}]$,  tends to 0 as $n\to \infty$ whenever
\begin{subequations}\label{eq:rates_piggyback_coding}
\begin{equation}
R_1 < I(X;Y_1).
\end{equation}
Moreover, by this covering lemma and because receiver~2 decodes both messages, the probability of error at receiver 2, $\Pr[{\hat{W}_2\neq W_2}]$, tends to 0 as $n\to \infty$ whenever
\begin{equation}\label{eq:sum_piggyback}
R_1+R_2 < I(X;Y_2).
\end{equation}
\end{subequations}
We have the following proposition.

\begin{proposition}
For a DMBC $P_{Y_1Y_2|X}$ with message side-information at receiver~1 shown in Figure~\ref{fig:piggyback_model}, piggyback coding achieves all nonnegative rate-pairs $(R_1,R_2)$ satisfying \eqref{eq:rates_piggyback_coding} for some {channel} input distribution $P_X$.
\end{proposition}

Specialising this proposition to the packet-erasure BC, we obtain the following.

\begin{proposition} \label{Prop:piggyback}
For the two-user {$F$~bit} packet-erasure BC, with erasure probabilities $\delta_1, \delta_2$ and message side information at receiver~1, piggyback coding achieves all rate-pairs $(R_1, R_2)$ \ssb{that satisfy} 
\begin{equation}\label{eq:piggyback_rate}
\max \Bigg\{\frac{R_1}{(1-\delta_1)F}, \, \frac{R_1+R_2}{(1-\delta_2)F}\Bigg\} \leq 1.
\end{equation}
\end{proposition}

\begin{figure}[t!]
\begin{center}
\includegraphics[width=0.32\textwidth]{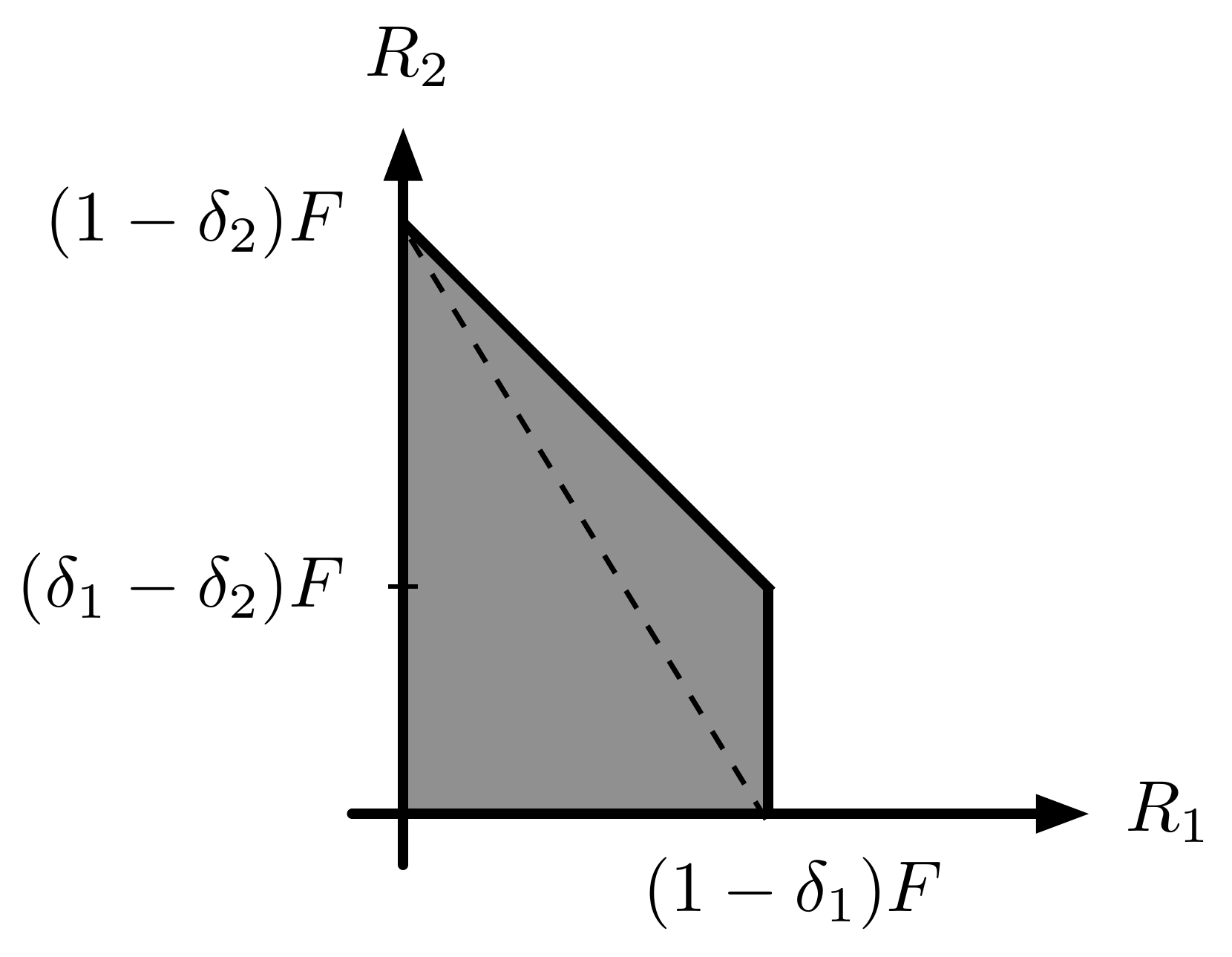}
\caption{The gray area depicts the rate-region achieved with piggyback coding over a two-user packet-erasure BC with message side-information at receiver~1 when receiver~2 has smaller erasure probability than receiver~1, i.e., $\delta_2<\delta_1$. The dashed line indicates the border of the capacity region without message side-information. The figure shows that the message side-information allows to \emph{piggyback} information to receiver~2 without reducing the achievable rate to receiver~1. }
\label{fig:piggyback_region}
\end{center}
\end{figure}

The achievable region in {Proposition~\ref{Prop:piggyback} is illustrated} in Figure~\ref{fig:piggyback_region}. This region covers the entire capacity region of the two-user $F$-bit packet-erasure BC with message side-information at receiver~1 whenever $\delta_2<\delta_1$.

{
\begin{remark}
The piggyback coding scheme does not achieve the entire capacity region of an arbitrary DMBC with message side-information at receiver~1. Consider, for example, a degenerate DMBC in which the channel to receiver~2 is useless (i.e., $P_{Y_2|X}$ has $0$ point-to-point capacity) and receiver~1's channel $P_{Y_1|X}$ has positive capacity. Over such a channel the piggyback coding achieves no positive rates because~\eqref{eq:sum_piggyback} constrains the sum-rate to 0.  It is clear that with a different scheme positive rates $R_1>0$ can be achieved. The optimal coding scheme splits message $W_{1}$ into $(W_{1,p}, W_{1,c})$ and combines piggyback coding with superposition coding: in the cloud center it sends $(W_{1,c}, W_2)$ as in piggyback coding, and in the satellite it sends~$W_{1,p}$.
\end{remark}
}

\section{New Results for Arbitrary Demands}\label{sec:results}
In this section we assume that the demand vector 
\[
\d \textnormal{ can take on every value in }\set{D}^K.
\]
Our main results are a general lower bound and a general upper bound on the capacity-memory tradeoff $C(M)$. The bounds are tight in certain regimes of $M$ when {$K_\w=1$}. 
We further present improved lower and upper bounds for the special case $K_\s=K_\w=1$ and $D=2$. These bounds match except for a small regime of $M$'s.
\subsection{General lower bounds}
Define  $K_\w+2$ rate-memory pairs $\{(R_t, \M_t);\ t=0,1,$ $\ldots,K_\w+1\}$ as follows:
\begin{enumerate}[(i)]
\item 
\begin{align}\label{eq:extremepoin1}
R_0 := 
F \left( \frac{K_\w}{1-\delta_\w}+\frac{K_\s}{1-\delta_\s} \right)^{-1}, \qquad \M_0:=0;
\end{align} 

\item For each $t \in \{1, \ldots, K_\w\}$:
\small{
\begin{align}
\notag
{R}_t := 
& \frac{\F (1-\delta_\w)\left(1+ \dfrac{K_\w-t+1}{t K_\s} \ \dfrac{\delta_\w-\delta_\s}{1-\delta_\w}\right) }{ \dfrac{K_\w-t+1}{t}\left(1+ \dfrac{K_\w-t}{(t+1) K_\s} \ \dfrac{\delta_\w-\delta_\s}{1-\delta_\w}\right) + K_\s\dfrac{1-\delta_w}{1-\delta_\s}}, \nonumber \\
\label{eq:pointt}
\M_t &: =R_t \ \frac{D}{K_\w}\! \left(\! t\!-\! \left(\!1\!+\!\frac{K_\w\!-\!t\!+\!1}{t K_\s} \ \frac{\delta_\w\!-\!\delta_\s}{1-\delta_\w}\! \right)^{\!-1}\right);
\end{align}
}

\item
\begin{equation}
\label{eq:extremepoin2}
R_{K_\w+1}
:= F \frac{1-\delta_\s}{K_\s}, 
\qquad 
\M_{K_\w+1} := DF\  \frac{1-\delta_\s}{K_\s}.
\end{equation}
\end{enumerate}

\begin{theorem}[Lower bound]\label{thm:ach}
The upper convex hull of the  $K_\w+2$ rate-memory pairs $\{(R_t,  M_t);\ t=0,1,\ldots, K_\w+1\}$ in \eqref{eq:extremepoin1}--\eqref{eq:extremepoin2} forms a lower bound on the capacity-memory tradeoff:% $C(M)$:
\begin{align}
C(M)\geq \textnormal{upper hull} \big\{(R_t,  M_t) \colon \; t=0, \ldots, K_\w+1\big\}.
\end{align}
\end{theorem}
\begin{IEEEproof}[Proof outline]
The pair $(R_0, \ \M_0=0)$ corresponds to the case without caches, and the achievability of $R_0$ follows from the usual capacity {result} for packet-erasure BCs, see Proposition~\ref{Prop:CapacityRegion} in \mw{the previous section~\ref{sec:capacity}}.
Achievability of the pair  $(R_{K_\w+1}, \ \M_{K_{\w}+1})$ follows from~\eqref{eq:ClargeM}. %corresponds to the case where $M$ is large enough so that  each receiver in $\Kw$ can store the entire library in its cache memory, see the discussion after defin %The delivery phase thus only needs to serve receivers in~$\Ks$ and achievability of $(R_{K_\w+1},$ $ \M_{K_{\w}+1})$ follows \mw{again from the usual capacity of} packet-erasure BCs (where we only consider strong receivers).

The remaining pairs $(R_t, \M_t)$, $t=1,\ldots,K_\w$, are more interesting and are achieved by the joint cache-channel coding  scheme in section~\ref{sec:joint_cache_channel} \mw{ahead}. The upper convex hull of $\{(R_t, \M_t); t=0,1,\ldots,K_\w+1\}$, finally,   is achieved by time-sharing.
\end{IEEEproof}

To better illustrate the strength of our lower bound, and thus of our joint cache-channel coding scheme in section~\ref{sec:joint_cache_channel},  consider the following separation-based lower bound: 
\begin{proposition}~\label{prop:separate}
The upper convex hull of 
the rate-memory pairs $\{(R_{t,\textnormal{sep}}, \M_{t,\textnormal{sep}}); t=\{0,\ldots,K_\w\}\big\}$ is achievable, where
\begin{subequations}\label{eq:ach_rm}
 \begin{IEEEeqnarray}{rCl}
 {R}_{t,\textnormal{sep}} & := &   \F  \left(\frac{K_\w-t}{(t+1)(1-\delta_\w)} +\frac{K_\s}{(1-\delta_\s)}  \right)^{-1},\IEEEeqnarraynumspace\\
%{F(1-\delta_\s)} \cdot \left[ \left( \frac{K_\w+1}{t+1}-1\right)\cdot \frac{1-\delta_\w+1-\delta_\s}{1- \delta_\w+K_\s(1-\delta_\s)}+K_\s \right]^{-1}\\
 \M_{t,\textnormal{sep}}& :=& D \frac{t}{K_\w}R_{t,\textnormal{sep}}.
 \end{IEEEeqnarray}
\end{subequations}
\end{proposition}
\begin{IEEEproof}
For $t=0$, no cache memory is used and the achievability of $R_{0,\textnormal{sep}}$ follows simply by the standard capacity region of the packet-erasure BC, Proposition~\ref{Prop:CapacityRegion} \mw{in  section~\ref{sec:capacity}}. For $t\in\{1,\ldots, K_\w\}$ achievability of the pair $(R_{t,\textnormal{sep}}, \M_{t,\textnormal{sep}})$ can be shown by trivially combining the Maddah-Ali \& Niesen coded caching \cite[Algorithm 1]{maddahali_niesen_2014-1}  with a capacity-achieving scheme for the packet-erasure BC without caching, see the previous section~\ref{sec:separate}.\end{IEEEproof}

{Of special interest is the regime of small cache size~$M$. In this regime, Theorem~\ref{thm:ach} specializes as follows:
\begin{corollary}\label{cor:small_memory}
For small cache memory sizes, $M \leq \M_1$, the capacity-memory tradeoff is lower bounded as
\begin{equation}
C(M) \geq R_0 + \frac{\M}{D} \cdot \gamma_{\textnormal{local}} \cdot \gamma_{\textnormal{global,sep}}\cdot \gamma_{\textnormal{global,joint}}, \qquad \M \leq \M_1,
\end{equation}
where $R_0$ is defined in \eqref{eq:extremepoin1} and
\begin{subequations}\label{eq:gains}
\begin{IEEEeqnarray}{rCl}
\gamma_{\textnormal{local}} &:= &  \frac{K_\w(1-\delta_\s)}{K_\w(1-\delta_\s)+K_\s(1-\delta_\w)},\\
 \gamma_{\textnormal{global,sep}} &:= &\frac{1+K_\w}{2}, \\
 \gamma_{\textnormal{global,joint}}&:= & 1 +  \frac{2 K_\w}{1+K_\w} \cdot\frac{K_\s(1-\delta_\w)}{K_\w (1- \delta_\s)}.\label{eq:joint_gain}
\end{IEEEeqnarray}
\end{subequations}
%For $K_\w=1$ we have $ \gamma_{\textnormal{sep}} \cdot \gamma_{\textnormal{joint}}=1$. 
\end{corollary}
If in the above lower bound one replaces the product $\gamma_{\textnormal{global,sep}}\cdot\gamma_{\textnormal{global,joint}}$ by $1$, then one obtains the lower bound that corresponds to a coding scheme with only local caching gain. If only the factor $\gamma_{\textnormal{joint}}$ is replaced by 1, then one obtains the lower bound implied by Proposition~\ref{prop:separate}. The factor $\gamma_{\textnormal{global,sep}}$ is thus due to the separation-based Maddah-Ali \& Niesen coded caching idea. In contrast, the last factor $\gamma_{\textnormal{global,joint}}$ is due to our joint cache-channel coding scheme. Notice that this factor $\gamma_{\textnormal{global,joint}}$ is unbounded when one increases the number of strong receivers $K_\s$ or more generally the ratio $\frac{K_\s (1-\delta_\w)}{K_\w(1-\delta_\s)}$.}

\subsection{General upper bound}
We now present our upper bound.
%For each pair of integers $(k_\w,k_\s)$ satisfying
%%\begin{subequations}
%\begin{IEEEeqnarray}{C}
%0 \leq k_\w \leq K_\w; \quad 
%0 \leq k_\s \leq K_\s; \quad \textnormal{and} \quad
%1 \leq k_\w + k_\s;\quad\label{eq:upperbound}
%\end{IEEEeqnarray}
%%\end{subequations}
Define for each $k_\w\in\{0, \ldots, K_\w\}$
\begin{align}
R_{k_\w}(M)
\notag
&:= \F \left( \frac{k_\w}{1-\delta_\w} + \frac{K_\s}{1-\delta_\s}\right)^{-1} 
+\frac{k_\w  M}{D}\cdot
\end{align} 
\begin{theorem}[Upper bound]\label{thm:converse}
The capacity-memory tradeoff $C(M)$ is upper bounded as
\begin{equation}\label{eq:u}
C(M)\leq \min_{k_\w\in\{0,\ldots, K_\w\}} R_{k_\w}(M).
\end{equation}
\end{theorem}
\begin{IEEEproof} {In section~\ref{sec:general_upperbound} we  derive an upper bound on the capacity-memory tradeoff for a general degraded BC with arbitrary cache sizes at the receivers, see Theorem \ref{thm:general_upperbound}. We then specialize this upper bound to packet-erasure BCs with arbitrary cache sizes and erasure probabilities in Corollary~\ref{cor:BEC}, and we show how the upper bound in \eqref{eq:u} is obtained  from this corollary. }
\end{IEEEproof}

%\mw{
%The upper bound corresponding to a specific value of $k_\w$ is obtained by considering only $k_\w$ weak receivers and their cache memories, see Section~\ref{sec:upperThm}. Without caches, the best upper bound is always received for $k_\w=K_\w$. With increasing cache sizes, smaller values of $k_\w$ lead to tighter bounds because some of the cache memories can be ignored.
%}

\begin{figure}[htb!]
\centering
\begin{tikzpicture} [every pin/.style={fill=white},scale=.9]
  \begin{axis}[scale=1,
width=0.5\columnwidth,
scale only axis,
xmin=0,
xmax=25,
xmajorgrids,
xlabel={Memory $M$},
ymin=0.2,
ymax=.5,
ymajorgrids,
ylabel={Capacity $C(M)$},
axis x line*=bottom,
axis y line*=left,
legend pos=south west,
legend style={draw=none,fill=none,legend cell align=left, font=\normalsize}
]
%\addplot +[mark=none] table [x index=0, y index=1] {\datatable};

 \addplot[color=blue,solid,line width=1.0pt,mark=*]
 table[row sep=crcr]{
   0    0.2500\\
    2.0548    0.3836\\
    6.8681    0.4505\\
   12.6386    0.4789\\
   18.7500    0.4926\\
   25.0000    0.5000\\};
    \addlegendentry{Joint Cache-Channel Coding (Theorem \ref{thm:ach})}
 \addplot[color=red,solid,line width=1.0pt]
 table[row sep=crcr]{
         0    0.2500\\
    0.5000    0.3000\\
    1.0000    0.3500\\
    1.5000    0.4000\\
    2.0000    0.4436\\
    2.5000    0.4678\\
    3.0000    0.4836\\
    3.5000    0.4940\\
    4.0000    0.5000\\
   25.0000    0.5000\\
};
    \addlegendentry{Upper Bound of Theorem \ref{thm:3}}
     \addplot[color=green!50!black,dashed,line width=1.0,mark=+]
 table[row sep=crcr]{  
             0    0.2500\\
    4.5455    0.3636\\
   10.7143    0.4286\\
   17.6471    0.4706\\
   25.0000    0.5000\\   };
    \addlegendentry{Separate Cache-Channel Coding (Proposition \ref{prop:separate})}

\end{axis}

\end{tikzpicture}
\caption{Bounds on  the capacity-memory tradeoff $C(M)$ for $K_\w=4$, $K_\s=16$, $D=50$,  $\delta_\w=0.8$, $\delta_\s=0.2$, and $F=10$.}
\label{fig:boundsasym}
\end{figure}
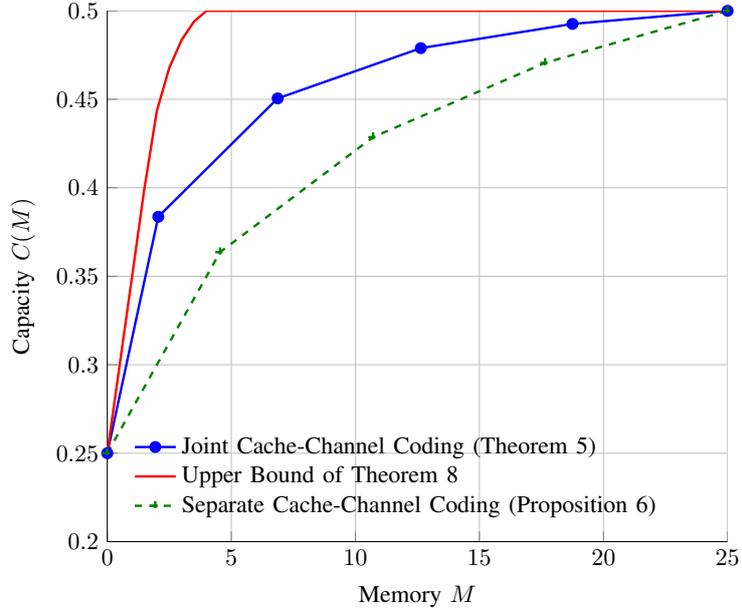

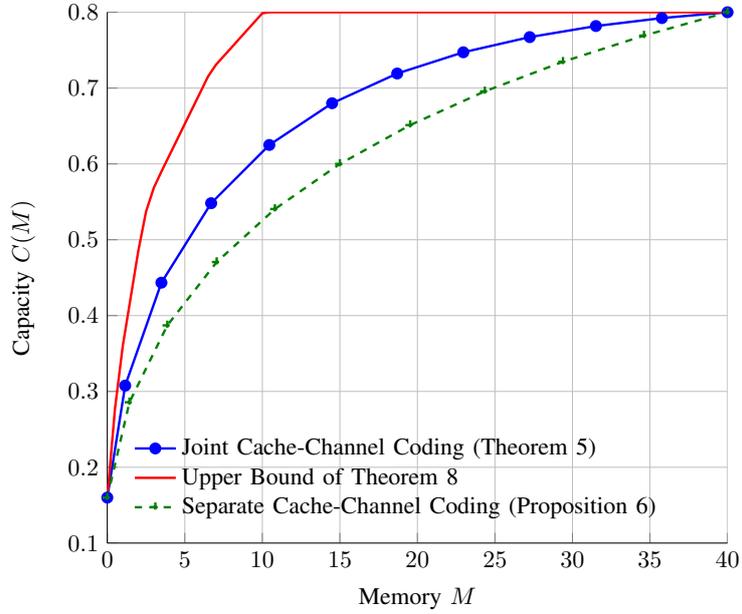
\begin{figure}[htb!]
\centering
\begin{tikzpicture} [every pin/.style={fill=white},scale=.9]
  \begin{axis}[scale=1,
width=0.5\columnwidth,
scale only axis,
xmin=0,
xmax=40,
xmajorgrids,
xlabel={Memory $M$},
ymin=0.1,
ymax=.8,
ymajorgrids,
ylabel={Capacity $C(M)$},
axis x line*=bottom,
axis y line*=left,
legend pos=south west,
legend style={draw=none,fill=none,legend cell align=left, font=\normalsize}
]
%\addplot +[mark=none] table [x index=0, y index=1] {\datatable};

 \addplot[color=blue,solid,line width=1.0pt,mark=*]
 table[row sep=crcr]{
           0    0.1600\\
    1.1538    0.3077\\
    3.4906    0.4434\\
    6.7005    0.5482\\
   10.4508    0.6250\\
   14.5000    0.6800\\
   18.6986    0.7192\\
   22.9613    0.7472\\
   27.2414    0.7672\\
   31.5147    0.7818\\
   35.7692    0.7923\\
   40.0000    0.8000\\    };
    \addlegendentry{Joint Cache-Channel Coding (Theorem \ref{thm:ach})}
 \addplot[color=red,solid,line width=1.0pt]
 table[row sep=crcr]{
 0    0.1600\\
    0.5000    0.2781\\
    1.0000    0.3603\\
    1.5000    0.4228\\
    2.0000    0.4853\\
    2.5000    0.5372\\
    3.0000    0.5685\\
    3.5000    0.5903\\
    4.0000    0.6111\\
    4.5000    0.6319\\
    5.0000    0.6528\\
    5.5000    0.6736\\
    6.0000    0.6944\\
    6.5000    0.7153\\
    7.0000    0.7305\\
    7.5000    0.7419\\
    8.0000    0.7532\\
    8.5000    0.7646\\
    9.0000    0.7760\\
    9.5000    0.7873\\
   10.0000    0.7987\\
   10.5000    0.8000\\
   40.0000    0.8000\\
};
    \addlegendentry{Upper Bound of Theorem \ref{thm:3}}
 \addplot[color=green!50!black,dashed,line width=1.0,mark=+]
 table[row sep=crcr]{  
               0    0.1600\\
    1.4286    0.2857\\
    3.8710    0.3871\\
    7.0588    0.4706\\
   10.8108    0.5405\\
   15.0000    0.6000\\
   19.5349    0.6512\\
   24.3478    0.6957\\
   29.3878    0.7347\\
   34.6154    0.7692\\
   40.0000    0.8000\\
        };
    \addlegendentry{Separate Cache-Channel Coding (Proposition \ref{prop:separate})}

\end{axis}

\end{tikzpicture}
\caption{Bounds on capacity-memory tradeoff $C(M)$ for $K_\w=K_\s=10$, $D=50$,  $\delta_\w=0.8$, $\delta_\s=0.2$, and $\F=50$.}
\label{fig:boundssym}
\end{figure}

We numerically compare our upper and lower bounds on the capacity-memory tradeoff in 
Figures~\ref{fig:boundsasym} and~\ref{fig:boundssym}.

%%%
%%%
{
\subsection{Special case  of $K_\w=1$}

We first evaluate our bounds for a setup with a single weak receiver and any number of strong receivers.  Let %For simplicity, we assume that $K_\s+1$ divides $D$. 

\begin{IEEEeqnarray}{rCl}
\Gamma_1&:= &F \frac{(1-\delta_\s)}{K_\s} \frac{(\delta_w-\delta_\s)}{\left(K_\s(1-\delta_w)+(1-\delta_\s)\right)},\\[5pt]
 \Gamma_2&:=&\frac{(1-\delta_\s)}{K_\s}F,\\[4pt]
\Gamma_3&:=&F \frac{(1-\delta_\s)}{K_\s}\frac{(1-\delta_\s)}{ \left(K_\s(1-\delta_w)+(1-\delta_\s)\right)}.
\end{IEEEeqnarray}
Notice that $0\leq \Gamma_1 \leq \Gamma_3\leq \Gamma_2$.
From Theorems~\ref{thm:ach} and \ref{thm:converse} we obtain the following corollary.
  \begin{corollary} \label{cor:1}
If $K_\w=1$ % and $K_\s=1$,
the capacity-memory tradeoff is lower bounded by:
 \begin{align}\label{eq:lowerbound1}
 C(\M) \geq \begin{cases}
 F\frac{(1-\delta_\w)(1- \delta_\s)}{K_\s(1 -\delta_\w)+(1 -\delta_\s)} + \frac{\M}{D}, & \textnormal{ if } \frac{\M}{D} \in[0,\Gamma_1]  \\
F\frac{(1- \delta_\s)}{1+K_\s} + \frac{\M}{(1+K_\s)D},   & \textnormal{ if } \frac{\M}{D} \in( \Gamma_1, \Gamma_2],%\\
  %F(1-\delta_\s) &  \textnormal{ if } \frac{\M}{D} = \Gamma_2,
 \end{cases}
 \end{align}
 and upper bounded by:
 \begin{align}\label{eq:upperbound1}
 C(\M) \leq \begin{cases}
 F\frac{(1-\delta_\w)(1- \delta_\s)}{K_\s(1 -\delta_\w)+(1 -\delta_\s)} + \frac{\M}{D}, & \textnormal{ if } \frac{\M}{D} \in[0, \Gamma_3]  \\
  F\frac{(1-\delta_\s)}{K_\s} &  \textnormal{ if } \frac{\M}{D} \in (\Gamma_3, \Gamma_2].
 \end{cases}
 \end{align}
  \end{corollary} 
Figure~\ref{fig:boundsK2} shows these two bounds and the bound in Proposition~\ref{prop:separate} for {{$K_\w=1$, $K_\s=10$, $D=22$,}} $D=10$,  $\delta_\w=0.8$, $\delta_\s=0.2$, and $\F=10$.
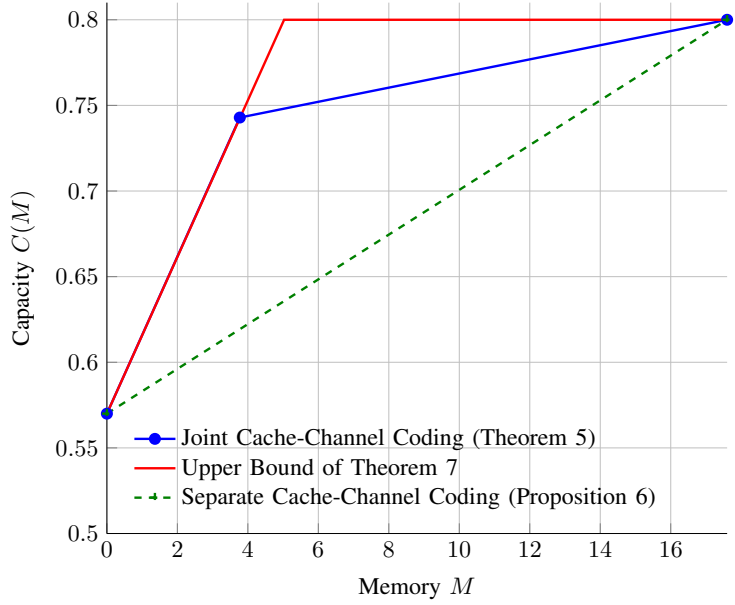
\begin{figure}[t!]
\centering
\begin{tikzpicture} [every pin/.style={fill=white},scale=.9]
  \begin{axis}[scale=1,
width=.5\columnwidth,
scale only axis,
xmin=0,
xmax=17.6,
xmajorgrids,
xlabel={Memory $M$},
ymin=0.5,
ymax=.81,
ymajorgrids,
ylabel={Capacity $C(M)$},
axis x line*=bottom,
axis y line*=left,
legend pos=south west,
legend style={draw=none,fill=none,legend cell align=left, font=\normalsize}
]
%\addplot +[mark=none] table [x index=0, y index=1] {\datatable};

 \addplot[color=blue,solid,line width=1.0pt,mark=*]
 table[row sep=crcr]{
    0    0.57\\
          3.7714 .7429\\
   17.6000    .8000\\
        };
    \addlegendentry{Joint Cache-Channel Coding (Theorem \ref{thm:ach})}
 \addplot[color=red,solid,line width=1.0pt]
 table[row sep=crcr]{
           0    0.57\\
        5.0286 .80  \\
   17.6000    .8000\\
  };
    \addlegendentry{Upper Bound of Theorem \ref{thm:converse}}

      \addplot[color=green!50!black,dashed,line width=1.0,mark=+]
 table[row sep=crcr]{  
                 0    0.57\\
   17.6000    .8000\\
};
    \addlegendentry{Separate Cache-Channel Coding (Proposition \ref{prop:separate})}

\end{axis}

\end{tikzpicture}
\caption{{Bounds on the capacity-memory tradeoff for $K_\w=1$, $K_\s=10$, $D=22$,  $\delta_\w=0.8$, $\delta_\s=0.2$, and $\F=10$.}}
\label{fig:boundsK2}
\vspace{-3mm}
\end{figure}

We identify two regimes. In the first regime $0\leq \frac{M}{D}\leq \Gamma_1$, the cache memory allows reducing the rate $R$ to each receiver by $\frac{M}{D}$. This is the same performance as when a naive uncoded caching strategy  is used in a setup where \emph{all $K_\s+1$ receivers} have cache memories of rate $\M$. %The single cache at the weak receiver thus seems to serve both receivers in the network. 
In the first regime, our joint cache-channel coding scheme thus enables all receivers to profit from the single cache memory and provides the best possible global caching gain.
In the second regime $\Gamma_1 < \frac{M}{D} \leq \Gamma_2$ the gains are not as significant as in the first regime, but  increasing the cache size still results in an improved performance. This won't be the case for $\frac{M}{D}>\Gamma_2$.% the weak receivers have all relevant information in their caches and thus the communication is only restricted by the communication to the stronger receivers. In this regime the performance is independent of the cache sizes. 

In the first regime, $0\leq \frac{M}{D} \leq \Gamma_1$, our 
joint cache-channel coding scheme of section~\ref{sec:joint_cache_channel} achieves the capacity-memory tradeoff~$C(M)$.

}

 \subsection{Special case $K_\w=K_\s=1$ and $D=2$}
For this special case we present tighter upper and lower bounds on $C(M)$. These new bounds meet for a large range of cache memory sizes $\M$. Let 
\begin{IEEEeqnarray}{rCl}
\tilde{\Gamma}_1 &:=&\F \frac{ (1-\delta_\w)^2+(1-\delta_\s)^2- (1-\delta_\w)(1-\delta_\s)}{(1-\delta_\w)+(1-\delta_\s)},\\
\tilde{\Gamma}_2&:=&  \frac{1}{2}\F\left((1-\delta_\s)+(\delta_\w-\delta_\s)\right).
\end{IEEEeqnarray}
\begin{theorem}\label{thm:3}
If $K_\w=K_\s=1$ and $D=2$, the capacity-memory tradeoff is upper bounded as: 
\begin{align}\label{eq:upper}
C(\M) \leq \begin{cases}
\F\frac{(1-\delta_\w)(1- \delta_\s)}{(1 -\delta_\w)+(1 -\delta_\s)} + \frac{\M}{2}, & \textnormal{ if } \frac{\M}{2} \in[0,\tilde{\Gamma}_1]  \\
\F\frac{1}{3} (2- \delta_\s-\delta_\w) + \frac{\M}{3},   & \textnormal{ if } \frac{\M}{2} \in( \tilde{\Gamma}_1, \tilde{\Gamma}_2]\\
  \F(1-\delta_\s) &  \textnormal{ if } \frac{\M}{2} \in( \tilde{\Gamma}_2, \Gamma_2].
 \end{cases}
\end{align}
and  lower bounded as: 
{\small
\begin{align}
C(\M) \geq\! \begin{cases}
\F\dfrac{(1-\delta_\w)(1- \delta_\s)}{(1 -\delta_\w)+(1 -\delta_\s)} + \dfrac{\M}{2}, & \dfrac{\M}{2} \in [0,{\Gamma}_1]  \\
\dfrac{(1-\delta_s)}{3(1-\delta_\s)-(1-\delta_\w)}\left(\F(1\!-\!\delta_\s)+M\right),  & \dfrac{\M}{2} \in( {\Gamma}_1, \tilde{\Gamma}_2]\\
  \F(1-\delta_\s) &  \dfrac{\M}{2} \in( \tilde{\Gamma}_2, \Gamma_2].
 \end{cases}\label{lower-thm4}
\end{align}
}
\end{theorem}
\begin{IEEEproof}
Lower bound \eqref{lower-thm4} coincides with the 
%The lower bound in Theorem~\ref{thm:4} may equivalently be written as the 
upper convex hull of the three rate-memory pairs: $(R_0, \ M_0)$ in \eqref{eq:extremepoin1};  $(R_1,\ M_1)$ in~\eqref{eq:pointt}; and $(\F(1-\delta_\s), \ 2 \tilde{\Gamma}_2)$.
Achievability of the former two pairs follows from Theorem~\ref{thm:ach}.  Achievability of the last pair  is proved in appendix~\ref{sec:coded_lowerbound}. The upper bound is proved in appendix~\ref{sec:upper2}. 
\end{IEEEproof}

Figure~\ref{fig:boundsD2} shows the bounds of Theorem~\ref{thm:3} for $\delta_\w=0.8$, $\delta_\s=0.2$, and $\F=10$. 
\begin{figure}[t!]
\centering
\begin{tikzpicture} [every pin/.style={fill=white},scale=.9]
  \begin{axis}[scale=1,
width=0.5\columnwidth,
scale only axis,
xmin=0,
xmax=16,
xmajorgrids,
xlabel={Memory $M$},
ymin=0,
ymax=8.5,
ymajorgrids,
ylabel={Capacity $C(M)$},
axis x line*=bottom,
axis y line*=left,
legend pos=south west,
legend style={draw=none,fill=none,legend cell align=left, font=\small}
]
%\addplot +[mark=none] table [x index=0, y index=1] {\datatable};

 \addplot[color=blue,solid,line width=1.0pt,mark=*]
 table[row sep=crcr]{
          0    1.6000\\
    9.6000    6.4000\\
   14.0000    8.0000\\        };
    \addlegendentry{Joint Cache-Channel Coding (Theorem \ref{thm:3})}
 \addplot[color=red,solid,line width=1.0pt]
 table[row sep=crcr]{
                0    1.6000\\
    0.5000    1.8500\\
    1.0000    2.1000\\
    1.5000    2.3500\\
    2.0000    2.6000\\
    2.5000    2.8500\\
    3.0000    3.1000\\
    3.5000    3.3500\\
    4.0000    3.6000\\
    4.5000    3.8500\\
    5.0000    4.1000\\
    5.5000    4.3500\\
    6.0000    4.6000\\
    6.5000    4.8500\\
    7.0000    5.1000\\
    7.5000    5.3500\\
    8.0000    5.6000\\
    8.5000    5.8500\\
    9.0000    6.1000\\
    9.5000    6.3500\\
   10.0000    6.6000\\
   10.5000    6.8333\\
   11.0000    7.0000\\
   11.5000    7.1667\\
   12.0000    7.3333\\
   12.5000    7.5000\\
   13.0000    7.6667\\
   13.5000    7.8333\\
   14 8\\
   16.0000    8.0000\\};
    \addlegendentry{Upper Bound of Theorem \ref{thm:3}}
    \addplot[color=green!50!black,dashed,line width=1.0,mark=+]
 table[row sep=crcr]{  
           0    1.6000\\
   16.0000    8.0000\\          };
    \addlegendentry{Separate Cache-Channel Coding (Proposition \ref{prop:separate})}
   
\draw[fill] (axis cs:{ 14, 8}) node[above right] {$B$};
\draw[fill] (axis cs:{ 9.5, 6.5}) node[above left] {$A$};

\end{axis}

\end{tikzpicture}
%\vspace{-4mm}
\caption{Bounds on the capacity-memory tradeoff for $K_\w=1$, $K_\s=1$, $D=2$,  $\delta_\w=0.8$, $\delta_\s=0.2$, and $\F=10$.}
\label{fig:boundsD2}
\end{figure}
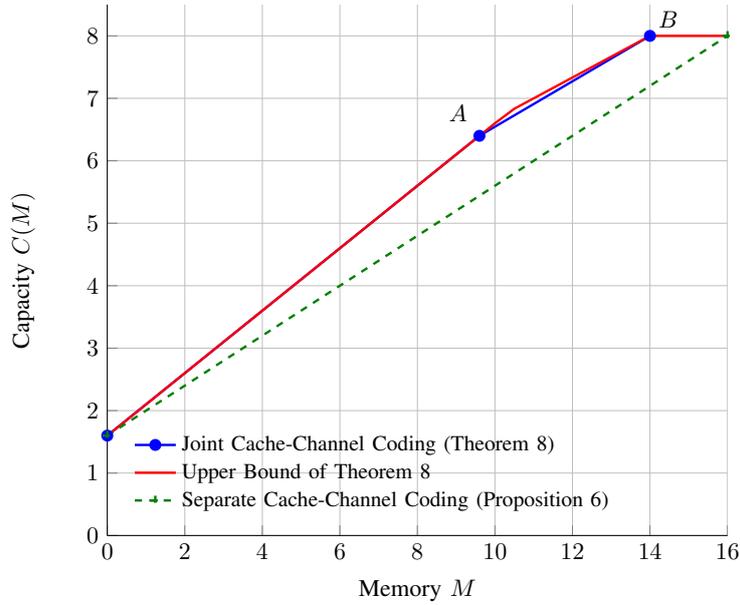
\begin{corollary}
The minimum cache size $M$ for which communication is possible at the maximum rate $F(1-\delta_\s)$ is  $2\tilde{\Gamma}_2$.
\end{corollary}

Upper and lower bounds of Theorem~\ref{thm:3}  coincide in the case of equal erasure probabilities $\delta_\w=\delta_\s$:
\begin{corollary} 
\label{cor:equaldelta}
If $K_\w=K_\s=1$, $D=2$ and $\delta_\w=\delta_\s=\delta$: % the upper bound in Theorem~\ref{thm:3} is tight and  the capacity-memory tradeoff  equals
\begin{align}
C(\M) = \begin{cases}
\F\frac{1}{2} (1-\delta)+\frac{\M}{2}, & \textnormal{if } \frac{\M}{2} \in\big[0, \ \frac{1}{2} \F (1-\delta)\big]  \\ 
 \F(1-\delta) &  \textnormal{if } \frac{\M}{2}\in \big(\frac{1}{2} \F (1-\delta), \ \Gamma_2\big].
 \end{cases}
\end{align}
\end{corollary} 
\begin{IEEEproof}
It follows from Theorem~\ref{thm:3} because for $\delta_\w=\delta_\s$: $\tilde{\Gamma}_1=\tilde{\Gamma}_2=\frac{1}{2} \F (1-\delta)$, and in the regime $\frac{M}{2} \in(\Gamma_1, \tilde{\Gamma}_2]$  lower bound~\eqref{lower-thm4} specialises to $C(M) \geq \F\frac{1}{2} (1-\delta)\!+\! \frac{\M}{2}$.
\end{IEEEproof}

%%%%
%%%%
%%%%

\section{A Joint Cache-Channel Scheme for Arbitrary Demands} \label{sec:joint_cache_channel}
%\mw{The scheme is built on three key ideas: 1) the capacity of the packet-erasure BC; 2) a coding idea that we recently introduced for the BC with feedback \cite{wuwigger-2016} and that we term here  \emph{piggyback} coding; and 3) the \emph{coded caching} idea that Maddah-Ali and Niesen used for cache-aided communication over a common noise-free pipe \cite{maddahali_niesen_2014-1}.}

{We describe a joint cache-channel coding scheme  parameterised by}
\begin{equation}\label{eq:t}
t\in\{1,\ldots, K_\w\}.
\end{equation}
 {\ssb{We show in} subsection~\ref{sec:analysis} that}, for parameter $t$, this scheme  achieves the rate-memory pair $(\R_t,\M_t)$ in~\eqref{eq:pointt}. 

%%%
%%%
%%%

\subsection{Preparations}

For each $d\in\{1,\ldots, D\}$, split message $W_d$ into two parts:
\begin{equation}\label{eq:submessage}
W_{d} =\big(W_{d}^{(t-1)}, \ W_{d}^{(t)}\big)
\end{equation}
of rates
\begin{subequations}\label{eq:t_rates}
\begin{IEEEeqnarray}{rCl}
R^{(t-1)} &= & R \left( 1+ \frac{K_\w-t+1}{t K_\s} \cdot \frac{\delta_\w-\delta_\s}{1-\delta_\w}\right)^{-1}\label{eq:Rtminus}\\
R^{(t)}&=& R \left( 1+ \frac{t K_\s}{K_\w-t+1} \cdot \frac{1-\delta_\w}{\delta_\w-\delta_\s}\right)^{-1}
\label{eq:Rt}
\end{IEEEeqnarray}
\end{subequations}
 {where $R^{(t-1)}+R^{(t)}=R$}.
  
%%%
%%%
%%%  
  
\subsection{Caching phase}  

 {The first step is to cache information about both parts of each message in~\eqref{eq:submessage} at the weak receivers. Specifically, we first apply Method} \texttt{Ca} with $\tilde{K}=K_\w$ and $\tilde{t}=t$ to messages $W_{1}^{(t)}, \ldots, W_{D}^{(t)}$.  {We then} apply Method \texttt{Ca} with $\tilde{K}=K_\w$ and $\tilde{t}=t-1$ to messages $W_{1}^{(t-1)}, \ldots, W_{D}^{(t-1)}$.

In the following, we  {will use the} superscript $(t)$ to  {identify the outputs of Methods} \texttt{Ca}, \texttt{En}, and \texttt{De$_i$} with $\tilde{t}=t$.  {Similarly, we will use the superscript ${(t-1)}$ to identify these outputs for $\tilde{t} = t - 1$}.  \ssb{We use the notation that we introduced in Section~\ref{sub:MAN}.}

 {Consider an arbitrary weak receiver $i\in\Kw$.} The total cache content  {at this receiver is}
 \begin{IEEEeqnarray}{rCl}\label{eq:cache_content}
{V}_i& =& V_{i}^{(t)} \cup V_i^{(t-1)} \nonumber \\
& = & \Big\{ W_{d,\set{G}_{\ell}^{(t)}}^{(t)} \colon \ d\in\{1,\ldots, D\}\; \textnormal{and}\; k \in \set{G}_{\ell}^{(t)}\bigg\} \nonumber \\
& & \bigcup  \bigg\{ W_{d,\set{G}_{\ell}^{(t-1)}}^{(t-1)} \colon \ d\in\{1,\ldots, D\}\;  \textnormal{and} \; k \in \set{G}_{\ell}^{(t-1)}\Big\}. \IEEEeqnarraynumspace
\end{IEEEeqnarray}

%%%
%%%
%%%
 
\subsection{Delivery phase} 

The delivery phase takes place in three subphases consisting of $\beta_1 n$, $\beta_2  n$, and $\beta_3  n$ channel uses, where $\beta_1, \beta_2, \beta_3\geq 0$ and
\begin{equation}\label{eq:nsum}
\beta_1+\beta_2+\beta_3=1.
\end{equation}

\subsubsection{ {Delivery subphase~1}}

 {Here we  \ssb{only} consider the  weak receivers, and we  communicate the ``$t$ parts'' of their demanded messages (see~\eqref{eq:submessage}) using separate source and channel coding. (The strong receivers will not participate in this subphase.)} 

The transmitter proceeds in two steps:
\begin{itemize}
\item[\textit{T1:}]  {The transmitter} applies Method \texttt{En} with $\tilde{K} = K_\w$ and $\tilde{t} = t$ to demand vector $(d_1, \ldots, d_{K_\w})$ and to messages
\begin{equation*}
\Big\{W_{d_i}^{(t)} \colon \ i\in\set{K}_\w\Big\}.
\end{equation*}
 {
Let 
\begin{equation}\label{Eqn:XORSubPhase1}
\Big\{ W^{(t)}_{\text{XOR},\set{S}} : \set{S} \subseteq \{1,\ldots,K_\w\},\ |\set{S}| = t + 1\Big\}
\end{equation}
denote the output of Method \texttt{En}.
}
\item[\textit{T2:}]  {The transmitter} uses a capacity-achieving code for the packet-erasure BC to send the XORs  {in~\eqref{Eqn:XORSubPhase1}} to the weak receivers.
\end{itemize}

Each weak receiver $i\in \set{K}_{\w}$ decodes in two steps: 
\begin{itemize} 
\item[\textit{R1:}]  {Receiver $i$ recovers all transmitted XOR messages in~\eqref{Eqn:XORSubPhase1} using an appropriate channel decoder.} 
\item[\textit{R2:}]  {Receiver $i$} applies method \texttt{De$_i$} with  $\tilde{K} = K_\w$ and $\tilde{t} = t$ to demand vector $(d_1, \ldots, d_{K_\w})$ and to the XOR messages produced in step~R1. 
 {For $i \in \Kw$, let
\begin{equation}\label{eq:dec1}
\hat{W}_{d_i}^{(t)} = \Big(\hat{W}_{d_i,\set{G}^{(t)}_1},\hat{W}_{d_i,\set{G}^{(t)}_2},\ldots,\hat{W}_{d_i,\set{G}^{(t)}_{\binom{K_\w}{t}}} \Big)
\end{equation}
denote the output produced by \texttt{De$_i$}.}
\end{itemize}

%%%
%%%
%%%

\subsubsection{{Delivery subphase~2}}

{Here we consider \ssb{all} receivers. To the strong receivers: We communicate the ``$t$ parts'' of their demanded messages. To the weak receivers: We communicate the ``$(t-1)$ parts'' of their demanded messages. Both communications will be done simultaneously using joint cache-channel coding (via piggyback coding).} 

The transmitter proceeds in two steps:
\begin{itemize}

\item[\emph{T1:}] {The transmitter} applies Method \texttt{En} with $\tilde{K}=K_\w$ and $\tilde{t}=t-1$ to demand vector $(d_1,\ldots, d_\w)$ and messages 
\begin{equation*}
\Big\{ W_{d_i}^{(t-1)} \colon \  i\in\set{K}_\w\Big\}.  
\end{equation*} 
{Method \texttt{En} outputs} an XOR message for each size-$t$ subset of $\{1,\ldots, K_\w\}$, {for example}, see \eqref{eq:xor}. {We denote} these XOR messages by\footnote{{The messages in~\eqref{Eqn:JointCacheChannelXORMessages} have the superscript $(t-1)$, because they correspond to the output of Method \texttt{En} with the parameter $\tilde{t}=t-1$. In contrast, $\{\set{G}_\ell^{(t)}\}$ have superscript $(t)$ because they correspond to subsets of size $\tilde{t}=t$.}}
\begin{equation}\label{Eqn:JointCacheChannelXORMessages}
\left\{ W_{\text{XOR},\set{G}_\ell^{(t)}}^{(t-1)}\colon \; \ell=1, \ldots, {{K_\w} \choose t} \right\}.
\end{equation}

\item[\emph{T2:}] Time-sharing is performed over ${K_\w} \choose t$ {different} periods, where each period is associated with a size-$t$ subset of $\{1,2,\ldots, K_\w\}$.  {Consider the $\ell$-th subset $\set{G}_\ell^{(t)}$.}  {First recall that the subset $\set{G}_\ell^{(t)}$ of weak receivers has 
\begin{equation*}
\Big\{ W_{d_j, \mathcal{G}_\ell^{(t)}}^{(t)} \colon \  j\in\set{K}_\s\Big\}
\end{equation*}
stored as ``side information'' in their cache memories.} The transmitter uses piggyback coding to send 
\begin{subequations}\label{eq:single}
\begin{equation}
\Big\{ W_{d_j,  \mathcal{G}_\ell^{(t)}}^{(t)} \colon \  j\in\set{K}_\s\Big\}
\end{equation}
to all strong receivers $\Ks$ and the XOR message
\begin{IEEEeqnarray}{rCl}\label{eq:di}
W_{{\text{XOR},}\mathcal{G}_\ell^{(t)}}^{(t-1)}
\end{IEEEeqnarray}  
\end{subequations}
to all weak receivers in $\set{G}_\ell^{(t)}$.
\end{itemize}

Each strong receiver $j\in\Ks$ performs piggyback decoding (for the receiver without side-information) for all ${K_\w} \choose t$ transmission periods. It forms the ${K_\w}\choose t$-tuple \ssb{estimate}
\begin{equation}\label{eq:dec2}
\hat{W}_{d_j}^{(t)}:=\Big(\hat{W}_{d_j,\set{G}_1^{(t)}}^{(t)}, \ldots, \hat{W}_{d_j,\set{G}^{(t)}_{\binom{{K_\w}}{t}}}^{(t)}\Big), \quad j\in \Ks.
\end{equation}
 
Each weak receiver~$i\in\Kw$ proceeds in two steps: 
\begin{itemize}
\item[\emph{R1:}] {Receiver $i$ considers each subset $\set{G}_\ell^{(t)} $ of size $t$ to which it belongs (i.e., each $\set{G}_\ell^{(t)} \subseteq \{1,\ldots, K_\w\}$ such that  $\set{G}_\ell^{(t)} \ni i$), and it} decodes the XOR message $W_{{\text{XOR},}\set{G}_\ell^{(t)}}^{(t)}$ by applying piggyback decoding (for the receiver with side-information) to  the {channel} outputs of the period {associated with} $\set{G}_{\ell}^{(t)}$.% and to the  side-information $\{W_{d_j,\mathcal{G}_{\ell}^{(t)}}^{(t)} \colon j\in\Ks\}$ stored in its cache. 

\item[\emph{R2:}] {Receiver $i$ then} applies Method \texttt{De$_i$} with $\tilde{K}=K_\w$ and $\tilde{t}=t-1$ to demand vector $(d_1,\ldots, d_\w)$, the XOR messages decoded in step~R1 and its cache content $V_i$. {Let
\begin{equation}\label{eq:dec2b}
\hat{W}_{d_i}^{(t-1)}, \quad i\in\Kw
\end{equation}
denote the output of Method \texttt{De$_i$}.}
\end{itemize}

\subsubsection{{Delivery subphase~3}} 
{Here we consider only the strong receivers, and we will communicate the remaining ``$(t-1)$ parts'' of their demanded messages. (The weak receivers will not participate in this subphase.)} The transmitter {communicates} 
\begin{equation*}
\Big\{ W_{d_j}^{(t-1)} \colon \  j\in\set{K}_\s\Big\}
\end{equation*}
to the strong receivers using a  capacity-achieving code for the packet-erasure BC. Each receiver uses an optimal decoding method to produce the \ssb{estimate}
\begin{equation} \label{eq:dec3}
\hat{W}_{d_j}^{(t-1)},  \  j\in\set{K}_\s.
\end{equation}

\subsubsection{{Final decoding}}
 
Each receiver~$k\in\{1,\ldots, K\}$ {outputs}
\begin{equation}
\hat{W}_{d_k}=\Big(\hat{W}_{d_k}^{(t-1)}, \hat{W}_{d_k}^{(t)}\Big).
\end{equation}

%%%
%%%
%%%

\subsection{Analysis}\label{sec:analysis}

{Fix} $t\in\{1,\ldots, K_\w\}$. {We show that the above scheme} achieves {the} rate-memory pair $(R_t, \M_t)$ in \eqref{eq:pointt}.

\subsubsection{{Caching phase}}
 
By \eqref{eq:rate_memory},  our caching strategy requires a cache memory size of
 \begin{IEEEeqnarray}{rCl} \label{eq:MM}
 \M& = &R^{(t)}\cdot D \frac{t}{K_\w} + R^{(t-1)} \cdot D \frac{t-1}{K_\w} \nonumber \\
 & = &R \cdot \frac{D}{K_\w} \left( t- \left(1+\frac{K_\w-t+1}{t K_\s} \cdot \frac{\delta_\w-\delta_\s}{1-\delta_\w} \right)^{-1}\right). \IEEEeqnarraynumspace
 \end{IEEEeqnarray} 
 
We now analyse the probability of decoding error. We present conditions under which the probability that the \ssb{estimates} produced in Subphases~1--3, \eqref{eq:dec1}, \eqref{eq:dec2}, \eqref{eq:dec2b}, and \eqref{eq:dec3} \ssb{are not equal to the corresponding} submessages in \eqref{eq:submessage} tends to 0 as $n\to \infty$.

\subsubsection{{Delivery subphase~1}} 

Proposition~\ref{Prop:CapacityRegion} combined with Lemma~\ref{lem:MN} and \eqref{eq:Rt}, 
prove that  the probability that the \ssb{estimates} in~\eqref{eq:dec1} are \ssb{incorrect} tends to 0 as $n\to \infty$, whenever
\begin{equation}\label{eq:phase1_rate}
\frac{  R\cdot\frac{K_\w-t}{t+1}}{\left(1+\frac{t K_\s}{K_\w-t+1} \cdot \frac{1-\delta_\w}{\delta_\w-\delta_\s}\right)\cdot F(1-\delta_\w)} < \beta_1.
\end{equation}

\subsubsection{{Delivery subphase~2}}
Consider a single period with the transmission of message in \eqref{eq:single}. Since all weak receivers and all strong receivers are statistically equivalent, the probability that the \ssb{estimates} in \eqref{eq:dec2} and \eqref{eq:dec2b} are \ssb{incorrect} is at most $t\cdot K_\s$ times larger \ssb{than the probability of error in \eqref{eq:dec2} and \eqref{eq:dec2b} for a single weak and a single strong receiver}. %, \eqrdecoding error at a specific of the receivers is at most $t \cdot K_\s$ times the probability that a specific weak receiver in $\set{G}_\ell^{(t)}$ errs or a specific strong receiver in $\Ks$ errs. 
By  Corollary~\ref{Prop:piggyback} and Lemma~\ref{lem:MN}, this latter probability of error tends to 0 (and thus also the original probability of error tends to 0) as  $n\to \infty$, whenever %These results show that the 
%Notice that there are ${K_\w \choose t}$ sets $\set{S}$ and each message $W_{\mathcal{S}}^{(t-1)}$ is of rate
%\begin{equation}
%\tilde{R}^{(t-1)}= R^{(t-1)} \cdot {K_\w \choose t-1}^{-1}. 
%\end{equation} 
%Moreover,  the sum-rate of all messages in \eqref{eq:dj} summed over all possible sets $\set{S}$ equals $R^{(t)}\cdot K_s$. 
%Corollary~\ref{Prop:piggyback} combined with Lemma~\ref{lem:MN} proves that the
%Since the Maddah-Ali \& Niesen decoding is error-free, the only way the guessed messages can be wrong is when the piggyback coding failed. The
% probability of error  for a given period of delivery phase~2 tends to 0 as $n\to \infty$, whenever 
\begin{equation} 
\max\left\{\frac{R^{(t-1)} \cdot \frac{K_\w -t+1}{t}}{F(1-\delta_\w)},  \frac{R^{(t-1)}\cdot  \frac{K_\w -t+1}{t}+R^{(t)} K_\s}{F(1-\delta_\s)}\right\} < \beta_2.
\end{equation} 
By our choice \eqref{eq:t_rates} the two terms in the maximization are equal, and thus by \eqref{eq:Rtminus} we conclude that the probability of producing an error in \eqref{eq:dec2} or \eqref{eq:dec2b}  tends to 0 as $n\to \infty$, whenever
\begin{IEEEeqnarray}{rCl} \label{eq:phase2_rate}
%\frac{R^{(t-1)}  \frac{K_\w -t+1}{t}}{F(1-\delta_\w)}=
 \frac{R \frac{K_\w-t+1}{t} }{\left(1+ \frac{K_\w-t+1}{t K_\s} \cdot \frac{\delta_\w-\delta_\s}{1-\delta_\w} \right) F(1-\delta_\w)}< \beta_2.
\end{IEEEeqnarray}

\subsubsection{{Delivery subphase~3}}
 
 Proposition~\ref{Prop:CapacityRegion} combined with \eqref{eq:Rtminus} prove that the probability of producing a wrong guess in \eqref{eq:dec3} tends to 0 as $n\to \infty$, whenever 
 \begin{equation}\label{eq:phase3_rate}
\frac{R K_\s}{\left(1+ \frac{K_\w-t+1}{t K_\s} \cdot \frac{\delta_\w-\delta_\s}{1-\delta_\w} \right) F(1-\delta_\s)}< \beta_3.
 \end{equation}

\subsubsection{{Overall scheme}}

Combining  \eqref{eq:phase1_rate}, \eqref{eq:phase2_rate}, and \eqref{eq:phase3_rate} and using~\eqref{eq:nsum}, after some algebraic manipulations, we see that the probability of decoding error tends to 0 as $n\to \infty$, whenever
  \begin{IEEEeqnarray}{rCl}
R <  F (1-\delta_\w)\cdot\frac{1+ \frac{K_\w-t+1}{t K_\s} \cdot \frac{\delta_\w-\delta_\s}{1-\delta_\w} }{ \frac{K_\w-t+1}{t}\left(1+ \frac{K_\w-t}{(t+1) K_\s} \cdot \frac{\delta_\w-\delta_\s}{1-\delta_\w}\right) + K_\s\frac{1-\delta_\w}{1-\delta_\s}}.\nonumber
 \end{IEEEeqnarray}
Together with \eqref{eq:MM}, this proves achievability of the rate-memory pair $(R_t, \ \M_t)$ in \eqref{eq:pointt}.

   \section{Extensions of our Joint Cache-Channel Coding Scheme}\label{sec:extension}
The scenario in Section~\ref{sec:definition} allowed for a compact exposition of our new joint cache-channel coding idea. 
This idea however extends also to more general scenarios. In the following subsections we present some ideas.

\subsection{Weak receivers have different erasure probabilities:}   For simplicity, assume that the weak receivers are ordered so that $\delta_{1} \geq \delta_{2} \geq \ldots \geq \delta_{K_\w}$ holds. \ssb{The scheme of section~\ref{sec:joint_cache_channel} may be modified} as follows: For each XOR message sent in delivery phase~1, set the rate of the codebook \ssb{to be} equal  to the capacity of the weakest receiver to whom the XOR message is intended.% 2) For each application of piggyback coding, set the rate of the piggyback codebook  codewords for sending the combined messages in delivery phase 2 are also adapted to the intended set of weak receivers. 

\subsection{Strong receivers have different erasure probabilities:} For simplicity, assume that the strong receivers are ordered so that $\delta_{K_\w+1} \geq \delta_{K_\w+2} \geq \ldots \geq \delta_{K}$. We split the set of strong receivers into a set of moderately strong receivers $K_\w+1,\ldots, j^\star$ and a set of very strong receivers $j^\star, \ldots, K$, where $j^\star$ is chosen depending on the various erasure probabilities and cache sizes. 
We now time-share two schemes whose lengths need to be optimized: In the first period,  a standard capacity-achieving  coding scheme for the packet-erasure BC is used to serve only the moderately strong  receivers $K_\w+1,\ldots, j^\star$. In the second period,   our joint cache-channel coding scheme  is used to serve all other receivers. 

\subsection{Some weaker receivers do not have caches:} We time-share two schemes whose lengths need to be optimized. In the first period, a standard capacity-achieving  code for the packet-erasure BC
is used to serve the
 weak receivers without caches. In  the second  period, our joint cache-channel coding scheme is used to serve all other receivers. 
\subsection{Weak receivers have different cache sizes:} For simplicity, assume that the weak receivers are ordered so that $\M_1 \geq M_2 \geq \ldots \geq \M_{\Kw}$, where $M_i$ denotes the cache memory size at receiver~$i\in\Kw$. 

We time-share~$K_\w$ schemes of equal length.  In period~$i\in\Kw$,  \ssb{we treat the  weak receivers $i+1,\ldots, K_\w$  assuming that they have no zero cache memories and we treat the weak receivers~$1,\ldots, i$ assuming that they have cache memories of size $(\M_i-\M_{i+1})$.} We thus apply the coding scheme that we  sketched in the previous subsection.

\subsection{Strong receivers have cache memories} For simplicity, assume that all strong receivers in $\Ks$ have cache memories of equal size $\M> \M_\s>0$. We time-share two schemes.  In the first period of length $n\big(1- \frac{\M_\s}{\M}\big)$  we suppose that the strong receivers have no caches: we thus use our joint cache channel coding scheme. In the second period of length $n\frac{\M_\s}{\M}$ we suppose that all receivers in the network have cache memory $\M$: we apply separate cache-channel coding combining Maddah-Ali \& Niesen coded caching  over all $K$ receivers with capacity-achieving code for the packet-erasure~BC.

\subsection{Strong receivers have cache memories and {are served additional data}} The strong receivers have cache memories  of size $\M_\s$ as in the previous subsection. There are two libraries now: 
\[\textnormal{library~A: \quad files }W_1^{(\textnormal{A})}, \ldots, W_D^{(\textnormal{A})}\]
 and 
 \[\textnormal{library~B: \quad  files }W_1^{(\textnormal{B})}, \ldots, W_D^{(\textnormal{B})}.\] Each weak receiver~$i\in\Kw$ demands only file $W_{d_i}^{(\textnormal{A})}$ from library A, whereas each strong receiver demands a file $W_{d_j}^{(\textnormal{A})}$  from library A and a file $W_{d_j}^{(\textnormal{B})}$  from library B.
 
We time-share two schemes whose lengths need to be optimized. In the first period, we suppose that the strong receivers have no cache memories and use our joint cache-channel coding scheme for library~A. In the second period, we only serve the strong receivers from library~B. To this end, we apply separate cache-channel coding combining Maddah-Ali \& Niesen coded caching  over all $K$ receivers with capacity-achieving code for the packet-erasure~BC. 

In the scheme that we propose, the caches at the strong receivers are used only to cache messages from library B, but not  from library~A. The idea is that when the strong receivers are sufficiently strong, then our joint cache-channel coding scheme is as performant as if all receivers in the network had caches. One might therefore choose to dedicate  the caches at the strong receivers entirely to the transmission of files from library~B.
 
 \subsection{General DMBCs}
   
   Packet-erasure BCs are simpler than \ssb{general} BCs because time-sharing of optimal point-to-point codes \ssb{for} various receivers achieves capacity without caches.   For our joint cache-channel coding scheme to be effective on more general DMBCs, we will have to partially replace time-sharing by superposition coding (and more generally Marton coding). More specifically, we use superposition coding and superpose the codewords sent in delivery subphase~3 on the piggyback codeword sent in delivery subphase~2 and on the codewords sent in delivery subphase~1. When there is no clear notion of ``weaker" and ``stronger", i.e., the channel is neither \emph{degraded}, \emph{less noisy}, \emph{more capable}, or \emph{essentially less noisy}, then we use Marton coding where the piggyback codewords serve as cloud centers and the other codewords as satellites.

   %\section{Proofs of Upper Bounds}\label{sec:upper}
 %  \input{Sec-Upp-old_v12}

\section{Upper Bound for General Degraded BCs under Arbitrary Demands}\label{sec:general_upperbound}
{We consider a more general setup for the upper bound, where each receiver $i$ has a cache of size $M_i$, and where the broadcast channel is a} discrete memoryless
\emph{degraded BC} with 
 input alphabet $ \set{X}$ and equal output alphabets $ \set{Y}_1, \ldots, \set{Y}_K$. The joint transitional law of the memoryless BC is given by $P_{Y_1Y_2\cdots Y_K|X}(y_1, \ldots, y_K|x)$.  
{We assume that the BC is \emph{degraded}, i.e., the transition law satisfies \ssb{the Markov chain}
 \begin{equation}\label{eq:equation}
 X - Y_K - Y_{K-1} - \cdots - Y_1.
 \end{equation}
 For our problem setup, only the marginal transition law is relevant. Therefore our upper bound holds also for \emph{stochastically degraded BC}, i.e., for transition laws $P_{Y_1,\ldots, Y_K|X}$  for which \ssb{there exists a conditional probability distribution $\tilde{P}_{Y_2|Y_1}, \tilde{P}_{Y_3|Y_2}, \ldots, \tilde{P}_{Y_{K-1}|Y_{K}}$ that satisfies 
\begin{align}\label{eq:degr}
&P_{Y_1Y_2\cdots Y_K|X}(y_1, \ldots, y_K|x)\nonumber\\& = P_{Y_K|X}(y_k|x) \tilde{P}_{Y_{K-1}|Y_{K}}(y_{k-1}|y_{k}) \ldots  \tilde{P}_{Y_1|Y_2}(y_1|y_2)
\end{align}
for all $(x, y_1, y_2,\ldots, y_K)\in\mathcal{X}\times\mathcal{Y}_1\times \mathcal{Y}_{2} \times \cdots \times \mathcal{Y}_K$.} Note that the packet-erasure BC that we study falls in the class of (stochastically) degraded BCs.

\begin{figure}[t!]
\begin{center}
\includegraphics[width=0.42\textwidth]{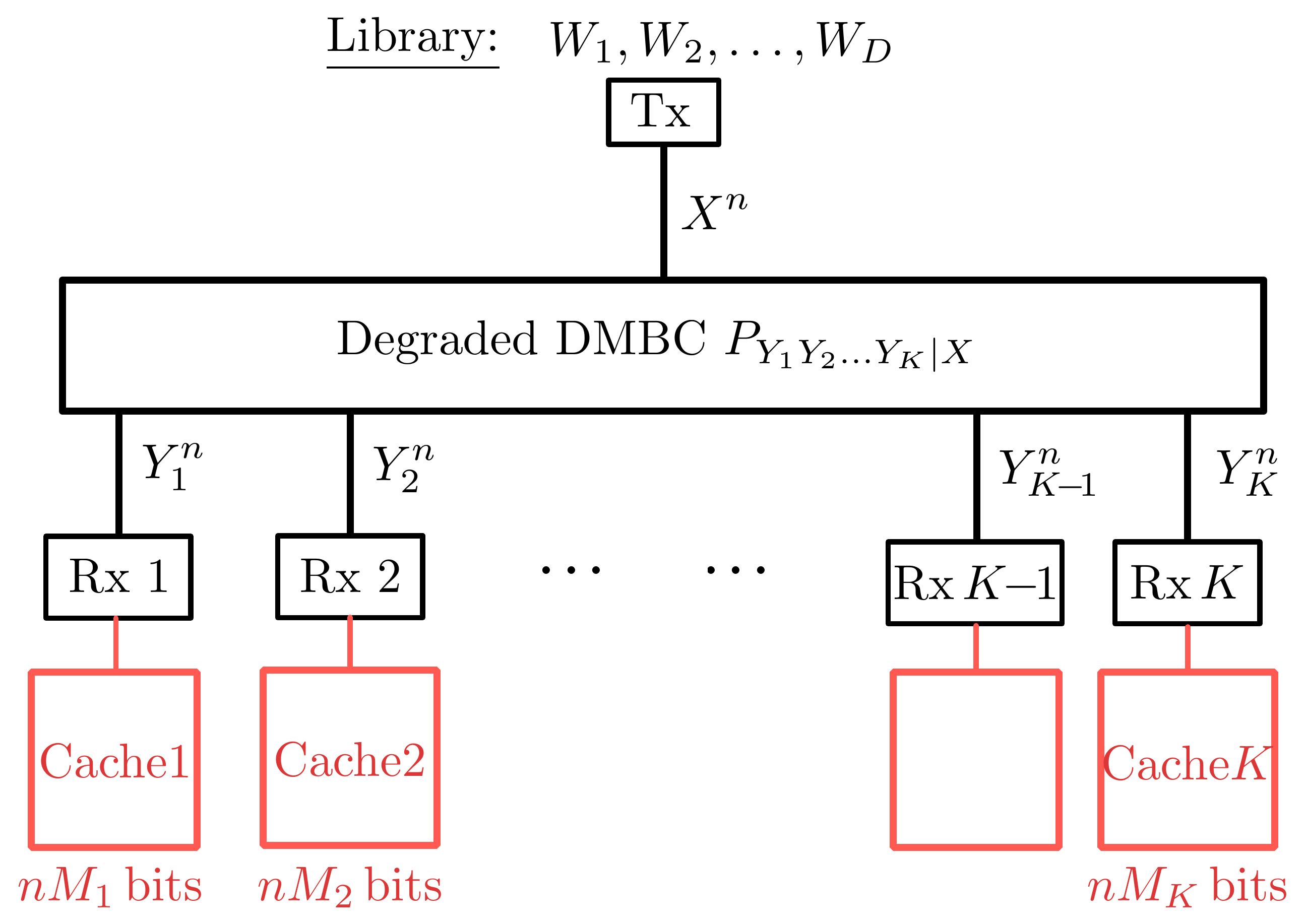}
\caption{Degraded $K$-user BC $P_{Y_1Y_2\cdots Y_K|X}$ where each Receiver~$k\in\set{K}$ has cache memory of size $n\M_k$ bits.}
\label{fig:deg}
\end{center}
\vspace{-5mm}
\end{figure}
The library and the probability of worst-case error $\Pe^{\!\!\textnormal{worst}}$ is defined as before. A rate-memory tuple $(R,\ \M_1, \ldots, \M_K)$ is said \emph{achievable} if for every $\epsilon >  0$ there exists a sufficiently large blocklength $n$ and caching, encoding and decoding functions as in \eqref{eq:caching}--\eqref{eq:decoding} such that $\Pe^{\!\!\textnormal{worst}}<\epsilon$. 
The capacity-memory tradeoff $C(M_1, \ldots, M_K)$ is  defined as the supremum over all rates~$R$ so that $(R, M_1, \ldots, M_K)$ are achievable. 

For each $\set{S}\in\mathcal{K}$, let $R_{\textnormal{sym},\set{S}}$ denote the largest equal-rate that is achievable over a BC with receivers in $\set{S}$ {when there are no cache memories}. We prove the following upper bound on $C(M_1, \ldots, M_K)$: 
\begin{theorem}\label{thm:general_upperbound}
The capacity-memory tradeoff $C(M_1, \ldots, M_K)$ of a degraded BC is upper bounded as \mw{follows:}
\begin{equation*}
C(M_1, \ldots, M_K)\leq \min_{\set{S}\subseteq \{1,\ldots, K\} }\left({R_{\textnormal{sym},\set{S}}}+ \frac{M_\set{S}}{D}\right),
\end{equation*}
{where $M_{\set{S}}=\sum_{k\in\set{S}} M_k$ is the total cache size at receivers in~$\set{S}$.}
\end{theorem}

\begin{remark} 
Theorem~\ref{thm:general_upperbound} also holds for stochastically degraded BCs.
\end{remark}

Before proving Theorem \ref{thm:general_upperbound}, let us specialize it to packet-erasure BCs by using Proposition~\ref{Prop:CapacityRegion} (to find ${R_{\textnormal{sym},\set{S}}}$):
%\textcolor{red}{why do you index $R_{\textnormal{sym},\set{S}}(M_1,\ldots, M_K)$ with the cache sizes?}
\begin{corollary}\label{cor:BEC}
The capacity-memory tradeoff $C(M_1, \ldots, M_K)$ of the packet-erasure BC with packet size $F$, erasure probabilities $\delta_1, \ldots, \delta_K\geq 0$, and cache memory sizes $M_1,\ldots, M_K$, is upper bounded as {\mw{follows:}}
\begin{equation}\label{eq:1}
C(M_1, \ldots, M_K)\leq \min_{\set{S}\subseteq \{1,\ldots, K\} }\Bigg(\mw{F}\bigg( \sum_{k\in\set{S}} \frac{1}{1-\delta_k}\bigg)^{-1}\!\!+ \frac{M_\set{S}}{D}\Bigg).
\end{equation}
\end{corollary}

\mw{\ssb{We now consider our original setup of section~\ref{sec:definition}}, where all weak receivers $i\in \set{K}_{\w}$ have {the} same erasure probability $\delta_\w$ and the same cache memory $M_i=M$ and where all strong receivers $j\in\Ks$ have the same erasure probability $\delta_\s\leq \delta_\w$ and no cache memory \ssb{$M_j=0$}.  In this case,} %the setup where we have two sets of weak and strong receivers, we have  $M_i=M$, $\delta_i=\delta_\w$ for all $i\in \set{K}_{\w}$, and $\delta_j=\delta_{\s}$, $M_j=0$ for all $j\in\set{K}_{\s}$. So
 \eqref{eq:1} simplifies to 
\begin{align}
C(M)\leq \min_{\substack{k_\w\in\{0,\ldots,K_\w\}\\k_\s\in\{0,\ldots,K_\s\}}}\Bigg(\ssb{F}\bigg(  \frac{k_\w}{1-\delta_\w}+\frac{k_{\s}}{1-\delta_\s}\bigg)^{-1}\!\!+ \frac{k_\w M}{D}\Bigg).\label{2-set}
\end{align}
Since the right-hand side of inequality~\eqref{2-set} is decreasing in $k_\s$, the tightest upper bound is \ssb{given by} $k_\s=K_\s$. \mw{This also concludes the proof of Theorem \ref{thm:converse}.}

\mw{Notice that} the choice of $k_\w$ \mw{in \eqref{eq:1}}  that leads to the tightest upper bound depends on the cache memory size~$M$. For small values of~$M$ the choice $k_\w=K_\w$ leads to the tightest bound, and for increasing cache sizes smaller values of $k_\w$ lead to tighter bounds.

\subsection{Proof of Theorem \ref{thm:general_upperbound}}
For ease of exposition, we only prove the bound corresponding to {$\set{S}=\mathcal{K}$}:
\begin{equation}\label{eq:wtp}
C(M_1, \ldots, M_K)\leq  \left( R_{\textnormal{sym},{\mathcal{K}}} + \frac{1}{D}\sum_{k=1}^K M_k\right),
\end{equation}
where here $R_{\textnormal{sym},{\mathcal{K}}}$  denotes the largest symmetric rate that is achievable over the BC  $P_{Y_1Y_2\cdots Y_K|X}$ when there are no caches.
The inequalities  in the theorem that correspond to other subsets $\set{S}\subseteq\{1,\ldots,K\}$ can be proved in an  analogous way.

We start the proof of \eqref{eq:wtp}. {Fix} the rate of communication 
\[
R < C(M_1,\ldots, M_K).
\] 
Since $R$ is achievable, for each sufficiently large blocklength $n$ and for each demand vector $\d$, there exist $K$  caching functions~$\big\{g_i^{(n)}\big\}$, an encoding function~$\{f_{\d}^{(n)}\}$, and $K$ decoding functions~$\big\{\varphi_{i,\d}^{(n)}\big\}$ so that the probability of worst-case error $\Pe^{(n)}(\mathbf{d})$ tends to 0 as $n\to \infty$. For each $n$ let
\[V_k^{(n)}=g_k^{(n)}(W_1,\ldots,W_D),\qquad k\in\{1,\ldots,K\},\]
denote the cache contents for the chosen caching functions. 
%We start our proof  with the following lemma. 
\begin{lemma} \label{lem:upperbound}
For any $\epsilon>0$, any demand vector $\d=(d_1,\ldots, d_K)$ with all different entries, and any blocklength $n$ that is sufficiently large (depending on $\epsilon$), there exist 
random variables $(U_{1,\d},\ldots, U_{K, \d}, X_{\mathbf{d}},Y_{1, \mathbf{d}}, \ldots, Y_{K, \mathbf{d}})$ such that% satisfy the Markov chain
\begin{equation}\label{eq:MarkovU}
U_{1,\d} - U_{2,\d} - \cdots - U_{K, \d} - X_{\mathbf{d}} - Y_{K, \mathbf{d}} -  Y_{K-1, \mathbf{d}}\cdots  - Y_{1, \mathbf{d}}
\end{equation}
forms a Markov chain, and  {given $X_{\mathbf{d}}=x\in\set{X}$:}
\[
(Y_{1, \mathbf{d}},  Y_{2, \mathbf{d}}, \ldots, Y_{K, \mathbf{d}})\sim P_{Y_1\cdots Y_K|X}(\cdots|x),
\]
and so that the following $K$ inequalities hold:
\begin{subequations}\label{eq:d_inequa}
\begin{align}
R-\epsilon&\leq \frac{1}{n} I\big(W_{d_1};V_{1}^{(n)}, \ldots, V_K^{(n)}\big) +I\big(U_{1,\d};Y_{1,\mathbf{d}} \big),\\
R-\epsilon&\leq\frac{1}{n} I\big(W_{d_k};V_{1}^{(n)}, \ldots, V_K^{(n)}| W_{d_1},\ldots, W_{d_{k-1}}\big)\nonumber \\& \qquad +I\big(U_{k,\d};Y_{k,\mathbf{d}} | U_{k-1,\d}), \,\, \hspace{1cm}\forall k\in\{2,\ldots, K\}.
\end{align}
\end{subequations}
\end{lemma}
\begin{IEEEproof}
The proof is similar to the converse proof of the capacity of degraded BCs without caching \cite[Theorem 5.2]{elgamalkim10}. It is deferred to Appendix \ref{app:lem_upperbound}.
\end{IEEEproof}
\vspace{2mm}
%
%and real numbers $\alpha_1, \ldots, \alpha_K$ satisfying 
%\begin{subequations} \label{eq:cons}
%\begin{IEEEeqnarray}{rCl}
%{\alpha}_{k} &\geq &0, \qquad k\in\{1,\ldots, K\}\label{eq:cons1} \\
%{\alpha}_{k'} &\leq&{\alpha}_{k}, \qquad k, k' \in\{1,\ldots, K\}, \ k' \leq k,\label{eq:cons2}\\
%\label{eq:cons3prime}
%\sum_{k=1}^K{\alpha}_{k} &  \leq &\frac{K}{D}\sum_{k\in\{1,\ldots, K\}} M_k,
%\end{IEEEeqnarray} 
%\end{subequations}
%so that the following  $K$ inequalities hold:
%\begin{subequations}\label{eq:c}
%\begin{align}
%C(M_1, \ldots, M_K)&\leq I\big(U_{1};Y_{1} \big)+{\alpha}_{1},\\
%C(M_1, \ldots, M_K)&\leq I\big(U_{k};Y_{k} \big| U_{1}, \ldots, U_{k-1})+{\alpha}_{k},\nonumber\\&\hspace{2cm}\qquad \forall k\in\{2,\ldots, K\}.
%\end{align}
%\end{subequations}
%\end{lemma}
%\begin{IEEEproof}
%See Appendix~\ref{app:lem_upperbound}.
%\end{IEEEproof}
%\vspace{2mm}

{Fix $\epsilon>0$ and a blocklength $n$ (depending on this $\epsilon$) so that Lemma \ref{lem:upperbound} holds for all demand vectors $\d$ that have all different entries.}
%We now fix such a large blocklength $n$ and 
We average the bound obtained in \eqref{eq:d_inequa} over different demand vectors. Let $\mathcal{Q}$ be the set of all the  ${D \choose K}{K!}$ demand vectors whose $K$ entries are all different. Also, let $Q$ be a uniform random variable over the elements of $\set{Q}$ and independent of all other random variables. 
Define: $U_1:=(U_{1, Q}, Q)$;  $U_k:=U_{k,Q}$, for $k\in\{2,\ldots, K\}$; $X_{k}:=X_{Q}$; and $Y_k:=Y_{k,Q}$ for $k\in\{1,\ldots, K\}$.  Notice that they form the Markov chain
\begin{equation}\label{eq:mm}
U_1 \to U_2 \to \cdots \to  U_K \to X \to (Y_1, \ldots, Y_K)
\end{equation}
and {given $X=x\in\set{X}$ satisfy}
\begin{equation} \label{eq:distribution}
(Y_{1},  Y_{2}, \ldots, Y_{K})\sim P_{Y_1\cdots Y_K|X}(\cdots|x).\end{equation}
Averaging inequalities \eqref{eq:d_inequa} over the demand vectors in~$\set{Q}$ and using standard {arguments to take care of the time-sharing random variable $Q$}, we obtain:
\begin{subequations}\label{eq:d_inequa2}
\begin{align}
R-\epsilon\leq&\alpha_1 +I\big(U_{1};Y_{1} \big),\\
R-\epsilon\leq& \alpha_k +I\big(U_{k};Y_{k} | U_{k-1}), \quad \forall k\in\{2,\ldots, K\},\IEEEeqnarraynumspace
\end{align}
\end{subequations}
where we defined {$\alpha_1, \ldots, \alpha_K$}  as follows:
\begin{subequations}\label{defgen}
\begin{align}
{\alpha}_{1}&:= \frac{1}{{D \choose K}{{K!}}}\sum_{\d \in \mathcal{Q}} \frac{1}{n} I(W_{d_1};V_{1}^{(n)}, \ldots, V_{K}^{(n)}), \\
{\alpha}_{k}&:= \frac{1}{{D \choose K}{{K!}}}\sum_{\d \in \mathcal{Q}}\frac{1}{n} I(W_{d_k};V_{1}^{(n)}, \ldots, V_{K}^{(n)}|W_{d_1}, \ldots, W_{d_{k-1}}).
\end{align}
\end{subequations}
%We show that $\alpha_1,\ldots, \alpha_K$ satisfy the conditions in Lemma \ref{lem:alphas} below.
\begin{lemma}\label{lem:alphas}
Parameters $\alpha_k$, $k=1,\ldots,K$,  defined in \eqref{defgen}, satisfy the following constraints:
\begin{subequations} \label{eq:cons}
\begin{align}
{\alpha}_{k} &\geq 0, \qquad k\in\{1,\ldots, K\}\label{eq:cons1} \\
{\alpha}_{k'} &\leq{\alpha}_{k}, \qquad k, k' \in\{1,\ldots, K\}, \ k' \leq k,\label{eq:cons2}\\
\label{eq:cons3prime}
\sum_{k\in\mathcal{K}} {\alpha}_{k} &  \leq \frac{K}{D}\sum_{k\in\mathcal{K}} M_k.
\end{align} 
\end{subequations}
\end{lemma}
\begin{IEEEproof}
See Appendix \ref{app:lemalphas}.
\end{IEEEproof}
Taking $\epsilon \to 0$, {by \eqref{eq:d_inequa2} and \eqref{defgen} and by Lemma~\ref{lem:alphas},} we conclude that the capacity-memory tradeoff $C(M_1, \ldots, M_K)$ is upper bounded by the following $K$ inequalities:
\begin{subequations}\label{eq:d_inequa}
\begin{align}
C(M_1, \ldots, M_K)&\leq \alpha_1 +I\big(U_{1};Y_{1} \big),\\
C(M_1, \ldots, M_K) &\leq \alpha_k+ I\big(U_{k};Y_{k} \big|  U_{k-1}), \qquad \forall k\in\{2,\ldots, K\},
\end{align}
\end{subequations}
for some $\alpha_1,\ldots,\alpha_K$ satisfying \eqref{eq:cons} and some  $U_1, \ldots, U_K, X, Y_1, \ldots, Y_K$ satisfying~\eqref{eq:mm} and \eqref{eq:distribution}.

%We have thus proved that for any achievable rate $R$ and $\epsilon>0$, there exist random variables $(U_1,\ldots,U_K,X,Y_1,\ldots,Y_K)$ satisfying \eqref{eq:mm} and \eqref{eq:distribution}, and real-valued parameters $\alpha_1, \ldots, \alpha_K$ satisfying \eqref{eq:cons}, so that the inequalities in \eqref{eq:d_inequa} hold. 
\begin{lemma}\label{lem:aa}
Replacing each and every real number ${\alpha}_{1}, \ldots,{\alpha}_{K}$ in \eqref{eq:d_inequa} by $\frac{1}{D} \sum_{k\in\{1,\ldots, K\}} M_k$ leads to a relaxed upper bound on $C(M_1,\ldots, M_K)$. 
\end{lemma}
\begin{IEEEproof}
See Appendix~\ref{app:lem_aa}.
\end{IEEEproof}
\vspace{2mm}

%Using Lemma \ref{lem:aa} and letting $\epsilon \to 0$, we conclude: The maximum achievable rate $C(M_1,\ldots,M_K)$ satisfies the following $K$  bounds 
Thus,
\begin{subequations}\label{eq:cd}
\begin{align}
C(M_1, \ldots, M_K)\!-\! \frac{1}{D}\!\!\!\sum_{k\in\{1,\ldots, K\}} \!\!\!\!\!M_k&\leq I\big(U_{1};Y_{1} \big),\\
C(M_1, \ldots, M_K) \!-\! \frac{1}{D}\!\!\!\sum_{k\in\{1,\ldots, K\}} \!\!\!\!\!M_k&\leq I\big(U_{k};Y_{k} \big|  U_{k-1}), \quad \forall k\in\{2,\ldots, K\},
\end{align}
\end{subequations}
for some  $U_1, \ldots, U_K, X, Y_1, \ldots, Y_K$ satisfying~\eqref{eq:mm} and \eqref{eq:distribution}.

All $K$ constraints in \eqref{eq:cd} have the same LHS, and their RHSs coincide with the rate-constraints of a degraded BC without caches. Therefore, the choice of the auxiliaries $(U_1, \ldots, U_K)$ that leads to the most relaxed constraint on $C(M_1, \ldots, M_K)$ coincides  with the choice of auxiliaries that determines the largest symmetric rate-point of the degraded BC without caches. This establishes the equivalence of \eqref{eq:cd} with the desired bound in \eqref{eq:wtp}, and thus concludes the proof.

%%%%

\section{New Results for All-Equal Demands}\label{sec:common}
In this section we consider the optimistic case where all receivers demand the same message. This corresponds to
\begin{equation*}\label{eq:common}
%\set{D}= \big\{ (d_1, \ldots, d_K) \in\{1,\ldots, D\}^{K}\colon 
d_1=d_2=\cdots=d_K \in \set{D}.%\big\}.
\end{equation*} 
Restricting to such ``all-equal demands" allows us to treat arbitrary (unequal) message rates and arbitrary DMBCs. So, we consider the scenario in figure~\ref{fig:equal_demands} where the transmitter communicates with  $K$ receivers over a DMBC $P_{Y_1\ldots Y_K|X}$ and where each receiver~$k\in\{1,\ldots, K\}$ has a cache memory of size $M_k \geq 0$. The messages $W_1, \ldots, W_D$ are independent of each other and each message $W_d$, for $d\in\{1,\ldots, D\}$, is uniformly distributed over the set $\{1,\ldots, \lfloor 2^{nR_d}\rfloor\}$ for some positive rate $R_d\geq 0$. 
\begin{figure}[h!]
\begin{center}
\includegraphics[width=0.49\textwidth]{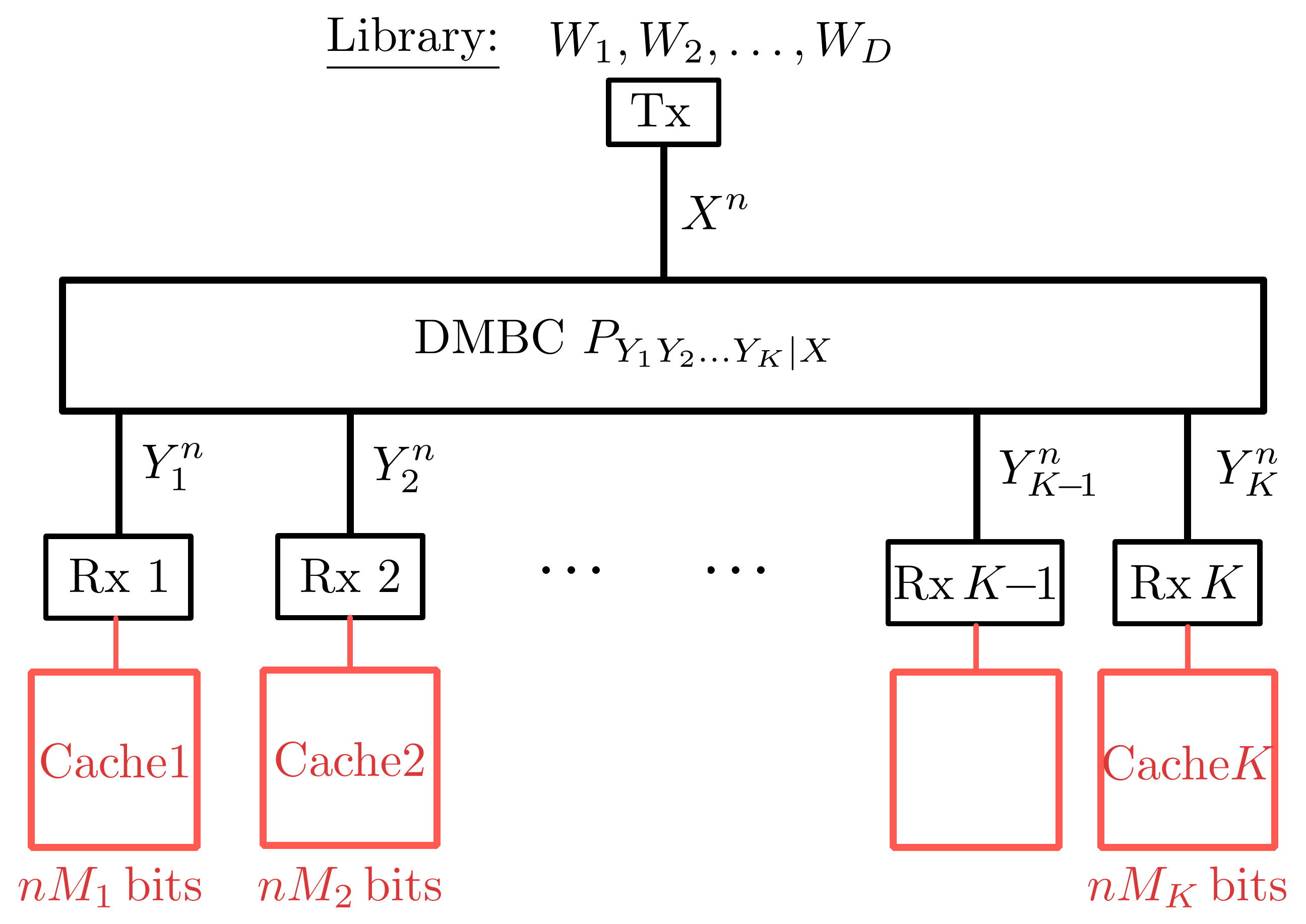}
\caption{$K$ user DMBC where each Receiver~$k\in\{1,\ldots, K\}$ has cache memory of size $n\M_k$ bits.}
\label{fig:equal_demands}
\end{center}
\end{figure}

The probability of worst-case error $\Pe^{\!\!\textnormal{worst}}$ is defined as before. A rate-memory tuple $(R_1, \ldots, R_D,\ \M_1, \ldots, \M_K)$ is said \emph{achievable} if for every $\epsilon >  0$ there exists a sufficiently large blocklength $n$ and caching, encoding and decoding functions as in \eqref{eq:caching}--\eqref{eq:decoding} (\ssb{accounting for the different rates}) such that $\Pe^{\!\!\textnormal{worst}}<\epsilon$.

%It can be shown that in this setup, a joint cache-channel code based on Tuncel's virtual binning technique~\cite{Tuncel-Apr-2006-A}  achieves the capacity-memory region.
%\subsection{Result and Discussion}
Under \ssb{the assumption that all receivers demand the same message,} we can completely \ssb{characterize} the \ssb{set of all achievable} rates-memory tuples.
\begin{theorem}\label{thm:common} A rate-memory tuple  $(R_1, \ldots, R_D, \M_1,\ldots,$ $ \M_K)$ is achievable under all-equal demands if and only if 
\begin{IEEEeqnarray}{rCl}\label{eq:RdM}
R_d \leq \max_{P_X}\min_{k\in\{1,\ldots, K\}}\big( I(X;Y_k) + \M_{k,d} \big), \ \  d\in\{1,\ldots, D\}\nonumber%\\
\end{IEEEeqnarray}
 for some nonnegative  real numbers $\{\M_{k,d}\}$  that satisfy
\begin{equation}
\sum_{d=1}^D \M_{k,d} \leq \M_k, \quad k\in\{1,\ldots, K\}.
\end{equation} 
\end{theorem} 
% We  see that when we are allowed to choose the cache sizes %at the receivers subject to an overall memory-constraint 
%  \begin{equation}
% \sum_{k=1}^K \M_k \leq \M_{\Sigma},
% \end{equation}
Clearly, one wishes to allocate small cache sizes to strong receivers  and large cache sizes to  weak receivers.

% in subsection~\ref{sec:ach}. It employs a joint source-channel coding scheme in the delivery phase that essentially employs Tuncel's \emph{Slepian-Wolf coding over broadcast channels}
If we used separate cache-channel codes, constraint~\eqref{eq:RdM} had to be replaced by
\begin{IEEEeqnarray}{rCl}
\max_{k\in\{1,\ldots, K\}} (R_d - \M_{k,d})\leq \max_{P_X} \min_{k\in\{1,\ldots, K\}}  I(X;Y_k), % \;  d\in\{1,\ldots, N\},
\end{IEEEeqnarray}
and the benefit of having unequal cache sizes $\{\M_{k}\}$ at the different receivers would disappear.

\subsection{Proof of Achievability of Theorem~\ref{thm:common}}\label{sec:ach}
We propose the following scheme. 
\ssb{For} each $d\in\{1,\ldots, D\}$ and each $k\in\{1,\ldots, K\}$, fix a positive integer number 
\begin{equation}
M_{k,d} \leq R_k
\end{equation}
so that for every $k\in\{1,\ldots, K\}$:
\begin{equation}\label{eq:cc}
\sum_{d=1}^N \M_{k,d} \leq \M_k.
\end{equation}
Fix a probability distribution $P_X$ over the channel input alphabet $\set{X}$. Construct a codebook $\mathcal{C}$ containing $\lfloor 2^{nR}\rfloor$ codewords of length $n$ by drawing all symbols of all codewords i.i.d. according to the probability distribution $P_X$. Reveal codebook~$\set{C}$ to the transmitter and to all receivers. \vspace{1mm}

\noindent\textit{Caching phase:} 
For each receiver~$k\in\{1,\ldots, K\}$, cache the first $\lfloor n\M_{k,d}\rfloor$ bits of each Message~$W_d$ in receiver~$k$'s cache. By \eqref{eq:cc} this caching strategy satisfies the cache memory constraint. \vspace{1mm}

\noindent\textit{Delivery phase:} Let $d^\star=d_1=\ldots = d_K$. The transmitter picks the codeword from codebook $\set{C}$ that corresponds to message $W_{d^\star}$ and sends it over the channel. 

Each receiver~$k$ decodes the same desired message $W_{d^\star}$. Since it knows the first $n\M_{k,d^{\star}}$ bits of $W_{d^\star}$, in its decoding it restricts attention to the part of the codebook corresponding to  messages starting with these bits. %For receiver $k$ it is thus as if the transmitter had sent only its missing bits over the channel. \\

\noindent\textit{Error analysis:}
The probability of error (averaged over codebooks and messages) at receiver~$k$ is the same as if only the last $n(R-M_{k,d^\star})$ bits of message $W_{d^\star}$ had been sent. Thus, by the packing lemma \cite{elgamalkim10}, the probability of decoding error at receiver~$k$ tends to 0 as $n\to \infty$, whenever
\begin{equation*}
R_{d^\star} - \M_{k,d^\star} \leq I(X;Y_k), \quad d^\star\!\in\{1,\ldots, D\}, \; k\in\{1,\ldots, K\}.
\end{equation*}
This proves achievability of Theorem~\ref{thm:common}.

\subsection{Proof of Converse to Theorem~\ref{thm:common}}\label{sec:con}
 % Let $d\in\{1,\ldots, D\}$ and $k\in\{1,\ldots, K\}$. %, be such that there is strictly positive probability 
%\begin{equation}\label{eq:supp}
%\Pr[\Theta_k=d].
%\end{equation}  

%Let $\{\epsilon_n\}$ be an arbitrary sequence of small positive numbers that tends to 0 as $n\to \infty$. 
Let   $(R_1, \ldots, R_N, \M_1, \ldots, \M_K)$ be an achievable rate-memory tuple. \ssb{For each sufficiently large $n$ and demand $d\in\{1,\ldots,D\}$ fix $K$ caching functions~$\big\{g_i^{(n)}\big\}$, an encoding function~$\{f_{\d}^{(n)}\}$, and $K$ decoding functions~$\big\{\varphi_{i,\d}^{(n)}\big\}$ so that the probability of worst-case error $\mw{\Pe^{\textnormal{worst}}}$ tends to 0 as $n\to \infty$.}  %By condition \eqref{eq:supp}, 
Fix now a blocklength $n$, a demand $d^\star\in\{1,\ldots,D\}$, and a receiver~$k\in\{1,\ldots, K\}$. 
We have
\begin{IEEEeqnarray}{rCl}
R_{d^\star} & \leq & \frac{1}{n} H(W_{d^\star}) \nonumber \\
& = & \frac{1}{n} I(W_{d^\star}; Y_k^n, V^{(n)}_k) + \frac{1}{n} H(W_{d^\star}|Y_k^n, V^{(n)}_k) \nonumber\\
& \leq & \frac{1}{n} I(W_{d^\star}; Y_k^n| V^{(n)}_k) + \frac{1}{n} I(W_{d^\star}; V^{(n)}_k) +   \epsilon_n\nonumber \\
& = & \frac{1}{n} \sum_{t=1}^n \big( H(Y_{k,t} | V^{(n)}_k, Y_{k}^{t-1}) - H(Y_{k,t} | W_{d^\star,} Y_{k}^{t-1}, V^{(n)}_k) \big)\nonumber \\
 & &+  \M_{k,d^\star} +\epsilon_n\nonumber \\ 
& \leq & \frac{1}{n} \sum_{t=1}^n \big( H(Y_{k,t}) - H(Y_{k,t} | X_{k,t})\big)+\M_{k,d^\star} + \epsilon_n\nonumber \\
& = & \frac{1}{n} \sum_{t=1}^n I(Y_{k,t}; X_{k,t})+ \M_{k,d^\star} +\epsilon_n,%\nonumber \\
%& \leq & (1- \delta_k)F+ \M_{k,d} +\epsilon_n,
\end{IEEEeqnarray} 
where we defined
\begin{IEEEeqnarray}{rCl}
\M_{k,d} \triangleq\frac{1}{n} I(W_d; V^{(n)}_k), \quad k\in\{1,\ldots, K\}, \;d\in \{1,\ldots, D\}.\nonumber
\end{IEEEeqnarray}
The second inequality above follows by Fano's inequality; the third inequality \ssb{holds} because conditioning \ssb{does not} increase entropy and because  $(M_d, Y_{k}^{t-1}, V^{(n)}_k)\to X_{t} \to Y_{k,t}$ \ssb{forms a Markov chain}.%; and the equalities follow by the definition and the chain rule of mutual information. 

Moreover, for each $k\in\{1,\ldots, K\}$, 
\begin{IEEEeqnarray}{rCl}
\sum_{d=1}^N \M_{k,d} & = &\frac{1}{n} \sum_{d=1}^D I(W_d; V^{(n)}_k) \nonumber \\
%&=&\frac{1}{n}\sum_{d=1}^D \big( H(W_d) -H(W_d | V_k) \big) \nonumber \\
% & = &\frac{1}{n}\big(  H(W_1, \ldots, W_D) -\sum_{d=1}^N H(W_d | V_k)\big)  \nonumber \\
 & \leq &\frac{1}{n}  \sum_{d=1}^D I(W_d; V^{(n)}_k| W_1,\ldots, W_{d-1})  \nonumber \\
 & =&\frac{1}{n} I(W_1, \ldots, W_D; V^{(n)}_k) \nonumber \\
  &\leq &\frac{1}{n} H(V^{(n)}_k) \leq \M_k,
\end{IEEEeqnarray}
where the first  inequality follows  because messages $W_1,\ldots, W_D$ are independent.

Letting $n\to \infty$, and thus $\epsilon_n\to 0$, establishes the desired converse.

\appendices

%\section{Proof of Theorem~\ref{thm:general_upperbound}}\label{app:upperbound}
%
\section{Proof of Lemma~\ref{lem:upperbound}}\label{app:lem_upperbound}
\mw{ Fix a small $\epsilon >0$ and a demand vector $\d$ with all different entries. Then, let the blocklength $n$ be sufficiently large as will be come clear in the following. Also, let}
%for each blocklength~$n$
\begin{IEEEeqnarray}{rCl}
V^{(n)}_i&= &g_i^{(n)}(W_1, \ldots, W_D), \qquad i\in \{1,\ldots, K\}, \\
X_\d^n&=&f_{\d}^{(n)}(W_1,\ldots, W_D) 
  \end{IEEEeqnarray}
  denote cache contents and the input of the degraded BC \mw{for demand vector $\d\in\mathcal{D}^K$ and for above chosen caching and encoding functions.} 
  Also, let $Y_{k, \d}^{n}$  denote the corresponding channel outputs at Receiver~$k$.

\mw{By Fano's inequality, by  the independence of the messages $W_1, \ldots, W_D$, and because the caching, encoding, and decoding functions have been chosen so that the worst case probability of error tends to 0 for increasing blocklengths, we obtain that that for all sufficiently large $n$ the following $K$ inequalities hold:}% for any $\epsilon_n>0$ we have % $W_{d_1}, \ldots, W_{d_{K}}$, 
\begin{subequations}\label{eq:out1}
\begin{align}\label{eq:36a}
R\!-\!\epsilon &\leq\frac{1}{n} I\big(W_{d_1};Y_{k,\mathbf{d}}^n,V_{1}^{(n)}, \ldots, V_K^{(n)}\big) \nonumber\\
& =  \frac{1}{n} I\big(W_{d_1};V_{1}^{(n)}, \ldots, V_K^{(n)}\big)\!+\! \frac{1}{n} I\big(W_{d_1};Y_{k,\mathbf{d}}^n\big|V_{1}^{(n)}, \ldots, V_K^{(n)}\big)
\end{align}
and 
 for all $k\in\{2,\ldots, K\}$:
\begin{align}\label{eq:36b}
R\!-\!\epsilon_n \leq&\frac{1}{n} I\big(W_{d_k};Y_{k,\mathbf{d}}^n,V_{1}^{(n)}, \ldots, V_K^{(n)}\big| W_{d_1},\ldots, W_{d_{k-1}}\big)\nonumber \\
 =& \frac{1}{n} I\big( W_{d_k}; V_{1}^{(n)}, \ldots, V_K^{(n)} \big| W_{d_1},\ldots, W_{d_{k-1}}\big) \nonumber\\&+ \frac{1}{n} I\big(W_{d_k};Y_{k,\mathbf{d}}^n \big| V_{1}^{(n)}, \ldots, V_K^{(n)}, W_{d_1},\ldots, W_{d_{k-1}}\big) 
\end{align}
\end{subequations}
We further develop the second summands in \eqref{eq:36a} and \eqref{eq:36b}. For  the second summand in \eqref{eq:36a} we obtain
\begin{align}
&\frac{1}{n} I\big(W_{d_1};Y_{k,\mathbf{d}}^n\big|V_{1}^{(n)}, \ldots, V_K^{(n)}\big)\nonumber\\
&\quad =   \frac{1}{n}\sum_{t=1}^n I\big(W_{d_1};Y_{k,\mathbf{d},t}\big|V_{1}^{(n)}, \ldots, V_K^{(n)}, Y_{k,\mathbf{d}}^{t-1}\big)\nonumber \\
 &\quad =   \frac{1}{n}\sum_{t=1}^n I\big( Y_{k,\mathbf{d}}^{t-1}, W_{d_1};Y_{k,\mathbf{d},t}\big|V_{1}^{(n)}, \ldots, V_K^{(n)}\big) \nonumber \\
 &\quad  =   I\big({U}_{1,\d,T};Y_{k,\mathbf{d},T}\big|V_{1}^{(n)}, \ldots, V_K^{(n)},T\big) \nonumber \\
&\quad    \leq   I\big({U}_{1,\d};Y_{k,\mathbf{d}}\big|V_{1}^{(n)}, \ldots, V_K^{(n)}\big),\label{eq:out0}
\end{align}
where $T$ denotes a random variable that is uniformly distributed over $\{1,\ldots, n\}$ and independent of all other random variables, and where we defined
\begin{IEEEeqnarray*}{rCl}
U_{1,\d,T}&:=&(V^{(n)}_1 \ldots, V_K^{(n)}, W_{d_1}, Y_{1, \mathbf{d}}^{t-1}), \\
U_{1,\d} & := & (U_{1,\d,T}, T), \\
Y_{k,\mathbf{d}} &:=&Y_{k,\mathbf{d},T},  \qquad k\in\{1,\ldots, K\}.
\end{IEEEeqnarray*}
We also define for $k\in\{2,\ldots, K\}$:
\begin{IEEEeqnarray*}{rCl}
U_{k,\d,T} & := & (\ssb{V^{(n)}_1 \ldots, V_K^{(n)}}, W_{d_1}, W_{d_2}, \ldots, W_{d_k}, Y_{1, \mathbf{d}}^{t-1}, \ldots, Y_{k, \mathbf{d}}^{t-1}), \\
U_{k,\d} & := & U_{k,\d,T},
\end{IEEEeqnarray*}
in order to expand the second summand in \eqref{eq:36b} as follows:
\begin{align}
&{\frac{1}{n} I\big(W_{d_k};Y_{k,\mathbf{d}}^n \big| V_{1}^{(n)}, \ldots, V_K^{(n)}, W_{d_1},\ldots, W_{d_{k-1}}\big) } \qquad \nonumber \\
 &\, =  \frac{1}{n} \sum_{t=1}^n I\big(W_{d_k};Y_{k,\mathbf{d},t} \big| V_{1}^{(n)}, \ldots, V_K^{(n)}, W_{d_1},\ldots, W_{d_{k-1}}, Y_{k,\mathbf{d}}^{t-1}\big) \nonumber \\
&\,=  \frac{1}{n} \sum_{t=1}^n I\big(W_{d_k};Y_{k,\mathbf{d},t} \big| V_{1}^{(n)}, \ldots, V_K^{(n)}, W_{d_1},\ldots, W_{d_{k-1}}, Y_{1,\mathbf{d}}^{t-1}, \ldots, Y_{k-1,\mathbf{d}}^{t-1}, Y_{k,\mathbf{d}}^{t-1}\big) \nonumber \\
&\,\leq \frac{1}{n}\sum_{t=1}^n I\big( W_{d_k}Y_{k,\mathbf{d}}^{t-1}\ ;\ Y_{k,\mathbf{d},t} \big| V_{1}^{(n)}, \ldots, V_K^{(n)}, W_{d_1},\ldots, W_{d_{k-1}},Y_{1,\mathbf{d}}^{t-1}, \ldots, Y_{k-1,\mathbf{d}}^{t-1}\big)\nonumber\\
&\,={ I\big(U_{k,\d,T};Y_{k,\mathbf{d},T} \big| U_{k-1,\d,T}, T)}\nonumber \\
&\,= {I\big(U_{k,\d};Y_{k,\mathbf{d}} \big| U_{k-1,\d})}\label{eq:out2}
\end{align}
where the second equality follows from the degradedness of the outputs, see \eqref{eq:equation}.

Notice that if we also define $X_{\mathbf{d}} :=X_{\mathbf{d},T}$, then  %the Markov chain
\mw{\eqref{eq:MarkovU} and \eqref{eq:d_inequa} hold.} Combining this observation with \eqref{eq:out1}--\eqref{eq:out2} concludes the proof.

%From \eqref{eq:out1} and \eqref{eq:out2} %and by taking the blocklength $n$ to infinity, 
%we conclude the following: For given small $\epsilon>0$, a sufficiently large blocklength $n$, and demand vector $\d=(d_1,\ldots, d_K)$ with all different entries, there exist random variables $(U_{1,\d},\ldots, U_{K, \d}, X_{\mathbf{d}},Y_{1, \mathbf{d}}, \ldots, Y_{K, \mathbf{d}})$ satisfying the Markov chain \eqref{eq:MarkovU} and  {given $X_{\mathbf{d}}=x\in\set{X}$:}\[
%(Y_{1, \mathbf{d}},  Y_{2, \mathbf{d}}, \ldots, Y_{K, \mathbf{d}})\sim P_{Y_1\cdots Y_K|X}(\cdots|x),
%\]
%so that \eqref{eq:d_inequa} holds. 

\section{Proof of Lemma~\ref{lem:alphas}}\label{app:lemalphas}

Constraint~\eqref{eq:cons1} follows by the nonnegativity of mutual information. 
To prove Constraint~\eqref{eq:cons2}, we fix a demand vector~$\d\in \mathcal{Q}$, and consider the cyclic shifts of this vector. For $\ell\in\{0,\ldots, K-1\}$, let $\overrightarrow{\d}^{(\ell)}$ be the vector obtained from $\overrightarrow{\d}$ when the elements are cyclically shifted $\ell$ positions to the right. (E.g., if $\d=(1, 2, 3)$ then $\overrightarrow{\d}^{(2)}=(2, 3, 1)$.) 
For each $\ell\in\{0,\ldots, K-1\}$ and $k\in\{1,\ldots, K\}$, let $\overrightarrow{d}_{k}^{(\ell)}$ denote the $k$-th index of demand vector $\overrightarrow{\d}^{(\ell)}$. So, 
\begin{equation}\label{eq:circ}
\overrightarrow{d}_{k}^{(\ell)} = d_{(k-\ell)\!\! \!\!\mod K}
\end{equation}
where for each positive integer $\xi$ the term $(\xi \mod K)$ takes value in $\{1,\ldots, K\}$ so that 
\mw{\begin{equation}
\xi \mod K =
\xi - b K \quad  \textnormal{ for some positive integer } b.
\end{equation}}

For each $\ell\in\{1,\ldots, K\!-\!1\}$ and $k, k'\in\{2,\ldots, K\}$, $k'\leq k$:
\begin{align}\label{eq:in1}
\lefteqn{I(W_{d_1};V_{1}^{(n)}\!\!, \ldots, V_K^{(n)}) {\stackrel{(a)}{=}  I(W_{\overrightarrow{d}_{k'}^{(k'-1)}};V_{1}^{(n)}\!\!, \ldots, V_K^{(n)})}}\quad\nonumber\\
 \stackrel{(b)}{\leq}  &  I(W_{\overrightarrow{d}_{k'}^{(k'-1)}};V_{1}^{(n)}\!\!, \ldots, V_K^{(n)}| W_{\overrightarrow{d}_1^{(k'-1)}},\ldots, W_{\overrightarrow{d}_{k'-1}^{(k'-1)}}) \nonumber \\
  \stackrel{(a)}{=}  &{  I(W_{\overrightarrow{d}_{k}^{(k-1)}};V_{1}^{(n)}\!\!, \ldots, V_K^{(n)}| W_{\overrightarrow{d}_2^{(k-1)}},\ldots, W_{\overrightarrow{d}_{k-1}^{(k-1)}}) }\nonumber \\
  \stackrel{(b)}{\leq} &  I(W_{\overrightarrow{d}_{k}^{(k-1)}};V_{1}^{(n)}\!\!, \ldots, V_K^{(n)}| W_{\overrightarrow{d}_1^{(k-1)}},\ldots, W_{\overrightarrow{d}_{k-1}^{(k-1)}}), %\IEEEeqnarraynumspace
\end{align}
{where (a) follows by \eqref{eq:circ} and (b) is by the independence of  messages.}  

{Fix a demand vector $\d\in\set{Q}$ and sum up the above  inequality~\eqref{eq:in1} over all $K$ cyclic shifts $\d^{(0)}, \d^{(1)}, \ldots,$ $\d^{(K-1)}$ of $\d$ to obtain:}
\begin{IEEEeqnarray}{rCl}
\lefteqn{\sum_{\ell=0}^{K-1} I(W_{\overrightarrow{d}_1^{(\ell)}};V_{1}^{(n)}, \ldots, V_K^{(n)})}\quad\nonumber\\& \leq &  \sum_{\ell=0}^{K-1} I(W_{\overrightarrow{d}_{k'}^{(\ell)}};V_{1}^{(n)}, \ldots, V_K^{(n)}| W_{\overrightarrow{d}_1^{(\ell)}},\ldots, W_{\overrightarrow{d}_{k'-1}^{(\ell)}}) \nonumber \\
 & \leq & \sum_{\ell=0}^{K-1}  I(W_{\overrightarrow{d}_{k}^{(\ell)}};V_{1}^{(n)}, \ldots, V_K^{(n)}| W_{\overrightarrow{d}_1^{(\ell)}},\ldots, W_{\overrightarrow{d}_{k-1}^{(\ell)}}). \label{eq:in2}\IEEEeqnarraynumspace
 \end{IEEEeqnarray}
Since the set $\set{Q}$ can be partitioned into subsets of demand vectors that are cyclic shifts of each others and all cyclic shifts of a demand vector in $\set{Q}$ are also in $\set{Q}$,   {we  conclude from~\eqref{eq:in2}:}
\begin{IEEEeqnarray}{rCl}
\lefteqn{\sum_{\d\in \set{Q}} I(W_{d_1};V_{1}^{(n)}, \ldots, V_K^{(n)})}\quad\nonumber\\& \leq &  \sum_{\d\in\set{Q}} I(W_{d_{k'}};V_{1}^{(n)}, \ldots, V_K^{(n)}| W_{d_1},\ldots, W_{d_{k'-1}}) \nonumber \\
 & \leq & \sum_{\d\in\set{Q}} I(W_{d_{k}};V_{1}^{(n)}, \ldots, V_K^{(n)}| W_{d_1},\ldots, W_{d_{k-1}}).
 \end{IEEEeqnarray}
{This proves~\eqref{eq:cons2}.}

 We proceed to prove Constraint~\eqref{eq:cons3prime}. 
 For each  $\d\in \set{Q}$: 
\begin{IEEEeqnarray}{rCl}
\lefteqn{I(W_{d_1}; V_1^{(n)}, \ldots, V_K^{(n)})}\qquad\nonumber\\ \lefteqn{+ \sum_{k=2}^K I(W_{d_k}; V_1^{(n)}, \ldots, V_K^{(n)}|W_{d_1}, W_{d_2}, \ldots, W_{d_{k-1}})} \quad \nonumber \\
&  = & I(W_{d_1},W_{d_2}, \ldots, W_{d_{K-1}}; V_1^{(n)}, \ldots, V_K^{(n)}).
\end{IEEEeqnarray}
So,
\begin{align}
\lefteqn{\sum_{\d\in\set{Q}} \bigg[ I(W_{d_1}; V_1^{(n)}, \ldots, V_K^{(n)})}  \nonumber\\&\qquad+\sum_{k=2}^K I(W_{d_{k}}; V_1^{(n)}, \ldots, V_K^{(n)}|W_{d_1}, W_{d_2}, \ldots, W_{d_{k-1}}) \bigg]  \qquad \nonumber \\
  = &  \sum_{\d \in \set{Q}} I(W_{d_1}, W_{d_2},\ldots, W_{d_{K}}; V_1^{(n)}, \ldots, V_K^{(n)}) \nonumber \\
 \stackrel{(a)}{=} & \sum_{\d \in \set{Q}}  \Big[ H(W_{d_1})+ H(W_{d_2})+ \ldots + H(W_{d_K})  \nonumber\\&\qquad\qquad\qquad- H( W_{d_1}, \ldots, W_{d_K}| V_1^{(n)}, \ldots, V_K^{(n)}) \Big]\nonumber \\
 \stackrel{(b)}{=} &  \frac{K}{D} |\set{Q}| H(W_1, \ldots, W_D)\nonumber\\& -  \sum_{\d \in\set{Q}}  H( W_{d_1}, \ldots, W_{d_K}| V_1^{(n)}, \ldots, V_K^{(n)})  \nonumber \\
 \stackrel{(c)}{\leq}  &   \frac{K}{D}{K!} {D \choose K}  H(W_1, \ldots, W_D)  \nonumber\\& -   \frac{K}{D}{K!}  {D \choose K} H( W_1, \ldots, W_D| V_1^{(n)}, \ldots, V_K^{(n)}) \nonumber \\
 \stackrel{(b)}{=} &   \frac{K}{D}{K!} {D \choose K} I( W_1, \ldots, W_D; V_1^{(n)}, \ldots, V_K^{(n)})\nonumber\\
\leq   & \frac{K}{D}{K!} {D \choose K}{n}\sum_{k=1}^K M_k, \nonumber
\end{align}
where (a) holds by the chain rule of mutual information,  (b) by the independence and uniform rate of messages $W_1,\ldots, W_D$ and the definition of the set $\set{Q}$, which is of size {${D \choose K} K!$}, and (c) by the generalized Han-Inequality (the following Proposition~\ref{prop:han}).
\vspace{1mm}

\begin{proposition} \label{prop:han}
Let $L$ be a positive integer and $A_1,\ldots, A_L$ be a finite random $L$-tuple. Denote by  $A_{\mathcal{S}}$ the subset $\{A_\ell,\ \ell\in\mathcal{S}\}$.
For every $\ell \in\{1,\ldots, L\}$:
\begin{align}
{1\over {L \choose \ell}}\sum_{\mathcal{S}\subseteq\{1,\ldots,L\}:|\mathcal{S}|=\ell}\frac{H(A_{\mathcal{S}})}{\ell}\geq \frac{1}{L}H(A_1,\ldots,A_L).\label{Han}
\end{align}
\end{proposition}
\begin{IEEEproof}
See~\cite[Theorem 17.6.1]{CoverThomas}.
\end{IEEEproof}

\section{Proof of Lemma~\ref{lem:aa}}\label{app:lem_aa}
Fix random variables $U_1, U_2, \ldots, U_K, X$ satisfying the Markov chain \eqref{eq:mm} and real numbers ${\alpha}_{1}, \ldots,{\alpha}_{K}$  satisfying \eqref{eq:cons}. We will show that if $\alpha_{\tilde{k}} \neq \alpha_{\tilde{k}+1}$ for some $\tilde{k}\in\mathcal{K}$, then we can find new random variables $\bar{U}_1, \bar{U}_2, \ldots, \bar{U}_K, \bar{X}$ satisfying the Markov chain \eqref{eq:mm} and real numbers $\bar{\alpha}_1, \ldots, \bar{\alpha}_{K}$  satisfying \eqref{eq:cons} so that the upper bound on $C(M_1, \ldots, M_K)$ in \eqref{eq:d_inequa} is  relaxed if we replace 
\[(U_1, U_2, \ldots, U_K, X) \qquad \textnormal{and} \qquad ({\alpha}_{1}, \ldots,{\alpha}_{K})\] by  
\[(\bar{U}_1, \bar{U}_2, \ldots, \bar{U}_K, \bar{X}) \qquad \textnormal{and} \qquad (\bar{\alpha}_1, \ldots, \bar{\alpha}_{K}).\] This proves that we obtain a relaxed upper bound on $C(M_1, \ldots, M_K)$ if in \eqref{eq:d_inequa} we replace all numbers $\alpha_1, \ldots, \alpha_K$ by the same number $\alpha$. By \eqref{eq:cons3prime} this number $\alpha \leq \frac{1}{D}\sum_{k\in\{1,\ldots, K\}} M_k$, and  by the monotonicity of the RHSs of \eqref{eq:d_inequa} in $\alpha_1, \ldots, \alpha_K$  the choice $\alpha=\frac{1}{D}\sum_{k\in\{1,\ldots, K\}} M_k$ leads to the most relaxed upper bound. This will conclude the proof.

Assume that $\alpha_{\tilde{k}} \neq \alpha_{\tilde{k}+1}$ for  some $\tilde{k}\in\{1,\ldots, K-1\}$. By \eqref{eq:cons2}, the strict inequality 
\begin{equation}\label{eq:strict}
\alpha_{\tilde{k}} < \alpha_{\tilde{k}+1}
\end{equation} must hold.
Choose
\begin{IEEEeqnarray}{rCl}
\bar{\alpha}_{k} & =&  \alpha_{k}, \qquad  k \in \mathcal{K}, \ k\notin \{\tilde{k}, \tilde{k}+1\}, \IEEEeqnarraynumspace\\
\bar{\alpha}_{\tilde{k}}=\bar{\alpha}_{\tilde{k}+1}& = &  \frac{1}{2} (\alpha_{\tilde{k}}+\alpha_{\tilde{k}+1}), \label{eq:alphabar}\\
\bar{U}_{k} &= & U_{k}, \qquad  k \in\mathcal{K}, \ k\neq \tilde{k}.
\end{IEEEeqnarray}
The choice of $\bar{U}_{\tilde{k}}$ depends on whether 
\begin{subequations}
\begin{equation}\label{eq:condition1}
I(U_{\tilde{k}};Y_{\tilde{k}}| U_{\tilde{k}-1}) \leq I(U_{\tilde{k}+1};Y_{\tilde{k}+1}|U_{\tilde{k}}),
\end{equation} 
or 
\begin{equation}\label{eq:condition2}
I(U_{\tilde{k}};Y_{\tilde{k}}|U_{\tilde{k}-1}) > I(U_{\tilde{k}+1};Y_{\tilde{k}+1}|U_{\tilde{k}}).
\end{equation}
\end{subequations}
If \eqref{eq:condition1} holds, choose
\begin{equation}
\bar{U}_{\tilde{k}} =  U_{\tilde{k}}.
\end{equation}
If \eqref{eq:condition2} holds, let $E \in \{0,1\}$ be a Bernoulli-$\beta$ random variable independent of everything else, where
 \begin{equation}\label{eq:beta}
 \beta := \frac{1}{2} + \frac{1}{2} \frac{I(U_{\tilde{k}+1};Y_{\tilde{k}+1}|U_{\tilde{k}})}{ I(U_{\tilde{k}};Y_{\tilde{k}}|U_{\tilde{k}-1})}.
 \end{equation}
 Choose
 \begin{equation}
 \bar{U}_{\tilde{k}} = \begin{cases} (U_{{\tilde{k}}}, E),& \ \textnormal{if } E=0\\
(U_{\tilde{k}-1},E), & \ \textnormal{if } E=1.\end{cases}\label{choiceofUsecond}
 \end{equation}
The proposed choice satisfies the Markov chain $\bar{U}_1 - \bar{U}_2 - \cdots \bar{U}_K - X$. Moreover, by \eqref{choiceofUsecond} and \eqref{eq:beta}:
 \begin{IEEEeqnarray}{rCl}\label{eq:ubareq}
 \lefteqn{ I(\bar{U}_{\tilde{k}};Y_{\tilde{k}}|\bar{U}_{\tilde{k}-1})} \nonumber \\
 &  = & \frac{1}{2}\left( I(U_{\tilde{k}+1};Y_{\tilde{k}+1}|U_{\tilde{k}})+ I(U_{\tilde{k}};Y_{\tilde{k}}| U_{\tilde{k}-1})\right).
 \end{IEEEeqnarray}

Trivially, for $k\notin\{\tilde{k},\tilde{k}+1\}$, constraint \eqref{eq:d_inequa} is unchanged if we replace 
$(U_1, U_2, \ldots, U_K, X)$ by  $(\bar{U}_1, \bar{U}_2, \ldots, \bar{U}_K, \bar{X})$ and $({\alpha}_{1}, \ldots,{\alpha}_{K})$ by  $(\bar{\alpha}_1, \ldots, \bar{\alpha}_{K})$. 

If \eqref{eq:condition1} holds, then the proposed replacement relaxes constraint~\eqref{eq:d_inequa}  for $k=\tilde{k}$ and it tightens it  for $k=\tilde{k}+1$. However, the new constraint for $k=\tilde{k}+1$ is less stringent than the original constraint for $k=\tilde{k}$. We conclude that when~\eqref{eq:condition1} holds,  the upper bound on $C(M_1, \ldots, M_K)$ in \eqref{eq:d_inequa} is relaxed if everywhere one replaces $(U_1, U_2, \ldots, U_K, X)$ and $({\alpha}_{1}, \ldots,{\alpha}_{K})$  by $(\bar{U}_1, \bar{U}_2, \ldots, \bar{U}_K, \bar{X})$ and $(\bar{\alpha}_1, \ldots, \bar{\alpha}_{K})$.

If \eqref{eq:condition2} holds, then the new constraint obtained for $k=\tilde{k}$  coincides with the average of the two original constraints for $k=\tilde{k}$ and for $k=\tilde{k}+1$, see \eqref{eq:alphabar} and \eqref{eq:ubareq}. This average constraint cannot be more stringent than the most stringent of the two original constraints. The new constraint obtained for $k=\tilde{k}+1$ is more relaxed than the new constraint obtained for $k=\tilde{k}$, because of \eqref{eq:alphabar} and because
\begin{IEEEeqnarray}{rCl}
\lefteqn{
I(\bar{U}_{\tilde{k}+1};Y_{\tilde{k}+1}| \bar{U}_{\tilde{k}}) }\quad \nonumber \\
& \stackrel{(a)}{=} & \beta I(U_{\tilde{k}+1};Y_{\tilde{k}+1}|{U}_{\tilde{k}})  + (1-\beta) I(U_{\tilde{k}+1};Y_{\tilde{k}+1}|{U}_{\tilde{k}-1}) \nonumber \\
& \stackrel{(b)}{=} &  \beta I(U_{\tilde{k}+1};Y_{\tilde{k}+1}|{U}_{\tilde{k}}) + (1-\beta) I(U_{\tilde{k}+1}, U_{\tilde{k}};Y_{\tilde{k}+1}|{U}_{\tilde{k}-1}) \nonumber \\
& \stackrel{(c)}{=}  &  I(U_{\tilde{k}+1};Y_{\tilde{k}+1}|{U}_{\tilde{k}})+ (1-\beta) I( U_{\tilde{k}};Y_{\tilde{k}+1}|{U}_{\tilde{k}-1}) \nonumber \\
& \stackrel{(d)}{\geq}  &  I(U_{\tilde{k}+1};Y_{\tilde{k}+1}|{U}_{\tilde{k}})+ (1-\beta) I( U_{\tilde{k}};Y_{\tilde{k}}|{U}_{\tilde{k}-1}) \nonumber \\
& \stackrel{(e)}{=}  & \frac{1}{2}  I(U_{\tilde{k}+1};Y_{\tilde{k}+1}|{U}_{\tilde{k}})+ \frac{1}{2} I( U_{\tilde{k}};Y_{\tilde{k}}|{U}_{\tilde{k}-1})  \nonumber \\
& \stackrel{(f)}{=}  &   I(\bar{U}_{\tilde{k}};Y_{\tilde{k}}| U_{\tilde{k}-1}),
\end{IEEEeqnarray}
where (a) follows by the definition of $\bar{U}_{\tilde{k}}$ and  $\bar{U}_{\tilde{k}+1}$; (b) by the Markov chain \eqref{eq:mm}; (c) by the chain rule of mutual information; (d) by the degradedness of the channel \eqref{eq:mm}; (e) by the definition of $\beta$ in \eqref{eq:beta}; and (f) by \eqref{eq:ubareq}.

We can thus conclude that also when \eqref{eq:condition2} holds, the upper bound on $C(M_1, \ldots, M_K)$ in \eqref{eq:d_inequa} is relaxed if one replaces $(U_1, U_2, \ldots, U_K, X)$ and $({\alpha}_{1}, \ldots,{\alpha}_{K})$  by $(\bar{U}_1, \bar{U}_2, \ldots, \bar{U}_K, \bar{X})$ and $(\bar{\alpha}_1, \ldots, \bar{\alpha}_{K})$.

\section{Achievability Proof for Rate-Memory Pair $(\F(1-\delta_\s), \ 2 \tilde{\Gamma}_2)$}\label{sec:coded_lowerbound}
The following scheme achieves the rate-memory pair 
\begin{equation}\label{eq:rrmm}
R=F(1-\delta_\s) \quad \textnormal{and} \quad \M=2 \tilde{\Gamma}_2.
\end{equation}

Split messages $W_1$ and $W_2$ into two independent submessages
\begin{equation*}
W_d =(W_d^{(1)}, W_{d}^{(2)}) , \quad d\in\{1,\ldots, D\}
\end{equation*}
of rates 
\begin{subequations}\label{eq:rrr}
\begin{IEEEeqnarray}{rCl}
R^{(1)}&:=&F(\delta_\w-\delta_\s)\\
R^{(2)}& :=& F(1-\delta_\w)-\epsilon,
\end{IEEEeqnarray}
\end{subequations}
for an arbitrarily small $\epsilon>0$.

\noindent\textit{Caching Phase:}
Cache the pair 
\begin{equation}
V_1:=\big(W_1^{(1)}, W_{2}^{(1)}, W_1^{(2)}\bigoplus W_2^{(2)}\big)
\end{equation} 
in the weak receiver's cache. 
 \vspace{1mm}
 
 \noindent\textit{Delivery Phase:} Use \emph{piggyback coding}, see subsection~\ref{sec:piggyback}, to send $W_{d_2}^{(1)}$  to the strong receiver and  $W_{d_2}^{(2)}$ to the weak receiver who already has side-information $W_{d_2}^{(1)}$.
 
The strong receiver applies piggyback decoding (for the receiver without side-information), where it in fact decodes both transmitted messages. This way it produces the estimate
\begin{equation}
\hat{W}_{d_2}:=\Big(\hat{W}_{d_2,\textnormal{Rx}2}^{(1)}, \hat{W}_{d_2,\textnormal{Rx}2}^{(2)}\Big).
\end{equation} 

 The weak receiver applies piggyback decoding (for the receiver with side-information), which produces $\hat{W}_{d_2,\textnormal{Rx}1}^{(2)}$. Using its cache content $V_1$, it produces the guess
 \begin{equation}
\hat{W}_{d_1}:=\begin{cases} \Big({W}_{d_1}^{(1)}, \hat{W}_{d_2,\textnormal{Rx}1}^{(2)}\Big) & \textnormal{if } d_1=d_2\\
\Big({W}_{d_1}^{(1)}, \hat{W}_{d_2,\textnormal{Rx}1}^{(2)} \bigoplus W_1^{(2)}\bigoplus W_2^{(2)} \Big),& \textnormal{if } d_1\neq d_2.
\end{cases}
 \end{equation} 

\noindent\textit{Analysis:} By Corollary~\ref{Prop:piggyback} and due to the choice of rates $R^{(1)}$ and $R^{(2)}$ in \eqref{eq:rrr}, the probability of error tends to 0 as the blocklength $n$ tends to infinity. Since $\epsilon>0$ can be chosen arbitrarily close to 0, we have proved achievability of the rate-memory pair in \eqref{eq:rrmm}. 

\section{Proof of Theorem \ref{thm:3}}\label{sec:upper2}
 %On this physically degraded channel, we have $$\Pr[y_1=\Delta \text{ and } y_2=\Delta|X=x]=\delta_\s.$$. 
The first and last terms in \eqref{eq:upper} are special cases of \eqref{thm:converse} with \mw{$k_\w=1$ and $k_\w=0$}, respectively. Here, we prove the second term by showing that for every achievable memory-rate pair $(R, \ \M)$,
\begin{align}
3R&\leq M+(1-\delta_\w)F+(1-\delta_\s)F. \label{feas3}
\end{align}

%We start with the following lemma, whose proof is omitted. % that allows us to exploit the degraded nature of broadcast erasure channels.% and is proved in Appendix \ref{ap-degraded}.
%\begin{lemma}
%\label{lem-degraded}
%The capacity-memory tradeoff $C(M)$ depends on $p(y_1,y_2,\ldots,y_K|x)$ only through its marginals $p(y_1|x), p(y_2|x), \ldots, p(y_K|x)$.
%\end{lemma}
%In our derivations of the upper bounds on $C(M)$ we assume that the BC is physically degraded in the sense that the strong receivers all observe  channel outputs $Y_\s^n:=(Y_{\s,1}, \ldots, Y_{\s,n})$, where
%\begin{subequations}\label{eq:degraded_BC}
%\begin{equation}
%Y_{\s,t}=\begin{cases} x_t, & \textnormal{ with probability } 1-\delta_\s \\
%\Delta, & \textnormal{ with probability } \delta_s
%\end{cases}, 
%\end{equation}
%and the weak receivers all observe  channel outputs $Y_\w^n:=(Y_{\w,1},\ldots, Y_{\w,n})$, where
%\begin{equation}
%Y_{\w,t}=\begin{cases} Y_{\s,t}, & \textnormal{ with probability } \frac{1-\delta_\w}{1-\delta_\s} \\
%\Delta, & \textnormal{ with probability } 1-  \frac{1-\delta_\w}{1-\delta_\s}
%\end{cases}.
%\end{equation}
%\end{subequations}

\mw{Since the capacity-memory tradeoff only depends on the conditional marginal distributions of the channel law~\eqref{eq:channel}, we will assume that the packet-erasure BC is physically degraded. So, for each $t\in\{1,\ldots, n\}$,
\begin{equation}\label{eq:mark}
X_t \to Y_{2,t} \to Y_{1,t}.
\end{equation}}

For all sufficiently large blocklengths $n$, choose caching functions~$\{g_i^{(n)}\}$ as in \eqref{eq:caching}, encoding functions~$f_{\d}^{(n)}$ as in \eqref{eq:encoding}, and decoding functions~$\{\varphi_{i,\d}^{(n)}\}$  as in \eqref{eq:decoding}  so that the probability of worst-case error $P_\textnormal{e}^{\textnormal{worst}}$ tends to 0 as the blocklength $n\to \infty$. 
Consider now a fixed blocklength $n$ that is sufficiently large for the purposes that we describe in the following. Let
\begin{IEEEeqnarray}{rCl}
V^{(n)}_1&= &g_1^{(n)}(W_1, \ldots, W_D),\\ 
X_\d^n&=&f_{\d}^{(n)}(W_1,\ldots, W_D)
  \end{IEEEeqnarray}
  denote cache contents and the input of the packet-erasure BC for a given demand vector $\d\in\mathcal{D}^2$ and for above chosen caching and encoding functions. Also, let $Y_{1, \d}^{n}$ and $Y_{2,\d}^n$  denote the corresponding channel outputs at the weak and strong receivers.
  
We focus on the two demand vectors 
\[ \d_1:=(1,2) \qquad \textnormal{and} \qquad  \d_2:=(2,1).
\] 
So,  $W_1$ \ssb{should be decodable} from $(Y_{1, \d_1}^n, V^{(n)}_1)$ and from $Y_{2, \d_2}^n$,  and $W_2$ \ssb{should be decodable} from $(Y_{1, \d_2}^n ,V^{(n)}_1)$. Thus,  by Fano's inequality, for all $\epsilon_1,\epsilon_2, \epsilon_3>0$ and sufficiently large blocklength $n$, we have
\begin{subequations}
\begin{IEEEeqnarray}{rCl}
n R & \leq & I(W_1; V^{(n)}_1, Y_{1, \d_1}^n)+ n \epsilon_1\\
nR&\leq& I(W_{1};Y_{2,\d_2}^n)+n\epsilon_2\\
nR & \leq & I(W_2; V^{(n)}_1, Y_{1, \d_1}^n, Y_{1, \d_2}^n|W_1) +n \epsilon_3,
\end{IEEEeqnarray}
\end{subequations}
where for the last inequality we also used the independence of messages $W_1$ and $W_2$.

We first develop the second constraint using the chain rule of mutual information and \cite[Lemma 1]{DanaHassibi05}:
\begin{align}  
nR &\leq \sum_{t=1}^nI(W_{1};Y_{\d_2,t}|Y_{2,\d_2}^{t-1})+n\epsilon_2 \nonumber\\
&\leq (1-\delta_\s)\sum_{t=1}^n  I(W_{1};X_{\d_2,t}|Y_{2,\d_2}^{t-1})+n\epsilon_2.% \nonumber \\
%& \leq n(1-\delta_\s) F + n \epsilon_2.
 \label{sum-second}
\end{align}

We then jointly develop the first and the third constraints, where we also define $\epsilon^\prime:=\epsilon_1+\epsilon_3$:
\begin{align}
\lefteqn{2nR }\; \nonumber \\
& {\leq} I(W_{1},W_{2};V^{(n)}_1,Y_{1,\d_1}^n)+I(W_{2};Y_{1,\d_2}^n|W_{1},V^{(n)}_1,Y_{1,\d_1}^n)+n\epsilon^\prime \nonumber\\
&\stackrel{(a)}{\leq}I(W_{1},W_{2};V^{(n)}_1)+I(W_{1},W_{2};Y_{1,\d_1}^n|V^{(n)}_1)\nonumber\\
&\quad+I(W_{2};Y_{2,\d_2}^n|W_{1},V^{(n)}_1,Y_{1,\d_1}^n)+n\epsilon^\prime\nonumber \\
& \leq I(W_{1},W_{2};V^{(n)}_1)+\sum_{t=1}^nI(W_{1},W_{2};Y_{1,\d_1,t}|V^{(n)}_1,Y_{1,\d_1}^{t-1})\nonumber\\
&\quad+\sum_{t=1}^nI(W_{2};Y_{2,\d_2,t}|W_{1},V^{(n)}_1,Y_{1,\d_1}^n,Y_{2,\d_2}^{t-1})+n\epsilon^\prime\nonumber\\
&\leq I(W_{1},W_{2};V^{(n)}_1)+(1-\delta_\w)\sum_{t=1}^nI(W_1,W_{2};X_{\d_1,t}|V^{(n)}_1,Y_{1,\d_1}^{t-1})\nonumber\\
&\quad+(1-\delta_\s)\sum_{t=1}^nI(W_{2};X_{\d_2,t}|W_{1},V^{(n)}_1,Y_{1,\d_1}^n,Y_{2,\d_2}^{t-1})+n\epsilon^\prime\nonumber \\
&\leq I(W_{1},W_{2};V^{(n)}_1)+(1-\delta_\w)\sum_{t=1}^nI(W_{1},W_{2};X_{\d_1,i}|V^{(n)}_1,Y_{1,\d_1}^{t-1})\nonumber\\
&\quad+(1-\delta_\s)\sum_{t=1}^nI(W_{1},W_{2},V^{(n)}_1,Y_{1,\d_1}^n;X_{\d_2,t}|W_{1},Y_{2,\d_2}^{t-1})+n\epsilon^\prime\nonumber\\
&\leq nM+n(1-\delta_\w)F\nonumber\\&\quad+(1-\delta_\s)\sum_{t=1}^nI(W_{2},V^{(n)}_1,Y_{1,\d_1}^n ;X_{\d_2,t}|W_{1},Y_{2,\d_2}^{t-1})+n\epsilon^{\prime}.\label{sum-first}
\end{align}
In $(a)$, we used that the physically degradedness of the channel in \eqref{eq:mark} implies the Markov chain
\begin{equation*}
(W_1, W_2, V^{(n)}_1, Y_{1, \mathbf{d}_1}^n) \to Y_{2, \d_2}^{n} \to Y_{1, \d_2}^n.
\end{equation*}
% fact that $W_{1}$ is decodable from $V_1$ and $Y_{\d_1}^n$. Step $(b)$ holds  by \cite[Lemma 1]{DanaHassibi05} and the following Markov chain:
%\begin{align}
%(Y_{1,\mathbf{d}_1,i},Y_{1,\mathbf{d}_2,i})-X_{\mathbf{d}_1,i}-(W_1,W_2,V_1,Y^{n}_{1,\mathbf{d}_1},Y^{n}_{1,\mathbf{d}_2}).
%\end{align}

Adding up \eqref{sum-second} and \eqref{sum-first} and letting $\epsilon_1, \epsilon_2, \epsilon_3$ tend to 0, we obtain the missing converse bound in \eqref{feas3}, because
\begin{IEEEeqnarray}{rCl}
\lefteqn{I(W_{2},V^{(n)}_1,Y_{1,\d_1}^n ;X_{\d_2,t}|W_{1},Y_{2,\d_2}^{t-1}) +   I(W_{1};X_{\d_2,t}|Y_{2,\d_2}^{t-1}) } \nonumber \\
 & = & I(W_{1},W_{2},V^{(n)}_1,Y_{1,\d_1}^n ;X_{\d_2,t}|Y_{2,\d_2}^{t-1})\hspace{3cm} \nonumber \\
 & \leq& H(X_{\d_2,t})\leq F.
\end{IEEEeqnarray} 

%\input{app-upp-v12}

%%%%
%%%%
%%%%
%%%%

\end{document}